%% file: main.tex
\documentclass[sigconf]{acmart}

\AtBeginDocument{%
  \providecommand\BibTeX{{%
    \normalfont B\kern-0.5em{\scshape i\kern-0.25em b}\kern-0.8em\TeX}}}

\setcopyright{acmcopyright}
\copyrightyear{2018}
\acmYear{2018}
\acmDOI{XXXXXXX.XXXXXXX}

\acmConference[]{}{}{}

\usepackage{amsmath,amssymb,amsfonts}
\usepackage{algorithmic}
\usepackage{graphicx}
\newcommand{\tabincell}[2]{\begin{tabular}{@{}#1@{}}#2\end{tabular}}
\usepackage{textcomp}
\usepackage{xcolor}
\usepackage{pgfplots} 
\usepackage{pgf-pie}
\usepackage{tabularx}
\usepackage{listings}
\usepackage{flushend}
\usepackage{multirow}
\usepackage[normalem]{ulem} 
\newcommand\hl{\bgroup\markoverwith
  {\textcolor{yellow}{\rule[-.5ex]{2pt}{2.5ex}}}\ULon}

\newcommand\past{\bgroup\markoverwith
  {\textcolor{orange}{\rule[-.5ex]{2pt}{2.5ex}}}\ULon}

\newcommand\new{\bgroup\markoverwith
  {\textcolor{lime}{\rule[-.5ex]{2pt}{2.5ex}}}\ULon}

\usepackage[T1]{fontenc}
\input{solidity_listing/solidity_listing}

\usepackage{url}
\usepackage[ruled, linesnumbered]{algorithm2e}
\def\BibTeX{{\rm B\kern-.05em{\sc i\kern-.025em b}\kern-.08em
    T\kern-.1667em\lower.7ex\hbox{E}\kern-.125emX}}

\pgfplotsset{compat=1.18}

\begin{document}
\balance

\title{Generative Software Engineering}

\author{Yuan Huang}
\email{huangyuan5@mail.sysu.edu.cn}
\affiliation{%
  \institution{School of Software Engineering, Sun Yat-sen University}
  \city{Zhuhai}
  \country{China}
}

\author{Yinan Chen}
\email{chenyn273@mail2.sysu.edu.cn}
\affiliation{%
  \institution{School of Software Engineering, Sun Yat-sen University}
  \city{Zhuhai}
  \country{China}
}

\author{Xiangping Chen}
\authornote{*Corresponding authors}
\email{chenxp8@mail.sysu.edu.cn}
\affiliation{%
  \institution{School of Journalism and Communication, Sun Yat-sen University}
  \city{Guangzhou}
  \country{China}
}

\author{Junqi Chen}
\email{chenjq253@mail2.sysu.edu.cn}
\affiliation{%
  \institution{School of Software Engineering, Sun Yat-sen University}
  \city{Zhuhai}
  \country{China}
}

\author{Rui Peng}
\email{pengr27@mail2.sysu.edu.cn}
\affiliation{%
  \institution{School of Computer Science and Engineering, Sun Yat-Sen University}
  \city{Guangzhou}
  \country{China}
}

\author{Zhicao Tang}
\email{tangzhc@mail2.sysu.edu.cn}
\affiliation{%
  \institution{School of Software Engineering, Sun Yat-sen University}
  \city{Zhuhai}
  \country{China}
}

\author{Jinbo Huang}
\email{huangjb63@mail2.sysu.edu.cn}
\affiliation{%
  \institution{School of Software Engineering, Sun Yat-sen University}
  \city{Zhuhai}
  \country{China}
}

\author{Furen Xu}
\email{xufr@mail2.sysu.edu.cn}
\orcid{0000-0001-6258-5233}
\affiliation{%
  \institution{School of Software Engineering, Sun Yat-sen University}
  \city{Zhuhai}
  \country{China}
}

\author{Zibin Zheng}
\email{zhzibin@mail.sysu.edu.cn}
\affiliation{%
  \institution{School of Software Engineering, Sun Yat-sen University}
  \city{Zhuhai}
  \country{China}
}





\input{text/abstract}

\ccsdesc[300]{Software and its engineering~Software maintenance tools}

\keywords{Software Engineering, LLM, Pre-Training Model, Generative Task}



\maketitle

\input{text/introduction}
\input{text/methodology}
\input{text/Challenge}
\input{text/relatedwork}

\input{text/conclusion}
\clearpage




\bibliographystyle{ACM-Reference-Format}
\bibliography{reference}

\end{document}

%% file: solidity_listing/solidity_listing.tex
\usepackage{listings, xcolor}
\definecolor{verylightgray}{rgb}{.97,.97,.97}
\lstdefinelanguage{Solidity}{
  keywords=[1]{anonymous, assembly, assert, balance, break, call, callcode, case, catch, class, constant, continue, constructor, contract, debugger, default, delegatecall, delete, do, else, emit, event, experimental, export, external, false, finally, for, function, gas, if, implements, import, in, indexed, instanceof, interface, internal, is, length, library, log0, log1, log2, log3, log4, memory, modifier, new, payable, pragma, private, protected, public, pure, push, require, return, returns, revert, selfdestruct, send, solidity, storage, struct, suicide, super, switch, then, this, throw, transfer, true, try, typeof, using, value, view, while, with, addmod, ecrecover, keccak256, mulmod, ripemd160, sha256, sha3}, 
  keywordstyle=[1]\color{blue}\bfseries,
  keywords=[2]{address, bool, byte, bytes, bytes1, bytes2, bytes3, bytes4, bytes5, bytes6, bytes7, bytes8, bytes9, bytes10, bytes11, bytes12, bytes13, bytes14, bytes15, bytes16, bytes17, bytes18, bytes19, bytes20, bytes21, bytes22, bytes23, bytes24, bytes25, bytes26, bytes27, bytes28, bytes29, bytes30, bytes31, bytes32, enum, int, int8, int16, int24, int32, int40, int48, int56, int64, int72, int80, int88, int96, int104, int112, int120, int128, int136, int144, int152, int160, int168, int176, int184, int192, int200, int208, int216, int224, int232, int240, int248, int256, mapping, string, uint, uint8, uint16, uint24, uint32, uint40, uint48, uint56, uint64, uint72, uint80, uint88, uint96, uint104, uint112, uint120, uint128, uint136, uint144, uint152, uint160, uint168, uint176, uint184, uint192, uint200, uint208, uint216, uint224, uint232, uint240, uint248, uint256, var, void, ether, finney, szabo, wei, days, hours, minutes, seconds, weeks, years},  
  keywordstyle=[2]\color{teal}\bfseries,
  keywords=[3]{block, blockhash, coinbase, difficulty, gaslimit, number, timestamp, msg, data, gas, sender, sig, value, now, tx, gasprice, origin},  
  keywordstyle=[3]\color{violet}\bfseries,
  identifierstyle=\color{black},
  sensitive=false,
  comment=[l]{//},
  morecomment=[s]{/*}{*/},
  commentstyle=\color{gray}\ttfamily,
  stringstyle=\color{red}\ttfamily,
  morestring=[b]',
  morestring=[b]"
}

%% file: text/abstract.tex
\begin{abstract}


The rapid development of deep learning techniques, improved computational power, and the availability of vast training data have led to significant advancements in pre-trained models and large language models (LLMs). Pre-trained models based on architectures such as BERT and Transformer, as well as LLMs like ChatGPT, have demonstrated remarkable language capabilities and found applications in Software engineering. Software engineering tasks can be divided into many categories, among which generative tasks are the most concern by researchers, where pre-trained models and LLMs possess powerful language representation and contextual awareness capabilities, enabling them to leverage diverse training data and adapt to generative tasks through fine-tuning, transfer learning, and prompt engineering. These advantages make them effective tools in generative tasks and have demonstrated excellent performance. However, there are limitations in the existing literature reviews on the development of pre-trained models and LLMs in the context of software engineering: (1) either only focus on the pre-trained models or only on LLMs, lacking the analysis of the development process from the pre-trained models to LLMs. (2) Software engineering tasks are not distinguished by type. (3) Lack of systematic analysis. In this paper, we present a comprehensive literature review of generative tasks in SE by applying pre-trained models and LLMs. We categorize software engineering generative tasks based on the software development life cycle and summarize the advanced pre-trained models and LLMs involved, as well as the datasets and evaluation metrics used. Additionally, we identify key strengths, weaknesses, and gaps in existing approaches, and propose potential research directions. This review aims to provide researchers and practitioners with an in-depth analysis and guidance on the application of pre-trained models and LLMs in generative tasks within SE.

\end{abstract}

%% file: text/introduction.tex
\section{Introduction}


Pre-trained models such as BERT \cite{devlin2019bert}, Transformer \cite{vaswani2023attention}, and LLMs, leveraging vast and diverse datasets, showcase formidable linguistic capabilities and find widespread applications across various domains, effectively bridging the gap between natural and machine languages\cite{Mosel,niu2022deep}. Pre-trained models based on architectures such as BERT and Transformer, as well as LLMs represented by ChatGPT, have access to vast and diverse training data, showcasing powerful language capabilities comparable to humans and being applied in various domains. With the ability to learn from and generate text from large-scale corpora, powerful pre-trained models and LLMs perform well on various generation tasks \cite{Tang2023TheSO}.


Software engineering is a discipline that focuses  on the development, implementation, and maintenance of software systems \cite{zan-etal-2023-large,10.1145/3572905}. It is also one of the important application areas for pre-trained models and LLMs. Downstream tasks in software engineering can be broadly categorized into two types: classification and generation, which are also the two major types of natural language processing tasks. Pre-trained models and LLMs possess powerful language representation and contextual awareness capabilities, enabling them to leverage diverse training data and adapt to generative tasks through fine-tuning \cite{Ding2023ParameterefficientFO}, transfer learning \cite{10.1007/978-3-030-01424-7_27}, and prompt engineering \cite{brown2020language}. These advantages make them effective tools in generative tasks and have demonstrated excellent performance in the field of software engineering. There have been some literature reviews on the relationship between SE (Software Engineering) and pre-trained models and LLM (Language Models). However, these studies have some limitations: (1) Some studies narrowly focus on either pre-trained models or LLMs, ignoring the development stages from pre-trained models to LLMs. For example, Mosel et.al. \cite{Mosel} focus on the ability of Transformer-based pre-trained models to understand words and sentences in the context of software engineering, while Hou et.al. \cite{hou2023large} lean towards LLM4SE, particularly emphasizing how to use LLM to optimize processes and results. (2) Some studies mix different types of SE tasks, failing to distinguish between the two major types of natural language processing tasks. For example, although Zheng et.al. \cite{zheng2023understanding} divides software engineering tasks into seven types, they do not classify these tasks into classification tasks or generation tasks. (3) Alternatively, some studies only explore the performance of LLM in various SE tasks through empirical experiments, without conducting systematic literature reviews. At the same time, considering the application of LLM in SE generation tasks from pre-trained models is a complex task. It requires considering representative models at different stages of development, understanding the unique features of different models, considering fine-tuning strategies of pre-trained models and prompt engineering of LLM, conducting thorough literature reviews and data analysis, and overcoming challenges in the implementation process. Currently, there is a lack of detailed scrutiny and review of pre-trained models and LLM specifically for generative tasks in software engineering in the existing literature. Therefore, our research aims to analyze and summarize the performance of pre-trained models and LLM in SE generation tasks, and provide development guidance in this direction.


In this paper, we provide a systematic literature review of generative tasks in software engineering (SE) based on pre-trained models and Large Language Models (LLM). By summarizing and comparing state-of-the-art technologies and the developmental history of models, we identify the key strengths, weaknesses, and gaps in existing generative task. Additionally, we propose potential directions for future research. Our review aims to provide researchers and practitioners with a comprehensive analysis and guide on the application of pre-trained models and LLM in generative tasks within SE. 
Our key contributions are as follows:

\begin{itemize}
\item[1] As far as we know, we present the first comprehensive literature review of pre-trained models and Large Language Models (LLM) applied to generative tasks in Software Engineering (SE).

\item[2] We categorize SE generative tasks into requirements generation, code generation, code summarization, test cases generation, patch generation, code optimization, and code translation, providing a detailed summary of each sub-direction.

\item[3] We summarize the datasets and evaluation metrics used for major generative subtasks in SE based on pre-trained models and LLM.

\item[4] We conduct a comparative analysis of the state-of-the-art SE generative task methods utilizing pre-trained models and LLM, along with their performance on publicly available datasets.

\item[5] We outline the key challenges in SE generative tasks addressed by pre-trained models and LLM, and propose potential research directions.
\end{itemize}

The organization of the article is as follows, Section~\ref{methodology} elaborates the method of our literature review, and the subsequent sections specifically review each sub-generation task. Section~\ref{relatedwork} introduced the related work, and Section~\ref{challenges} discussed the challenges faced by the existing methods and potential research directions. And we summarize our work in Section~\ref{conclusion}.

%% file: text/methodology.tex
\section{Methodology}
\label{methodology}

In this section, we will introduce the detailed method regarding literature search, selection, and data analysis, which adhered to the literature review methodology proposed by Kitchenham et.al \cite{kitchenham}. Specifically, for each generative task, we have assigned an experienced researcher to independently search and select literature for the respective generative subtask of software engineering.

\textbf{Literature Search:} To retrieve literature related to pre-trained models, LLM and generative software engineering, we selected 6 search engines (dblp, Google Scholar, arXiv, Elsevier Science Direct, IEEE Xplore Digital Library, and ACM Digital Library). These search engines enable us to quickly find papers published in journals, conferences, and workshops. Additionally, they provide a substantial number of preprints.

Referencing software engineering methodologies, we categorize generative tasks in software engineering into requirements generation, code generation, code summarization generation, test cases generation, patch generation, code optimization, and code translation. Consequently, for each task, we employed different keywords for retrieval, detailed as shown in Table \ref{keywords}. Through searches in the aforementioned databases, we acquired a substantial amount of literature. It should be noted that we discovered some relevant papers that we had previously read did not appear in the search results. Therefore, we conducted a round of inspection of major journals and conferences of software engineering to ensure comprehensive coverage of the literature. Detailed statistics are available in Table \ref{keywords}. Different keywords within the same search engine or the same keywords across different engines may yield numerous duplicate or irrelevant papers. Therefore, manual screening of these papers is necessary.

\begin{table*}[]
\caption{Retrieval keywords and statistics about SE generative tasks using Pre-trained Models and LLMs.}
\begin{tabular}{|c|c|c|c|c|c|c|c|}
\hline
Task & \begin{tabular}[c]{@{}c@{}}Requirements \\ Generation\end{tabular} & \begin{tabular}[c]{@{}c@{}}Code \\ Generation\end{tabular} & \begin{tabular}[c]{@{}c@{}}Code \\ Summarization\end{tabular} & \begin{tabular}[c]{@{}c@{}}Test Cases \\ Generation\end{tabular} & \begin{tabular}[c]{@{}c@{}}Patch \\ Generation\end{tabular} & \begin{tabular}[c]{@{}c@{}}Code \\ Optimization\end{tabular} & \begin{tabular}[c]{@{}c@{}}Code \\ Translation\end{tabular} \\ \hline
\begin{tabular}[c]{@{}c@{}}Search\\ Keywords\end{tabular} & \begin{tabular}[c]{@{}c@{}}Requirements\\ Generation \\ LLMs, \\Requirements\\ Generation \\Pre-trained\\ Model, \\ Requirements \\ Generation\\ Large\\ Language\\ Model,\\ Requirements \\ LLM\end{tabular} & \begin{tabular}[c]{@{}c@{}}Code \\ Generation \\ LLM, \\Code\\ Generation \\Pre-trained\\ Model, Code \\ Generation \\ Large \\ Language \\ Model\end{tabular} & \begin{tabular}[c]{@{}c@{}}Code \\ Sumamrization\\ LLM, \\Code\\ Sumamrization \\Pre-trained\\ Model,\\ Code Comment\\ Generation\\ LLM,\\ Large\\ Language\\ Model Code\\ Summarization\end{tabular} & \begin{tabular}[c]{@{}c@{}}Test Case \\Generation\\LLM, \\Test Case\\ Generation \\Pre-trained\\ Model,\\ Test Case \\Generation\\Large \\ Language \\ Model \end{tabular} & \begin{tabular}[c]{@{}c@{}} Patch \\ Generation\\ LLM, \\Patch\\ Generation \\Pre-trained\\ Model, \\ Automated\\ Program Repair\\ LLM, \\ Automated\\ Program Repair\end{tabular} & \begin{tabular}[c]{@{}c@{}}Code \\ Optimization\\ LLM, \\Code\\ Optimization \\Pre-trained\\ Model,\\ Code\\ Optimization, \\ Code \\ Refinement\end{tabular} & \begin{tabular}[c]{@{}c@{}}Code \\ Translation\\ LLM, \\Code\\ Translation \\Pre-trained\\ Model,\\ Code\\ Translation\end{tabular} \\ \hline
\begin{tabular}[c]{@{}c@{}}Search \\ Result\end{tabular} & 21 & 152 & 46 & 43 & 112 & 12 & 23 \\ \hline
Selected & 6 & 21 & 16  & 17 & 16 & 3 & 12 \\ \hline
\end{tabular}
\label{keywords}
\end{table*}

\textbf{Literature Selection:} The purpose of literature selection is to eliminate duplicate papers returned by search engines and those unrelated to our research content. The definition of generative tasks is explicit, referring to the automated generation of software requirements, code, code summaries, and so on using models. In this paper, these models should be advanced pre-trained models or based on LLM. Therefore, we applied the following criteria to aid in literature selection, retaining only those that satisfy all the criteria:

\begin{itemize}
\item The paper should be written in English.
\item It should not be a student thesis.
\item It should be related to generative software engineering.
\item It should be based on pre-trained or LLM models.
\item The research date should be after 2017.
\end{itemize}

To ensure the accuracy and efficiency of the literature screening process, we employed a closed card sorting method \cite{kitchenham} to categorize the collected literature quickly based on the above criteria into relevant and irrelevant papers. Specifically, for each generative subtask, the researcher responsible for literature collection applied the closed card sorting method and screening criteria to the papers returned by the search engine. In cases where the inclusion of a paper was ambiguous, these papers were compiled separately, and discussions were held to decide whether they met our requirements. Through this process, we ultimately identified 91 papers that met our research criteria.

\textbf{Data Analysis:} 
The statistical data for the screened literature, categorized according to different generative subtasks, are presented in Table \ref{keywords}. It is observed that code generation tasks receive the most attention from researchers, while there is relatively less research on tasks such as requirements generation and code optimization.

With the rapid development of LLM, an increasing number of researchers have started investigating its performance in software engineering tasks. Code generation, being the most popular downstream task in software engineering, has garnered significant attention, leading to numerous publications on evaluation datasets, methods, and related papers. The research on code summarization generation, test case generation, and patch generation is moderate, while there is relatively less research on tasks like requirements generation and code optimization, which only amounts to 6 and 3 publications, respectively.

Next, we will discuss each software engineering generative task in separate subsections.

\input{text/RequirementsGeneration}
\input{text/CodeGeneration}
\input{text/CodeSummarization}
\input{text/TestGeneration}
\input{text/PatchGeneration}
\input{text/CodeOptimization}

\input{text/CodeTranslation}

%% file: text/RequirementsGeneration.tex
\subsection{Requirements Generation}


Requirements engineering (RE)\cite{van2000requirements} is a systematic process in software development or system design that employs methods to identify, analyze, record, and manage project requirements, ensuring alignment with user needs. It involves aspects such as collection, analysis, specification, and validation. 

Presently, Large Language Models (LLMs)  can also be applied in the field of requirements engineering, facilitating automated requirements analysis and streamlining the collection and analysis of information.These models offer intelligent support in the specification, enabling the generation of clear and comprehensive documents and enhancing natural language interaction. Utilizing LLMs for swift prototype development accelerates validation and demonstration, ultimately enhancing overall efficiency.


\subsubsection{Datasets}

To comprehensively evaluate the effectiveness of LLMs in requirement generation tasks, researchers have gathered diverse datasets.


Considering that these datasets are constructed for requirement generation tasks, most are built on virtual application scenarios. These datasets typically include requirements generated by LLMs based on assumed application scenarios and those summarized by domain experts. By comparing them, researchers can explore the effectiveness of large models in the given task.



\textbf{Requirements dataset (user stories)\cite{Dalpiaz2018dataset}}. Dalpiaz et al. \cite{Dalpiaz2018dataset} proposes a dataset comprising 22 product backlogs and 1,679 user stories meticulously curated by the requirements engineering community. The dataset is published as a raw archive and includes 22 text files, each of which is dedicated to a specific product and contains associated user stories. Each story is presented in a single line. As far as we know, there is currently no publicly accessible expert-based annotation that establishes a definitive reference for the various concepts described within this dataset.



\textbf{Ace-design/qualified-user-stories dataset \cite{Arulmohan2023dataset}}.The dataset created by Arulmohan et al. \cite{arulmohan2023extracting} comprises an annotated version of Requirements dataset(user stories) \cite{Dalpiaz2018dataset}, encompassing 22 products and 1,679 user stories. They first start by manually annotating the Requirements dataset using the Doccano platform \footnote{https://doccano.herokuapp.com}. They load the labels (i.e., Persona, Action, Entity, Benefit) and two relations (i.e., triggers, targets) during their processing. To complement the ground truth, they run VN \cite{robeer2016automated} on top of the provided backlogs and extract the concepts by parsing the output files it produced. After this step, they obtain a new annotated dataset containing the results of a domain model extraction on top of the input corpus.

\textbf{Improved version of the PROMISE NFR dataset \cite{hey2020norbert}}. Hey et al. \cite{hey2020norbert}  gathers 249 non-functional requirements from the PROMISE NFR dataset \cite{nfr2007dataset}, spanning 15 projects. Each requirement statement is categorized into one of four NFR classes: usability, security, operational, and performance.

\textbf{App review NFR multi-Label classification dataset \cite{jha2019mining}}. Jha et al. \cite{jha2019mining} presents a dataset comprising 1800 reviews spanning various categories from both Google Play and the Apple App Store. Each review is tagged with at least one NFR label, including dependability, performance, usability, supportability, or a miscellaneous label indicating that the review does not mention NFR.

\subsubsection{Evaluation Criteria}

Requirement generation tasks typically rely on human eval rather than numerical metrics. It is probably because (i) Requirement generation is a complex task, and it is challenging to accurately measure the quality of generated requirements with specific numerical metrics. Human eval allows evaluators to subjectively assess aspects such as the generated requirements' naturalness, clarity, and utility. (ii) Generated requirements often involve broad domain knowledge and semantic understanding, aspects that automated numerical metrics may struggle to capture, including semantic accuracy and coherence of the generated text.


\textbf{Questionnaire Survey}. The Questionnaire survey is a research method for collecting information and opinions by presenting a series of questions to respondents to understand their views, attitudes, behaviors, or experiences. Questionnaire surveys are typically distributed in written form but can also be conducted through face-to-face interviews, telephone calls, or online surveys \cite{lethbridge2005studying}. Waseem et al. \cite{waseem2023using} conducted two rounds of surveys: entry and exit. Both the entry and exit surveys employed a cross-sectional survey design. It is suitable for collecting information at one specific point across a sample population \cite{kitchenham2008personal}. Ultimately, the collected data was the foundation for their research and analysis.


\textbf{Requirement quality assessment}. The requirement quality assessment based on the IEEE 29148:2011 standard \cite{iso2011ieee} is a systematic process designed to ensure that recorded requirements meet specific quality standards and criteria. This process involves evaluating various aspects such as consistency, completeness, clarity, traceability, verifiability, correctness, priority, and understandability. According to Schneider and Berenbach \cite{schneider2013literature}, it encompasses the fundamental principles of requirements engineering and implies how to achieve high-quality requirements. Through a comprehensive assessment of these aspects, the requirements engineering team can gain a holistic understanding of the quality of requirements and take appropriate measures to ensure that the requirement documents provide a clear, consistent, and verifiable foundation for subsequent system development.

\subsubsection{Methods}

For requirements generation, researchers conduct research from the following two aspects.

\begin{figure}[htbp]
\centerline{\includegraphics[scale=0.3]{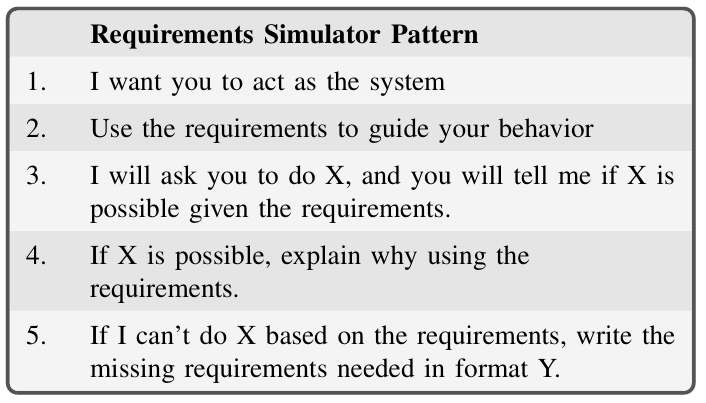}}
\caption{Examples of requirement simulator pattern}
\label{requirement patterns}
\end{figure}

\textbf{Study of using LLMs technology to improve requirements engineering}.White et al. \cite{white2023chatgpt} proposes a heuristic design technique for software engineering, presented in patterns, to address common challenges encountered when automating requirement engineering using LLMs such as ChatGPT. They explored several heuristic patterns that have already been applied to enhance requirements gathering in three different classifications: Requirements Simulator, Specification Disambiguation, and Change Request Simulation(example in Fig. \ref{requirement patterns}). The patterns have already been applied to enhance requirements gathering. Their work also contributes to the research on applying LLMs in requirements engineering.



\begin{figure}[htbp]
\centerline{\includegraphics[scale=0.2]{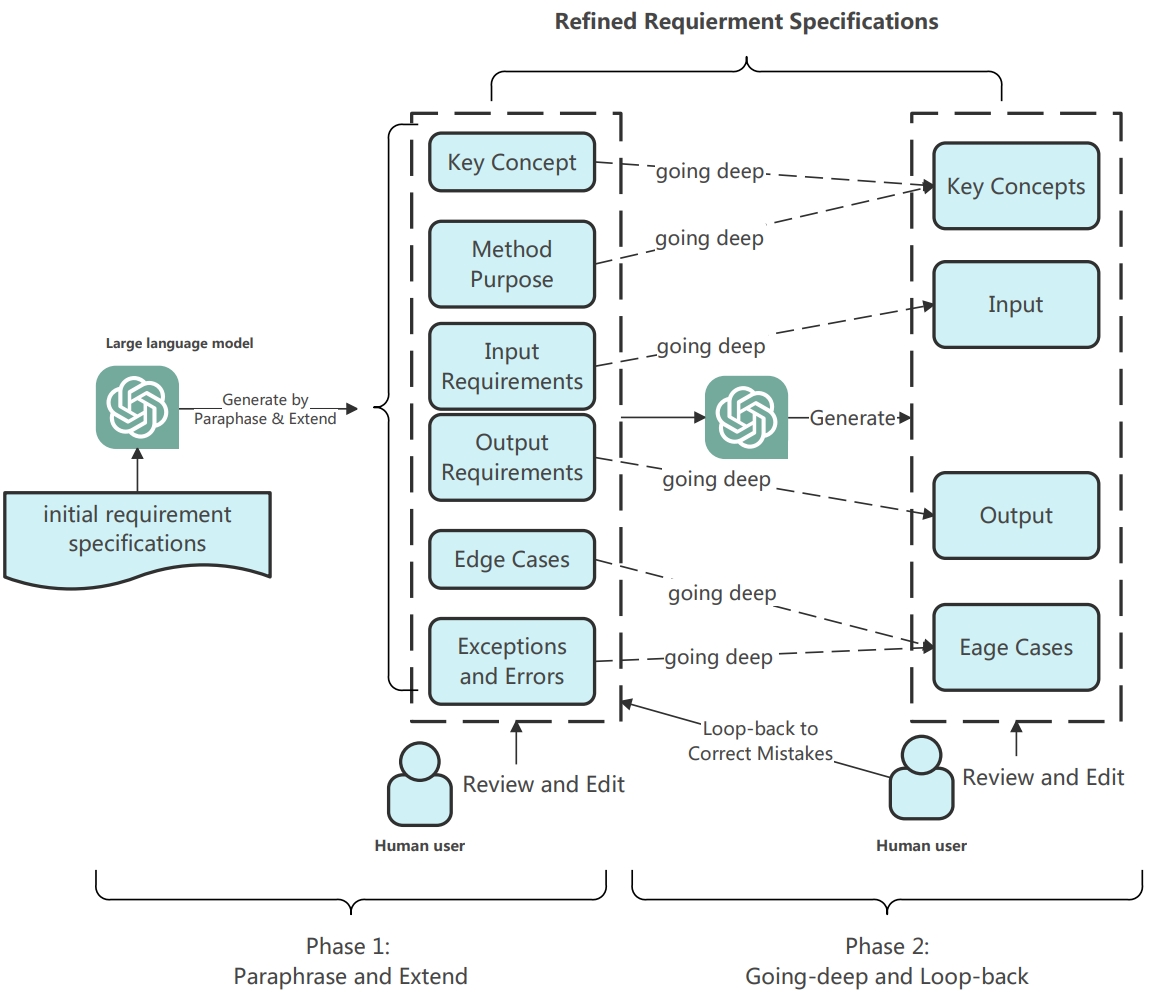}}
\caption{An example of Chatcoder}
\label{chatcoder}
\end{figure}

Wang et al. \cite{wang2023chatcoder} proposes a dialogue framework called Chatcoder between LLMs and users. Within the framework, LLMs analyze arguments to refine the user's original requirement expression, then return the refined arguments to users in an understandable format. The overall structure of ChatCoder comprises a two-round phase (Fig. \ref{chatcoder}). In the first phase, ChatCoder is tasked to rewrite the user's initial demand expression using a preset perspective extracted from existing requirements engineering research. Specifically, ChatCoder packages instructions for rewriting user requirements and preset angles for extension in a prompt to instruct large language models to perform rewriting. The prompt is then sent to the large language model to get response. During the second phase, the chatcoder asks the large language model to further refine the requirements based on the specifications obtained in the first phase. Then, the user is asked to review the refined requirements and check for any possible errors. Finally, the refined requirements are sent to the LLMs to generate the desired programs.

\begin{figure}[htbp]
\centerline{\includegraphics[scale=0.2]{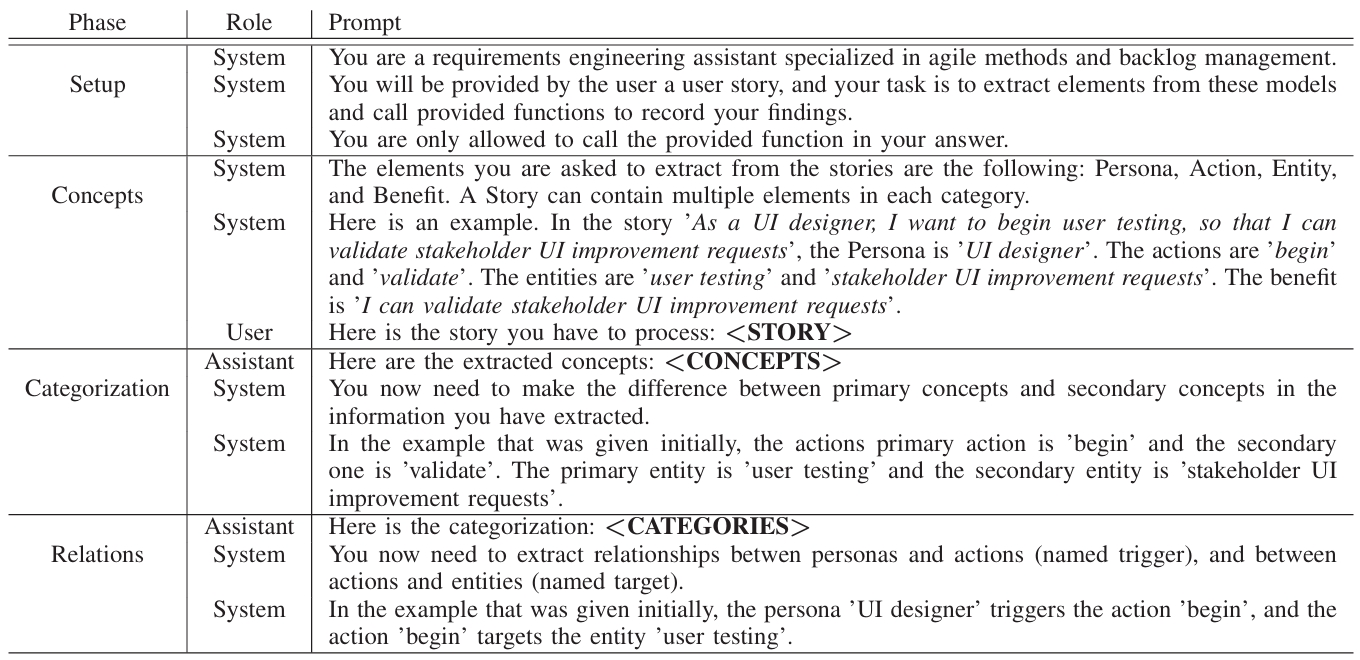}}
\caption{An example of work process for extracting domain model}
\label{Extracting Domain Models}
\end{figure}

Arulmohan et al. \cite{arulmohan2023extracting} proposes to investigate how LLMs and conversational agents can support model extraction from requirements, focusing on agile backlogs. Specifically, the process begins by setting up the system role, where they instruct the engine to adopt a persona and outline the overarching task for later request execution context setup. Moving on to the next phase, they elaborate on extracting personas, entities, actions, and benefits from a story. Subsequently, they present an illustrative example of such extraction using a manually annotated story. Transitioning to the user role, they then supply the <STORY> to be processed. Given the statelessness of the model, the answer from the previous phase needs to be injected into the conversation. Consequently, adopting the assistant role, they introduce a conversation entry detailing the <CONCEPTS> acquired earlier. Following a similar pattern as in the preceding phase, they articulate the task using the system role (categorizing primary and secondary actions and entities) and provide an example of such an execution. The final step mirrors the previous patterns. They inject the <CATEGORIES> as the assistant, followed by task description and example provision in the system role. The conversation they used is described in Fig. \ref{Extracting Domain Models}.

\textbf{Efficiency Analysis of LLMs on requirement engineering}. Arora et al. \cite{arora2023advancing} extensively explores the potential of LLMs in enhancing the efficiency and accuracy of software development RE processes. They specifically focus on applying LLMs in requirements elicitation, analysis, specification, and validation, conducting a thorough study of their preliminary evaluation in real-world systems. The researchers emphasize the advantages of LLMs in addressing ambiguities in requirements, highlighting dependencies and necessary conditions, and managing ethical and regulatory issues related to data storage. Simultaneously, they delve into the potential challenges of over-automation and regulatory issues when executing critical RE tasks, emphasizing the importance of maintaining consistency in terminology and style across requirement documents and balancing the level of detail. Overall, their research indicates that LLMs have the potential to provide continuous refinement of requirements, real-time feedback on quality and consistency, and automation of transforming raw requirements into structured formats, but it is crucial to be aware of their potential need for in-depth domain expertise.


Zhang et al. \cite{zhang4450322evaluation} conducts a comprehensive evaluation of the performance of ChatGPT in the task of requirement information retrieval. This assessment offers profound insights for the design of more efficient retrieval methods. The evaluation results indicate that ChatGPT successfully retrieves relevant requirement information. However, under zero-shot conditions, its precision in retrieving specific requirement information is subject to certain limitations. These findings provide valuable guidance for further optimizing the application of ChatGPT in the field of requirement information retrieval.



Waseem et al. \cite{waseem2023using} provides an in-depth exploration of the effects of students utilizing ChatGPT at different stages of the software development life cycle. Specifically, they provides students with a software project topic for development. During the requirements analysis phase, students gather functional and non-functional requirements with ChatGPT to understand project needs comprehensively. ChatGPT's assistance is crucial for subsequent development stages. Upon completion of the development, Waseem et al.\cite{waseem2023using} makes a survey to evaluate the effectiveness of ChatGPT, revealing a significant reduction in the time required for defining requirements during the requirements analysis phase. Additionally, students respond positively to the utility of ChatGPT, believing that its usage enhances requirement clarity. Also, ChatGPT aids in a better understanding of project needs and facilitates stakeholder communication. To conclude, ChatGPT demonstrated a significant positive impact during the requirements analysis phase of software development projects.

%% file: text/CodeGeneration.tex
\subsection{Code Generation}

In the recent field of natural language processing, large language models (LLMs) have made significant progress in code generation tasks. We have summarized multiple datasets and metrics used to evaluate the performance of LLMs in code generation tasks. Additionally, we have further summarized the information augmentation methods frequently employed by researchers to enhance the performance of LLMs in code generation. Next, we will provide details in three subsections: datasets, metrics, and information augmentation methods.

\subsubsection{Datasets}

To comprehensively assess and enhance the performance of LLMs in the field of code generation, researchers have utilized multiple datasets. These datasets cover testing benchmarks in various aspects, including single programming language, cross-programming languages, cross-prompt languages, and setups with different levels of robustness and complexity. Table \ref{datasets} shows a summary of the datasets.

\begin{table*}[]
\caption{The datasets used to evaluate the performance of LLM code generation.}
\begin{tabular}{|c|c|c|c|}
\hline
Type & Datasets & Languages & Intention \\ \hline
\multirow{2}{*}{\begin{tabular}[c]{@{}c@{}}Single programming\\ language\end{tabular}} & HumanEval & Python & \begin{tabular}[c]{@{}c@{}}To valuate the performance of LLMs in \\ code generation tasks\end{tabular} \\ \cline{2-4} 
 & MBPP & Python & \begin{tabular}[c]{@{}c@{}}To assess the ability of LLM to generate \\ concise Python code\\ from natural language requirements\end{tabular} \\ \hline
\multirow{2}{*}{\begin{tabular}[c]{@{}c@{}}Different programming\\ languages\end{tabular}} & MultiPL-E & \begin{tabular}[c]{@{}c@{}}18 additional programming \\ languages \\ such as Bash, C++, \\ Java, and Go.\end{tabular} & \begin{tabular}[c]{@{}c@{}}To extend LLM code generation benchmark\\ datasets to multiple languages\end{tabular} \\ \cline{2-4} 
 & CodeScope & 43 programming languages & \begin{tabular}[c]{@{}c@{}}To bridge the gap between existing benchmarks\\ and the expectations of multi-language\\ programming environments\end{tabular} \\ \hline
\begin{tabular}[c]{@{}c@{}}Different natural\\ languages\end{tabular} & MCoNaLa & Python & \begin{tabular}[c]{@{}c@{}}To evaluate the performance of LLMs using\\ non-English natural language prompts\end{tabular} \\ \hline
\multirow{2}{*}{\begin{tabular}[c]{@{}c@{}}Different levels of\\ robustness and \\ complexity\end{tabular}} & ClassEval & Python & \begin{tabular}[c]{@{}c@{}}To further investigate the performance of\\ LLMs in generating more\\  complex code\end{tabular} \\ \cline{2-4} 
 & ReCode & Python & To evaluate the robustness of code generation models \\ \hline
\end{tabular}
\label{datasets}
\end{table*}

\textbf{Single programming language.} Initially, to evaluate the performance of LLMs in code generation tasks, researchers have introduced a series of benchmark evaluation datasets. These datasets are typically focused on code generation in the Python programming language.

Chen et.al. \cite{chen2021evaluating} utilized a purely handwritten collection of programming problems called \textbf{HumanEval}\footnote{https://www.github.com/openai/human-eval}. This dataset is based on the Python programming language and consists of 164 programming problems. Each programming problem in HumanEval consists of a function name, a natural language statement describing the function of the function, and unit test cases used to evaluate the generated code. The average number of unit test cases for each programming problem is 7.1. These programming problems cover many types, including language comprehension, reasoning, algorithms, and simple mathematics.

To assess the ability of LLM to generate concise Python code, \textbf{MBPP}\footnote{https://github.com/google-research/google-research/tree/master/mbpp.} \cite{mbpp} (The Mostly Basic Programming Problems) is proposed. The dataset includes 974 short Python programming problems, and the difficulty of each programming problem is set at the entry-level programmer level. Among these problems, 58\% are mathematical calculation, 43\% involve list processing, 19\% require string operations, 9\% process integer sequences, and 2\% are centered on using other data structures. Similar to HumanEval, each programming problem sample consists of a function name, the natural language describing the target code, and three unit test cases used to evaluate the generated code.

\textbf{Different programming languages.} While the designs of HumanEval and MBPP have demonstrated good performance in evaluating LLM code generation tasks, they are limited by being set for a single programming language, specifically assessing code generation performance on Python. However, real-world development and LLM application scenarios indicate that code generation tasks often require adaptation to different programming language environments. To bridge this gap, researchers have introduced several LLM code generation benchmark datasets of different programming languages.

To extend LLM code generation benchmark datasets to multiple languages, Cassano et.al. \cite{mul-e} introduced \textbf{MultiPL-E}\footnote{https://github.com/nuprl/MultiPL-E.}, a system that transforms unit test-driven code generation benchmarks into new programming languages. By employing MultiPL-E, the authors converted two popular Python code generation benchmarks introduced above, HumanEval and MBPP, into 18 additional programming languages such as Bash, C++, Java, and Go. The key feature of MultiPL-E is its ease of extending with different programming languages. To automatically support different programming languages and benchmarks, they took into account the syntactical differences between the target programming language and Python and constructed 18 compilers to translate NL2Code benchmarks, written in Python, into 18 target languages. Each sample of HumanEval and MBPP programming problem after MultiPL-E conversion consists of a function name, a functional natural language description applicable to the target language, and converted target programming language test cases.

Similar to Multiple-E, Yan et.al. \cite{codescope} argue that most benchmarks are flawed as they focus on narrow-scope, popular programming languages, and specific tasks. Additionally, most benchmarks overlook the practical executability of generated code and the consistency of execution results. To bridge these gaps, they propose \textbf{CodeScope}\footnote{https://github.com/ WeixiangYAN/CodeScope.} \cite{codescope}, an execution-based, multi-language, multi-task, multi-dimensional evaluation benchmark for a comprehensive assessment of LLM capabilities in coding tasks. CodeScope encompasses 43 programming languages and 8 coding tasks. For the code generation task, they construct the Codeforces4LLM dataset by collecting data from the popular online algorithm competition platform Codeforces. The dataset includes problem descriptions and corresponding corrected submissions in 14 different programming languages such as C++, Java, Python, and Rust. Two difficulty levels, Easy ([800, 1600)) and Hard ([1600, 2800)), are set for each programming language based on the official difficulty standards of the Codeforces platform. They select 30 problems for each difficulty level, ensuring solutions exist that pass all test samples.

\textbf{Different natural languages.}
Actual development scenarios often involve developers who do not use English as natural language prompts. However, the benchmarks introduced earlier are typically based on English. Code generation is commonly centered around English, posing barriers for developers who are not proficient in English. As a result, researchers have proposed LLM code generation test datasets that utilize different natural language prompts.

Wang et.al. \cite{mconala} proposed a dataset called \textbf{MCoNaLa}\footnote{https://github.com/zorazrw/multilingual-conala.} with different natural languages as prompts. They annotated a total of 896 natural language-code pairs in three languages: Spanish, Japanese, and Russian. Specifically, they engaged annotators fluent in both the respective natural language and Python. The annotators manually collected 341, 210, and 345 natural language-code pairs from the Spanish, Japanese, and Russian subforums of StackOverflow, respectively.

\textbf{Different levels of robustness and complexity.} 
Although various LLM code generation benchmark datasets help to compare different LLMs, the existing evaluation mainly focuses on generating a single code unit (such as a function or statement), emphasizing function or statement-level code generation. However, there remains uncertainty about how LLMs perform in generating more complex code structures. Therefore, some researchers have proposed LLM code generation benchmark datasets that operate in more complex settings. Additionally, other researchers have developed testing benchmarks to assess the robustness of LLM-generated code.

To further investigate the performance of LLMs in generating more complex code, Du et.al. \cite{classeval} proposed a class-level code generation benchmark dataset. They manually build a benchmark for 100 class-level Python code generation tasks, named \textbf{ClassEval}\footnote{https://github.com/FudanSELab/ClassEval.}. Each coding task in ClassEval comprises a natural language description of the target class, a test suite used to validate the correctness of the generated code, and a specification solution serving as a reference implementation for the target class. Typically, LLMs generate code snippets based on the input description and use the provided test suite to verify correctness. They build test cases on two levels: method-level testing and class-level testing. To ensure that the generated code can be effectively checked according to the given test suite, they designed class-level and method-level descriptions for the target code. Method-level testing mainly involves independently calling each method to verify its correctness without calling any other method in the class. Class-level testing mainly evaluates correctness through sequential calls.

Although the LLM code generation model has achieved exciting performance, it is often fragile, as slight modifications to prompts can result in significantly different outputs. Most existing work on robustness in text or code tasks has focused on classification, lacking benchmarks for testing the robustness of code generation. Therefore, Wang et.al. \cite{recode} proposes \textbf{ReCode}, a comprehensive benchmark for evaluating the robustness of code generation models. They tailored over 30 perturbations specifically for code, including document strings, function and variable names, code syntax, and code formatting. These perturbations were applied to the HumanEval and MBPP datasets, forming a robustness benchmark test dataset for code generation.

\subsubsection{Metric}
Unlike the traditional deep learning code generation model, the evaluation of LLM code generation does not use token-based accurate or fuzzy matching metrics such as BLEU, because Rokon et.al. \cite{bleu} found that BLEU has problems in capturing semantic characteristics unique to code. More fundamentally, such metrics cannot evaluate the larger and more complex code solutions given by LLM. Therefore, the researchers proposed a series of new metrics to evaluate the performance of LLM code generation. Next, we will introduce these metrics.

\textbf{Pass@k:} Kural et.al. \cite{passk} proposes the use of unit tests to assess the correctness of generated code. Specifically, they utilize the pass@k metric, generally calculated by generating k candidate code solutions for each programming question and considering it correct if at least one passes the unit test. The pass@k metric is then computed as the proportion of programming questions for which at least one correct answer is generated. However, due to the high variance associated with this calculation, Chen et.al. \cite{chen2021evaluating} redefine a new way to compute pass@k shown in formula \ref{formula1}:
\begin{equation}
\label{formula1}
pass@k:=\underset{Problems}{\mathbf{E} } \left[ 1-\frac{\binom{n-c}{k} }{\binom{n}{k} } \right]      
\end{equation}
where they generate $n\geqslant k$ samples per task, count the number of correct samples $c\leqslant n$ that pass unit tests, and calculate the unbiased estimator.

\textbf{RP$_{s}$@k, RD$_{s}$@k, and RR$_{s}$@k:} To assess the robustness of generated code, Wang et.al. \cite{recode} defined robustness metrics for code generation models in ReCode: Robust Passs@k (RP$_{s}$@k), Robust Drops@k (RD$_{s}$@k), and Robust Relatives@k (RR$_{s}$@k). The calculation methods of these three metrics are shown as formulas \ref{formula2} to \ref{formula4}:

\begin{equation}
RP_{s}@k:=\mathbf{E}_{x} \left[ 1-\frac{\binom{n-rc_{s}(x)}{k} }{\binom{n}{k} } \right]  
\label{formula2}
\end{equation}

\begin{equation}
    RD_{s}@k:=\frac{Pass@k-Robust\  Pass_{s}@k}{Pass@k} 
\end{equation}
\begin{equation}
    RR_{s}@k:=\mathbf{E}_{x} \left[ 2-\frac{\binom{n-rc^{\left[ -\right]  }_{s}(x)}{k} }{\binom{n}{k} } -\frac{\binom{n-rc^{\left[ +\right]  }_{s}(x)}{k} }{\binom{n}{k} } \right]
\label{formula4}
\end{equation}

where, with $s$ random perturbations. For an original prompt $x$ and for each transformation, let the perturbed prompts be $x_{1},\cdot \cdot \cdot ,x_{s}$. They sample $n$ generations by the model for each prompt, and in total there are $n\cdot s$ generations $f_{x}(x_{j})$, where $1\leqslant i\leqslant n,\  1\leqslant j\leqslant s$. They consider the worst-case correctness across $f_{i}(x_{1}),\cdot \cdot \cdot ,f_{i}(x_{s})$ for $1\leqslant i\leqslant n$: Let $c_{i,s}(x)=1$ if $f_{i}(x_{1}),\cdot \cdot \cdot ,f_{i}(x_{s})$ are all correct and $c_{i,s}(x)=0$ otherwise. Let $rc_{x}(x)=\sum^{n}_{i=1} c_{i,s}(x)$. $rc^{\left[ -\right]  }_{s}(x)$ denote the number of correct-to-incorrect changes
under the worst-case measurement as discussed. Symmetrically, $rc^{\left[ +\right]  }_{s}(x)$ denote the number of incorrect-to-correct changes under best-case measurement.

\subsubsection{Methods}

LLM, while demonstrating impressive performance in code generation, has not fully realized its code generation potential. Different prompt types were tested for their impact on LLM code generation by Murr et.al. \cite{murr}, with the prompt format that only provides natural language requirements exhibiting the lowest performance. To further enhance LLM code generation performance, researchers have explored different information utilization methods, which we categorize as guiding and error feedback. In this section, we will elaborate on how information enhancement can be applied to LLM code generation tasks to boost its performance.


\textbf{Requirements-guided code generation:} Powerful capabilities have been demonstrated by language models such as ChatGPT in automatically generating code from provided natural language requirements. However, in real-world practice, user-written requirements may be ambiguous or incomplete. Currently, LLMs struggle to handle such unclear user requirements, and generating code solutions directly from ambiguous requirements may deviate from the user's original intent. To bridge this gap, Mu et.al. \cite{clarifygpt} introduce a method called \textbf{ClarifyGPT}, designed to enhance code generation by having LLM identify ambiguous requirements and propose targeted requirement clarification questions. Specifically, ClarifyGPT first inputs the requirements into LLM to generate multiple solutions. It then executes these solutions with pre-prepared test cases. If the output results of these solutions are consistent, the requirements are considered clear. Conversely, if the solutions output different results, indicating ambiguous requirements, ClarifyGPT guides LLM to generate targeted clarification questions. Upon receiving user responses, ClarifyGPT reconstructs the clarified requirements prompt and inputs it into LLM to generate code solutions.

To further refine user requirements as well, inspired by real-world software development practices where humans often collaborate as teams to handle complex tasks, Dong et.al. \cite{selfcollaboration} proposes a self-cooperative framework for code generation using LLM. Specifically, 1) they employ multiple LLMs acting as different "experts" each responsible for specific subtasks in the overall task; 2) they specify ways of collaboration and interaction so that different roles form a virtual team to facilitate each other's work, ultimately collaboratively solving code generation tasks without manual intervention. Following software development methodologies, they formed a team consisting of three ChatGPT roles (analyst, coder, and tester) responsible for the analysis, coding, and testing phases of software development. The analyst's goal is to formulate a high-level plan, focusing on guiding the coder in writing programs without delving into implementation details. For a given requirement, the analyst decomposes it into several easily solvable subtasks to facilitate the division of functional units and develops an outline of the major steps for implementation. Subsequently, the coder receives the analyst's plan and writes code that meets the specified requirements.

\textbf{Guiding programming thinking:} Programmers typically start implementing code by first writing high-level solution pseudocode, which is then refined into a concrete solution. Simultaneously, when faced with challenging code, most programmers seek similar solutions and improvements on programming forums like Stack Overflow. Inspired by this, researchers have also explored incorporating human programming thinking into Large Language Models.

A large-scale study \cite{prothink} has been published, asserting that programming requires programming thinking, which involves analyzing and implementing the logical requirements of programming (such as sequences, branches, and loops). Li et.al. \cite{TIP} discusses how to unlock programming thinking in LLMs during code generation and proposes a method called \textbf{TIP}. They decompose code generation into two steps, gradually guiding LLMs to analyze and implement the programming logic requirements. Specifically, TIP first generates a code sketch (i.e., pseudocode) that provides a high-level resolution process using programming logic but omits implementation details. Then, TIP transforms the sketch into a program using a specific programming language.

Another practice of introducing programming thinking into LLMs is code reuse. Human developers can identify content in similar code that is relevant to their requirements. This content can be considered a code sketch, which can be further edited into the desired code. Inspired by code reuse, Li et.al. \cite{skcoder} proposed a sketch-based code generation method named \textbf{SKCODER} to mimic developers' code reuse behavior. According to natural language requirements, SKCODER retrieves similar code snippets from programming forums like Stack Overflow, extracts relevant parts as a code sketch, and edits the sketch into the required code.

\textbf{Domain-specific code generation:} In real-world development scenarios, developers often customize APIs or classes in their projects, leading to a demand for code generation in such contexts. For instance, this includes generating code for specific library function calls or handling exceptions that may arise from specific APIs. Researchers have proposed some code generation methods under domain-specific settings.

Liu et.al. \cite{codegen4lib} found that code generation for library functions (i.e. using specific libraries or functions to generate high-quality code) is a more practical code generation scenario through empirical research. Consequently, they proposed a novel library-oriented code generation technique called \textbf{CodeGen4Libs}, comprising two phases: import generation and code generation. Specifically, the import generation phase generates import statements for a given third-party library based on natural language queries, while the code generation phase generates concrete code based on the generated imports and queries. Experiments indicated that their proposed CodeGen4Libs method showed promising results in generating high-quality code using specific libraries, thus enhancing the efficiency and effectiveness of software development.

Another practice of domain-specific code generation is to handle exceptions that may be generated by specific APIs. Due to the lack of robust programming practices, particularly in the domain of exception handling, generating high-quality and reliable code remains a challenging task for LLMs. Ren et.al. \cite{kpc} proposed a new knowledge-driven prompt chain-based code generation method named \textbf{KPC}. They initially constructed an API knowledge base from the Java official documentation, extracting knowledge patterns and conducting knowledge extraction. Subsequently, building upon ChatGPT, they designed a knowledge-driven prompt chain (KPC) code generation method to improve the specific exception-handling ability of code generation. Specifically, their approach involves three steps (generation, check, and rewrite). The generation phase obtains a code snippet solution from the natural language requirement, likely containing defects such as unhandled exceptions. In the check phase, they use the pre-generated API knowledge base to guide the LLM to check potential API exceptions in the generated code. After obtaining a list of exceptions, LLM addresses these in the rewrite phase, providing correct code solutions. Extensive experimental results indicate that the KPC-based approach has significant potential to improve the quality of code generated by LLMs.

\textbf{Error feedback:}
Based on the experience of human developers in debug code with error reporting information, researchers fed back the errors in the code generated by LLM to LLM, so that LLM found the defects in the generated code, and modified the defect code to generate the correct code.

Yan et al. \cite{yanrepair} found that ChatGPT may repair wrong programs by further providing ChatGPT with information about test failures when investigating ChatGPT's zero-shot ability to solve competitive programming problems.

Inspired by code debugging in the human programming process, Zhang et.al. \cite{selfedit} proposes a generation and editing method called \textbf{Self-Edit}, which uses the execution results of code generated by LLM to improve the code quality of competitive programming tasks. Specifically, they first use the programming problem requirements given by competitive programming websites as input-guided LLM generation solutions, then execute the generated code on the sample test cases provided in the problem, and wrap the execution results into supplementary comments. Use this comment to guide LLM to correct errors in the generated code. Dong et.al. \cite{selfcollaboration} obtain the code written by the coders also played by LLM through the testers played by LLM, and then record test reports including functions, readability, and maintainability. The coder repaired and improved the code after receiving test feedback.

%% file: text/CodeSummarization.tex
\subsection{Code Summarization}
A code summary is a concise description of the source code written in natural language, which contribute to enhancing developers' understanding of code \cite{xia2017measuring} and improving developer productivity. In reality, developers often neglect to incorporate an ample amount of comments in their code, primarily due to demanding workloads\cite{he2019understanding}. This has prompted researchers to explore methods for automatically generating comments in code, a process recognized as code summarization.

\subsubsection{Datasets}

\begin{table*}[]
\caption{Code summarization datasets}
\resizebox{\linewidth}{!}{
\begin{tabular}{|l|l|l|}
\hline 
\textbf{DATASET} & \textbf{Language}               & \textbf{Description} \\
\hline
TL-CodeSum       & Java                            &         \tabincell{c}{  A Java code summarization datasets contain 87k pairs  \\which collected from github projects from 2015 to 2016.}        \\
\hline
CodeSearchNet    & Ruby, Go, JS, Python, PHP, Java &            A multillingual dataset contain more than 2M paris          \\
\hline
Funcom           & Java                            &              A dataset contain more 2.1M Java methods.     \\ \hline        
\end{tabular}
}
\end{table*}
To more accurately assess the performance of large-scale language models in code summarization tasks, researchers have collected various datasets. These datasets typically contain a substantial amount of source code along with their corresponding document summaries' first sentences (which often summarize the functionality of the source code). The design of these datasets aims to reflect real-world programming scenarios and provide sufficient information to train and test the models' capabilities. In the following sections, we will introduce several datasets widely used in code summarization tasks.

\textbf{TL-CodeSum (TLC)}\cite{ijcai2018p314} is a dataset contained 87k code-summary pairs which is collected from open-source projects that were developed from 2015 to 2016 with at least 20 stars. To decrease noise introduced to the learning process, they only take the first sentence of the Javadoc comments since it typically describe the functionalities of Java methods.

\textbf{CodeSearchNet (CSN)}\cite{husain2019codesearchnet} is large-scale datasets with over 2M code-summary pairs in 6 programming languages. (Go, Java, Python, Ruby, JavaScript, PHP). CSN also has a filtered version proposed by CodeXGLUE. They filtered out code that couldn't be parsed into abstract syntax trees, documents with a length less than 3 and greater than 256, samples containing special characters (such as <img>), and non-English content. This filtered CSN dataset is now one of the most commonly used datasets for code summarization tasks.

\textbf{Funcom (FCM)}\cite{leclair2019neural} dataset is collected from 51 million Java methods and processed to form 2.1 million code-summary samples. The samples are divided into training, testing, and validation sets based on projects, with a ratio of 9:0.5:0.5, respectively.

\subsubsection{Evaluation Metrics}
There are many metrics used to assess the effectiveness and quality of generated code summaries. These metrics play a crucial role in objectively measuring how well an automatic code summarization model performs.

\textbf{BLEU} (Bilingual evaluation understudy)\cite{papineni2002bleu} is the most common assessment indicator. It measures the quality by comparing the n-gram overlap between the generated text and reference text. The more overlap there is, the higher the BLEU score, indicating better performance. The formula for BLEU is as follows:
$$
BLEU = BP*exp(\sum\nolimits_{n=1}^{N} \frac{1}{N}log\ p_{n}  )
$$
where $BP$ is the brevity penalty to account for shorter translations. $N$ is the maximum n-gram order considered (typically 4). $p_{n}$ is the precision for n-grams, and it is calculated as the number of matching n-grams in the candidate translation divided by the total number of n-grams in the candidate translation.

\textbf{ROUGE} (Recall Oriented Understudy for Gisting Evaluation) \cite{lin2004rouge} is a set of metrics that evolved from recall. ROUGE evaluates the quality of the generated text by comparing it to reference text in terms of overlapping n-grams and other measures. The most commonly used ROUGE metrics include ROUGE-1, ROUGE-2 and ROUGE-L and they are computed as:
$$
ROUGE-N=\frac{\sum\nolimits_{i=1}^{m} Count_{match}(n_{i})}{\sum\nolimits_{i=1}^{m}Count_{reference(n_{i})}} 
$$
where $m$ is the total number of distinct n-grams in the reference. $n_{i}$ represents the $i$-th distinct n-gram. $Count_{match}(n_{i})$ is the count of the $i$-th n-gram in the overlapping set of n-grams between the candidate and reference. $Count_{reference}(n_{i})$ is the count of the $i$-th n-gram in the reference.
$$
R_{lcs} = \frac{LCS(R, H)}{m},\ P_{lcs} = \frac{LCS(R,H)}{n} 
$$
$$
F_{lcs} = \frac{(1+\beta^2)R_{lcs}P_{lcs}}{R_{lcs}+\beta^2P_{lcs} } 
$$
where $R$ and $H$ are the reference and hypothesis respectively. $LCS(R,H)$ is the length of the longest common subsequence (LCS) between $R$ and $H$. $m$ is  the length of the reference and $n$ is the length of the hypothesis. $F_{lcs}$ represents the ROUGE-L $F1$ score. $\beta$ is a weight that adjusts the balance between precision and recall. ROUGE-L measures the longest common subsequence of words between the system and reference summaries.

\textbf{METEOR} (Metric for Evaluation of Translation with Explicit ORdering)\cite{2005METEOR} is a metric considering precision, recall, and the F-measure. It incorporates features such as penalizing stemming differences, recognizing synonymy, and aligning word order to provide a more comprehensive assessment. METEOR's design aims to offer a nuanced evaluation, addressing issues like word variations and order discrepancies in translations. It is computed as:
$$
F = \frac{(\alpha^2 + 1)P}{R +\alpha P} 
$$
$$
Penalty = \gamma(\frac{\#chunks}{\#unigrams\_matched} )^\theta 
$$
$$
METEOR = (1-Penalty)*F
$$
where $F$ calculates the F-measure, which is the harmonic mean of precision and recall, with the parameter $\alpha$ adjusting the balance between the two. $Penalty$ represents the penalty term in the context of the METEOR metric. $\#chunk$ refers to a sequence of contiguous words in a sentence. $\#unigrams_matched$ represents the number of unigrams that are matched between the candidate and reference translations.
\subsubsection{Method}
For code summarization tasks, researchers have proposed a variety of methods.

\textbf{Prompt-based methods\cite{sun2023prompt, sun2023automatic, gao2023makes,  jin2023binary}: } Sun et.al. \cite{sun2023automatic} provides an in-depth analysis of the capabilities of ChatGPT in automatic code summarization. The study focuses on evaluating ChatGPT's performance using the Python dataset from CSN and compares it with several state-of-the-art (SOTA) models. The paper explores how to choose the appropriate prompts to guide ChatGPT in generating concise summaries. They found that the optimal prompt is: \textit{Please generate a short comment in one sentence for the following function: <code>}. The results show that ChatGPT's performance in code summarization, in terms of metrics like BLEU and ROUGE-L, is notably lower than SOTA models. The study also addresses the potential of LLMs in this domain and underscores several challenges and opportunities in utilizing large language models for code summarization. These include designing more efficient prompt, pruning and templating comments generated by ChatGPT, constructing a high-quality dataset, and designing new evaluation metrics.

Gao et.al. \cite{gao2023makes} conduct a research on examineing the effectiveness of in-context learning (ICL) demonstrations in LLMs for code intelligence tasks, focusing on three aspects: selection, order, and number of demonstration examples. The study conducts extensive experiments on three code intelligence tasks: code summarization, bug fixing, and program synthesis. It finds that selecting demonstrations that showcase both similarity and diversity is crucial for ICL in code intelligence tasks. These factors not only boost overall performance but also contribute to more consistent predictions. The sequence in which demonstration examples are presented significantly influences ICL outcomes. Typically, arranging similar samples towards the end of a prompt yields superior results. Moreover, increasing the quantity of demonstration examples can benefit ICL, as long as these examples are not truncated due to the input length constraints of LLMs. This consideration is particularly important given that code tends to be lengthier than natural language.

\textbf{Fine-tuning-based methods\cite{shi2023sotana,su2023distilled, fried2022incoder, wang2023codet5+}: }
Shi et.al. \cite{shi2023sotana} presents SoTaNa, an open-source software development assistant that enhances LLMs for software engineering tasks. SoTaNa uses ChatGPT to generate high-quality, instruction-based data for software engineering and was trained by parameter-efficient fine-tuning methods on LLaMA. They  conduct extensive experiments to demonstrate the capabilities of SoTaNa in effectively
answering Stack Overflow questions, code summarization, and code generation.

While models like ChatGPT demonstrate powerful capabilities in code-related tasks, their status as closed-source models accessible only through APIs is unacceptable for organizations concerned about data privacy. Su et.al. \cite{su2023distilled} focusing on the challenges of using large models like GPT-3.5 due to data custody and closed nature, offering an alternative through knowledge distillation. The study presents three main contributions: comparing GPT-3.5's summaries with human-written references, studying the distillation process across various model sizes and data volumes, and evaluating the distilled model against GPT-3.5 with human experts. The results show that the distilled model, while smaller, can closely mimic GPT-3.5 in code summarization tasks. The paper also discusses the balance between model size, cost, and performance, highlighting the potential of smaller models in reproducing large model capabilities, especially in maintaining data custody.

Simultaneously, many large models specifically designed for code-related tasks have been proposed\cite{feng2020codebert, wang2021codet5, guo2022unixcoder, guo2020graphcodebert, bui2021self, niu2022spt}, and they have achieved excellent performance in code summarization tasks. 

Feng et al. proposed CodeBERT\cite{feng2020codebert} which is the first large NL-PL pre-trained programming languages model. It learns general-purpose representations that support downstream NL-PL applications such as natural language code search, code documentation generation, et. They pre-trained the model on a large NL-PL corpus by using Masked Language Modeling (MLM) and Replaced Token Detection (RTD). The experimental results show that CodeBERT perform well in code summarization.

Wang et.al. present codeT5\cite{wang2021codet5}, which  is a variant of the T5 model, specifically tailored for handling source code. They introduces identifier-aware pre-training tasks, enabling it to distinguish and recover masked code tokens, particularly identifiers. The model supports both code understanding and generation tasks, including defect detection, clone detection, and code summarization. It outperforms prior methods on various tasks, demonstrating its capability in capturing semantic information from code. The research underscores the significance of considering token types in code and exploiting code comments for improved natural language-programming language alignment. 

Guo et.al. present UniXcoder\cite{guo2022unixcoder}, which is a unified cross-modal pre-trained model for programming language. The model leverages multi modal contents, i.e. code comment and AST, to support code-related understanding, generation tasks and auto-regressive tasks. Experimental results show that UniXcoder provides significant improvement on code summarization task.

%% file: text/TestGeneration.tex
\subsection{Test Generation}

As software systems become more complex, manually writing test cases becomes more tedious and error-prone. In order to cope with this challenge, automatic test case generation technology based on large language models (LLM) has emerged in recent years. LLM automatically derives a series of test cases with high coverage through an in-depth understanding of the structure and logic of the software system. We have summarized multiple datasets and
metrics used to evaluate the performance of LLMs in test
generation tasks. Next, we will provide details in three subsections:
datasets, metrics, and methods.

\subsubsection{Datasets}  

In the field of test generation, there are the following widely used benchmark datasets:

\begin{table*}[]
\caption{The datasets used to evaluate the performance of test generation model.}
\begin{tabular}{|c|c|c|}
\hline
 Datasets & Languages & Intention \\ \hline
 \textbf{METHODS2TEST} \cite{tufano2020unit} & Java  & \begin{tabular}[c]{@{}c@{}}A supervised dataset consisting of Test Cases \\and their corresponding Focal Methods \\for test generation task \end{tabular} \\ \hline

 \textbf{CodeContests} \cite{li2022competition} & C++, Python and Java &  \begin{tabular}[c]{@{}c@{}}To evaluate the model's ability to \\generate test in multi-programming languages\end{tabular} \\ \hline

 \textbf{Defects4J version 2.0} \cite{just2014defects4j} & Java & \begin{tabular}[c]{@{}c@{}}To evaluate the performance of \\bug-reproduced by test generation model.\end{tabular} \\ \hline

 \textbf{Atlas-dataset} \cite{watson2020learning} & Java  & \begin{tabular}[c]{@{}c@{}}To give the generation model the ability to \\generate assert statements \\that closely resemble those \\created by developer and evaluate the model. \end{tabular} \\ \hline

\end{tabular}
\label{Test Generation Datasets}
\end{table*}

\textbf{METHODS2TEST} \cite{tufano2020unit} contains 780,944 test cases (method within a test class with the @Test annotation) mapped to their corresponding focal methods (the methods under test), extracted from 9,410 unique Java repositories (91,385 original repositories analyzed). METHODS2TEST splits the dataset in training (80\%), validaiton (10\%), and test (10\%) sets. The split is performed to avoid data leakage at repository-level. Duplicate pairs with same code representation have been removed.

\textbf{CodeContests} \cite{li2022competition} includes problems, solutions and test cases which are written in C++, Python and Java. CodeContests are scraped from the Codeforces platform along with existing public competitive programming datasets. CodeContests have 13,610 methods and each method has about 10 test cases. To avoid data leakage, CodeContests performed a strict temporal split: all pre-training and fine-tuning training data appeared online before any validation problems, and all validation problems before test ones.

\textbf{Defects4J version 2.0} \cite{just2014defects4j}  is a manually curated dataset of real-world bugs from 17 Java projects. Defects4J is a collection of reproducible bugs and a supporting infrastructure with the goal of advancing software engineering research. Each Defects4J bug is paired to a corresponding bug report, which makes the dataset ideal for evaluating the performance of bug-reproduced by test generation model.

\textbf{Atlas-dataset} \cite{watson2020learning} is collected from over 9k projects to extract 2,502,623 examples of developer-written assert statements being used within test methods.  Atlas-dataset is a corpus of test case prefixes, corresponding method units, and assertions. This data can be used to give the generation model the ability to generate assert statements that closely resemble those created by developers. 

\subsubsection{Evaluation Criteria}
The choice of evaluation metrics used for test case generation depends on the specific test objectives, system requirements, and project characteristics. A comprehensive assessment method often involves the comprehensive consideration of multiple indicators. Common evaluation indicators include the following:

\textbf{Code Coverage:} Code coverage measures the extent to which test cases cover the code under test and indicates how many different program paths or statements the generated test oracle or assertions cover. Higher coverage generally indicates more comprehensive testing. It includes metrics such as statement coverage, branch coverage, and condition coverage. Higher code coverage indicates that the test cases cover a significant portion of the code. However, high coverage doesn't guarantee test effectiveness or completeness.

\textbf{Path Coverage:} Path coverage evaluates whether test cases cover all possible execution paths within a software program. High path coverage enhances the thoroughness of test cases, but it may not cover all complex paths.

\textbf{Fault Detection Rate:} Fault detection rate measures the success of test cases in identifying defects or issues within the software system. A high fault detection rate suggests that the test cases are effective in detecting potential problems, though it doesn't guarantee to find all issues.

\textbf{Retesting Rate:} Retesting rate indicates whether running the same test cases multiple times produces consistent results. A low retesting rate implies good repeatability of test cases and stable test results.

\textbf{Test Case Execution Time:} Test case execution time is the time taken for a test suite to complete execution. Assessment: Shorter test case execution times may indicate efficient testing, but it's crucial to balance speed with the quality of test cases.

\textbf{Accuracy:} A measure of how well the generated test oracle or assertions match the actual expected results. High accuracy means that the generated content is more consistent with the actual desired results.

\subsubsection{Method.}   
 Many test case generation methods have been proposed in recently years.
   
\textbf{Methods for pre-training and fine-tuning training based on LLM:}
Tufano et al. \cite{tufano2020unit} proposed ATHENATEST (based on the BART transformer architecture), which aims to generate unit test cases by learning real-world focal-methods and test cases written by developers. They formulate unit test case generation as a sequence-to-sequence learning task, employing a two-step training process that includes denoising pre-training on a large unsupervised Java corpus and supervised fine-tuning of the downstream task of generating unit tests.
Li et al. \cite{li2022competition} proposed a large-scale code pre-training language model AlphaCode (encoder-decoder architecture based on transformer), which was pre-trained on a large-scale code corpus collected by GitHub. In the test case generation task, they fine-tune AlphaCode on the corresponding dataset.

Chen et al. \cite{chen2023codet} proposed a large-scale pre-trained language model CODET (based on the CodeX framework). CODET can not only generate the corresponding solution code based on natural language (i.e. the context of the original problem) but also input the original problem context and instruction into CodeX for generating test cases. Finally, CODET filters the generated solution codes based on the designed scoring rules and generated test cases.

\textbf{Methods based on LLM and prompt engineering:}
Patrick et al. \cite{bareiss2022code} used a large-scale code pre-training language model CodeX and few-shot learning for generating test cases. Their method is mainly divided into three steps: (1)instance extraction (i.e. some examples of test case generation are added to the prompt (i.e. few-shot learning)). (2) enter the designed prompt into the Codex. (3) post-process generated test cases, such as de-duplication and discarding test cases that do not pass compilation.

Siddiq et al. \cite{siddiq2023exploring} explored the ability of three LLMs: CodeGen, Codex and GPT-3.5 to generate test cases. They introduced the tested function and some natural language description information into the designed prompt, and then input the prompt into LLM for generating test cases.
Yu et al. \cite{yu2023llm} explored the application of LLM in the field of mobile application test script generation. They manually obtained comprehensive information of the entire testing process, including but not limited to basic configuration data (AUT package name, test device name, etc.), the details of the operation process (identifier of the target element, description of the operation, etc.), and introduce this information into the prompt. Finally, they input the prompt into ChatGPT to generate the corresponding test case program.

\textbf{Methods based on LLM, prompt engineering and strategy:}
Schäfer et al. \cite{schafer2023empirical} proposed TESTPILOT, an adaptive JavaScript test generation tool based on OpenAI's gpt3.5-turbo LLM, which can automatically generate unit tests for methods in a given project's API. TESTPILOT embeds contextual information about the function under test into the prompt, specifically its signature, additional documentation comments (if any), any usage examples in the project documentation, and the function's source code. Additionally, if a generated test fails, TESTPILOT will adaptively create a new prompt embedding the test and failure messages to guide the model in fixing the problematic test.

Lahiri et al. \cite{lahiri2022interactive} proposed an interactive test case generation framework ITDCG (interactive test driven code generation), which is based on CodeX LLM. ITDCG completes the function body of the given prefix in the file, the natural language description, and the function header/signature containing the method name, parameters and returns according to the user's request, and introduces this information into prompt. Then they input the prompt to LLM for generating a set of candidate codes and tests. ITDCG selects a test and queries the user, asking whether a test is consistent with the user's intent, and then uses the user response to prune, sort, and modify suggestions for existing code sets and test set suggestions. Once the interaction is terminated, ITDCG will output a user-approved set of tests and a ranked list of code suggestions consistent with user responses.

Lemieux et al. \cite{lemieux2023codamosa} proposed the CODAMOSA framework, which comprehensively utilizes CodeX LLM and the retrieval-based method MOSA to generate test cases. Specifically, CODAMOSA starts by running the SBST algorithm MOSA. When it reaches coverage stagnation, it requires a test case for an uncovered function in the test module. At this time, the designed prompt is directly input to CodeX to generate a test case.

Deng et al. \cite{deng2023large} proposed TitanFuzz for testing bugs in deep learning libraries (i.e. TensorFlow and Pytorch), which is a method based on CodeX LLM. TitanFuzz first uses the designed prompt to generate the initial seed program to LLM for fuzz testing. Then, multiple mutation operators and LLM are introduced in the prompt to automatically mutate the seed program and generate a new test program. Next, they designed an adaptation function to prioritize seed or mutation test programs based on data dependency depth and the number of unique library APIs. Finally, they perform differential testing on different backends on the generated test program to detect bugs.

Kang et al. \cite{kang2023large} proposed the LIBRO framework, which is the first automatic reproduction work for general defects. LIBRO builds a prompt based on the defect report, and the information used includes the title and description information of the defect report. In addition, LIBRO also adds test case instances to the prompt (i.e. few-shot learning) to guide CodeX LLM to generate test cases. LIBRO improves the query effect of large models through weighted random sampling and generates multiple test cases as alternative test cases. Finally, LIBRO uses a heuristic strategy to select and sort several test cases generated by large models, thereby giving priority to recommending higher-quality generated results.                                                            
\subsubsection{Method: }

There are many methods regarding the test oracle and assert statements generation in test case.

White et al. \cite{white2020reassert} proposed a method REASSERT to automatically generate JUnit test assertions based on the pre-trained model Reformer. REASSERT first collects test and tested method pairs from the target project via test-to-code traceability links. For each test and tested method pair, we extract assertion statements from the test method and concatenate them to generate a string of assertion statements associated with the method under test. The method and assertion strings under test are then processed into sequences of input and output tokens, called method sequences and assertion sequences respectively. These sequences are used to train the pre-trained model Reformer (a transformer-based model). The resulting sequence of assertions is processed into syntactically correct code that can be inserted directly into the method's test case.

Dinella et al. \cite{dinella2022toga} proposed TOGA (based on CodeBERT pre-trained model), which uses the context of focus methods to infer exceptions and assertion testing oracles. TOGA mainly contains two key modules: the Exceptional Oracle Classifier and the Assertion Oracle Ranker. The Exceptional Oracle Classifier is based on the CodeBERT pre-trained model framework. The model is trained by fine-tuning on the corresponding data, and finally it is decided whether an exception should be thrown based on the developer intent passed through the unit context. The Assertion Oracle Ranker is also based on the CodeBERT pre-trained model framework, which treats oracle reasoning as ranking a small set of possible common oracles. The model was fine-tuned on the ranking of the set of candidate assertions given a test prefix and unit context. Each assertion in the set is sorted and the highest-ranked candidate is selected as the assertion oracle. Finally, TOGA generates the corresponding test using the given test prefix and the inferred assertion oracle.

Tufano et al. \cite{tufano2022generating} proposed a method to generate accurate and useful assertion statements based on the BART transformer model to support developers in writing unit test cases. This method uses two pre-training stages: (1) pre-training: semi-supervised pre-training on a large corpus of English text and code pre-training on the Java source code dataset. (2) fine-tuning on the test cases and the labeled dataset.
Mastropaolo et al. \cite{mastropaolo2021studying} explored the performance of pre-training and fine-tuning for supporting code-related tasks based on T5 LLM. They pre-trained on the CodeSearchNet dataset, where an abstract version of each mined Java method was created using the src2abs tool. For the assertion statement generation task of test cases, they used the open source dataset \cite{watson2020learning} for fine-tuning training.

\textbf{Methods based on LLM, prompt engineering.}
Zhang et al. \cite{zhang2023algo} proposed the ALGO framework, using oracles generated by LLM to synthesize algorithm programs to guide generation and verify their correctness. The ALGO framework mainly contains two components: Coder and Verifier. Verfier is based on ChatGPT. Coder is based on Codex, CodeT, ChatGPT, PG-TD. Coder gets the problem description as a prompt, optionally gets the verification results of the previous generation, and generates a program that solves the problem. Verifier generates a reference oracle, and its output is used to verify that the candidate program generated by Coder. ALGO creates a Verifier once per question and uses it to guide any Coder allowing ALGO to be model agnostic.
Nashid et al. \cite{nashid2023retrieval} proposed the CEDAR framework based on CodeX LLM and few-shot learning, which is applied to two different programming languages including static typing and dynamic typing, and two different tasks (i.e. assertion generation of test cases and program fixes). The main goal of CEDAR is to design an efficient prompt that can help with LLM for different code-related tasks. Particularly, the example of few-shot learning in the prompt is embedded based on the st-codesearch-distilroberta-base pre-trained model and uses the BM-25 algorithm to retrieve. Finally, the designed prompt is input into LLM for assertion generation of test cases and program repair tasks.

%% file: text/PatchGeneration.tex
\subsection{Patch Generation}
Patches are used to fix bugs remaining in the program, and automated patch generation can be understood as automatic program repair (APR). In this section, we will introduce the research on APR from two aspects: datasets and main methods.

\subsubsection{Dataset}
In the field of APR, there are four widely used benchmarks: Defects4J \cite{just2014defects4j}, QuixBugs\cite{lin2017quixbugs}, BugAID\cite{hanam2016discovering} and ManyBugs\cite{le2015manybugs}, as shown in Table \ref{Patch Generation Datasets}.

\begin{table*}[]
\caption{The datasets used to evaluate the performance of LLM patch generation.}
\begin{tabular}{|c|c|c|c|}
\hline
Datasets & Type & Languages & Intention \\ \hline
 \textbf{Defects4J} \cite{just2014defects4j} & Single programming language & Java  & \begin{tabular}[c]{@{}c@{}}A Java patch generation dataset from \\ the bug collection of real-world open-source projects\end{tabular} \\ \hline

 \textbf{QuixBugs}  \cite{lin2017quixbugs} & multi-programming language & Python and Java &  \begin{tabular}[c]{@{}c@{}}A multilingual bug-fixing dataset with \\Python and Java program projects\end{tabular} \\ \hline

 \textbf{BugAID} \cite{hanam2016discovering} &Single programming language& Javascript & \begin{tabular}[c]{@{}c@{}}A Javascript dataset which contains \\ 219 bug fixing change types.\end{tabular} \\ \hline

 \textbf{ManyBugs} \cite{le2015manybugs} & Single programming language & C & \begin{tabular}[c]{@{}c@{}}A dataset used to study automatic program repair, \\containing 185 defects extracted from \\9 large open source C language projects. \end{tabular} \\ \hline

\end{tabular}
\label{Patch Generation Datasets}
\end{table*}

\textbf{Defects4J} includes two versions, namely Defects4J-v1.2 and Defects4J-v2.0. Among them, Defects4J-v1.2 comes from the bug collection of real-world open-source projects, containing 395 known and replicable bugs, each containing a buggy version and a fixed version. It also provides the corresponding test suite that triggers the bug for patch verification. And Defects4J-v2.0 provides 420 additional real-world bugs from 17 Java projects.

\textbf{QuixBugs} is a multilingual bug-fixing dataset with Python and Java program projects. Each language contains 40 small classic algorithms, each algorithm has only one bug in one line. In addition, QuixBug also provides a test suite for triggering bugs.

\textbf{BugAID} is a novel semi-automatic technique for discovering the most prevalent and detectable bug patterns. The authors build a javascript dataset which contains 219 bug fixing change types to evaluate BugAID.

\textbf{ManyBugs} is a dataset used to study automatic program repair, containing 185 defects extracted from 9 large open source C language projects. Each defect has at least one corresponding patch and test case written by the original developer. In total, the MANYBUGS benchmark programs include 5.9 million lines of code and over 10,000 test cases.

\begin{figure}[htbp]
\centerline{\includegraphics[scale=0.5]{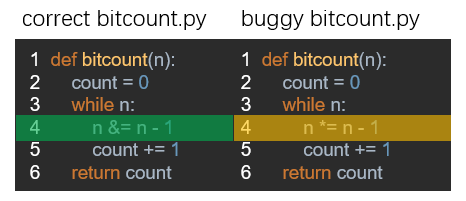}}
\caption{An example of QuixBugs benchmark}
\label{QuixBugs}
\end{figure}

In order to more intuitively display the datasets used in the APR field, Figure \ref{QuixBugs} shows a bug and repair example of bitcount.py from QuixBugs.

\subsubsection{Evaluation Criteria}

When verifying the effectiveness of patch generation methods, qualitative or quantitative indicators are often used to measure them. Evaluating the effectiveness of code patch generation techniques typically involves multiple metrics, and the following are some commonly used evaluation metrics:

\textbf{Repair accuracy:} Repair accuracy refers to the proportion of the number of bugs that the model can correctly fix in the provided defective code test set to the total number of defects in the test set. The calculation method is as follows:
\begin{equation}
\text{RA} = \frac{\text{Number of Successfully Repaired Defects}}{\text{Total Number of Attempted Defects}} \times 100\%
\end{equation}
Where RA is Repair Accuracy. "successfully repaired defects" means that for a defect test suite where each defect may have multiple test cases, only when the generated patch can pass all test cases without introducing new defects, can it be considered as successfully fixing this defect. This usually requires manual verification or additional verification steps to determine. The repair accuracy directly reflects the quality of the generated patches, and high accuracy means that the generated patches are more reliable in repairing defects.

\textbf{Overlap analysis:} Overlap analysis is a metric used in the field of APR to evaluate the diversity and originality of generated patches. During the process of APR, a system may generate multiple patches that fix the same defect. These patches may have overlapping fix strategies, i.e. they may modify the same lines of code or adopt similar fix patterns. Researchers often use this metric to analyze how well their proposed patch generation methods can complement existing methods. If a patch generated by one method can fix more defects that cannot be fixed by other methods, it indicates that it has more general repair capabilities. Overlap analysis is calculated by counting the number of bugs in the overlapping parts of bugs fixed by different technologies and displaying them visually using Venn diagrams.

\textbf{Compilation rate:} During the automated program repair process, a large number of candidate patches may be generated. Not all generated patches can pass compilation. Only patches that can be successfully compiled are likely to be effective fixes, and only successfully compiled patches need to undergo subsequent testing (for example, running test cases to check whether the defect is fixed). The patch compilation rate is a metric used to measure the extent to which the generated patches can be successfully compiled by the compiler. The compilation rate directly affects whether the patch can be further tested and deployed, so it is an important indicator to measure the quality of the patch. The compilation rate of a patch is usually calculated according to the following formula:
\begin{equation}
\text{CR} = \frac{\text{Number of Patches that Compile Successfully}}{\text{Total Number of Generated Patches}} \times 100\%
\end{equation}
Where CR is the Compilation rate.


\subsubsection{APR technology with LLM}
APR aims to help developers improve the reliability of software by automatically generating program patches to repair bugs in software to reduce manual participation in the manual debugging process.

Numerous APR technologies have emerged, categorized into traditional and learning-based methods. Traditional methods, including heuristic rule-based, template matching-based, and constraint-based methods, often use predefined fix templates for bug repair. However, they struggle with bugs outside the defined template. As code volume grows, traditional methods fall short. The rise of deep learning has advanced APR technology. Learning-based methods use deep learning models to generate patches for buggy software, typically using software programs as input. They build a neural network, optimize parameters using a training set, and evaluate the models on a test set. With their ability to learn different bug-to-patch patterns from large code corpora, learning-based methods often outperform traditional APR techniques.




Recently, large language models (LLM) have received increasing attention due to their powerful programming language processing capabilities in various software engineering (SE) tasks. These LLMs are usually trained using pre-training and fine-tuning mechanisms, that is, pre-training through self-supervised training methods on large-scale unlabeled corpora to obtain general knowledge, and finally fine-tuning through supervised training on limited labeled corpora to favor specific downstream tasks. In the field of APR, relevant researchers have tried to directly use LLM to generate correct patches and achieved excellent results. For example, Sobania et al.\cite{sobania2023analysis} evaluated ChatGPT's error repair capabilities and found that ChatGPT was able to fix 31 out of 40 errors on the QuixBugs benchmark. Xia et al.\cite{xia2023keep} used ChatGPT in a conversational manner to fix 114 and 48 bugs on the Defect 4J-v1.2 and Defect 4J-v2.0 benchmarks \cite{just2014defects4j}, as well as all 40 bugs on the QuixBugs benchmark \cite{lin2017quixbugs}, significantly outperforming the state-of-the-art APR technology. This chapter focuses on the technology of using LLM for patch generation in the APR field. 

The implementation of APR techniques using LLM can be mainly divided into three categories: pre-training model methods, LLM-based fine-tuning methods, and prompt engineering methods.

\textbf{LLM-based pre-training methods} aim to train the final model directly based on the APR training corpus. This type of method is more feasible because the model structure, training methods, training corpus, etc. can be modified and innovated. The more representative ones include the following work:

Zhu et al. \cite{zhu2021syntax} propose Recorder, a syntax-guided editing decoder model with placeholder generation. Its model structure is based on TreeGen, which is a syntax-guided code generation model. The Recorder uses the treeGen structure to easily extend non-terminal nodes in the AST to ensure the grammatical correctness of the generated program. Additionally, the Recorder's decoder component generates a sequence of edits rather than generating new code statements, and can copy sub-trees from code containing errors, so small edits can be represented efficiently.

Mashhadi et al. \cite{xia2022less} proposes AlphaRepair, which is the first cloze-style APR method that directly predicts what the correct patch code is based on the contextual information of the bug code (i.e. cloze or text-filling tasks). This method can be built on various pre-trained code models. The author implemented AlphaRepair based on the CodeBERT model. The evaluation results show that AlphaRepair can already outperform the state-of-the-art APR technology on the Defects4J benchmark without any historical bug repair.

Ye et al. \cite{ye2022neural} proposed an automatic patch generation method called RewardRepair. That is because most of the previous APR methods only use the character and token information of the program to optimize network parameters without considering the specific information of the program, which results in the patches generated by the model being of low quality (i.e. not compilable). Therefore,  when RewardRepair uses the loss function to optimize the neural network parameters, program compilation and test execution information are taken into account, so that the model can generate more high-quality patches (that is, compiled patches).

\textbf{LLM-based fine-tuning methods} mainly use the LLM that has been pre-trained on the code corpus to fine-tune the specialized APR corpus, so as to obtain a model that performs well in the APR field and further improves the upper limit of the model in the APR field.

Jiang et al. \cite{jiang2021cure} combines a GPT model pre-trained using a programming language with a translation model (Context-aware neural machine translation (CoNuT)) for patch generation. The pre-trained GPT model mainly learns the paradigm of a large number of correct patches in the pre-training stage, while CoNuT further enhances the learning of correct patches in the fine-tuning stage. This paper also designs a directional search strategy based on Code-aware. By focusing on patches that are compilable and similar to the original length, the search space is narrowed, the compilability rate of the generated patches is improved, and ultimately the high-quality patches are sorted more efficiently.
Mashhadi et al. \cite{mashhadi2021applying} propose an APR method based on CodeBERT. CodeBERT is a transformer-based neural architecture pre-trained on a large corpus of source code. The authors fine-tune CodeBERT on ManyStuBs4J small and large datasets to automatically generate patches.
Jiang et al. \cite{jiang2023impact} proposes to use APR training data to fine-tune CLM (code language model, including codeT5, Incode, codeGEN and other models). Through experiments, they concluded that fine-tuning can bring 31\%–1267\% improvement to CLM and enable them to fix 46\%–164\% more errors than existing DL-based APR technology.
Yuan et al. \cite{yuan2022circle} propose an APR framework based on the T5 model with continuous learning capabilities across multiple programming languages, namely the Continuous Repair Cross-Programming Language Model (CIRCLE). CIRCLE uses the prompt function to narrow the gap between natural language processing (NLP) pre-training tasks and APR tasks, and then obtains the final model by fine-tuning continuous APR corpus on difficult samples. It achieves state-of-the-art performance on the Defects4J benchmark.

\textbf{Prompt engineering method} relies on fine-tuned prompt words to improve the LLM model's ability to patch and repair buggy code. However, since the base LLM model is not further trained, the code processing capability of the base LLM determines the upper limit of the model's ability to perform patch generation.

Xia et al. \cite{xia2023keep} propose ChatRepair, a fully automated conversation-driven APR approach that interweaves patch generation with instant feedback to perform APR conversationally on chatGPT. ChatRepair first provides LLM with relevant test failure information and then learns from the failures and successes of earlier patch attempts for the same bug for more powerful automated program repair capabilities. For early patches that do not pass all tests, the incorrect patch is combined with its corresponding test failure information to build a new prompt for LLM to generate the next patch. In this way, LLM can avoid making the same mistakes and can further draw on its earlier experience of successfully fixing bugs to generate more reasonable patches.
Kolak et al. \cite{kolak2022patch} use Codex as the base model and explores the performance of LLMs of different sizes in patch generation. The results show a significant correlation between model size and test passing accuracy and patch quality rankings.
Prenner et al. \cite{prenner2022can} investigate whether Codex can locate and generate patches, and explore the impact of different prompts on model performance. The final conclusion is that although Codex is not trained with APR, Codex achieves competitive results compared with the state-of-the-art techniques.
Fan et al. \cite{fan2023automated} use Codex, a large language model, as the base model to systematically study whether APR technology can repair the error solutions generated by the language model in the LeetCode competition. The goal of this paper is to study whether APR technology can improve the reliability of code generated by LLMs. This research shows that APR technology has the potential to fix automatically generated code bugs.
Based on prompt engineering technology, Xia et al. \cite{xia2023automated} explore the ability of 9 latest and most advanced LLMs (ranging in size from 125M to 20B) to generate patches under 3 different repair settings designed: 1) Generate the entire patch of the bug, 2) Given the context of the wrong line of bug code, hollow out the bug line code and let the model fill it in, 3) Directly output the single-line repair result.
Sobania et al. \cite{sobania2023analysis} use the method of continuous prompts to evaluate ChatGPT's ability to fix bugs on the standard benchmark——QuixBugs. The results show that the bug repair performance of ChatGPT is competitive with common deep learning methods CoCoNut and Codex, and is significantly better than the reported results of standard program repair methods.

Zhang et al. \cite{zhang2023gamma} proposes GAMMA, an APR technology that combines LLM and repair templates. GAMMA converts the patch generation task into a cloze prediction task and does not require training on repair-defect data pairs. Specifically, GAMMA first uses existing template-based APR technology to extract corresponding repair templates, then converts these repair templates into cloze styles. Finally, based on prompt engineering technology, an existing LLM (ChatGPT) is used to directly predict the correct code fragment in the repair template.







%% file: text/CodeOptimization.tex
\subsection{Code Optimization}

Software optimization refers to improving a program to use fewer resources, such as time, memory, CPU, and energy while retaining its original functionality. There are many methods to achieve software optimization, including algorithm optimization, compiler optimization, and code syntax structure optimization, among others. Generally, this task is performed by developers or code compilers. Developers optimize at the source code level, while compilers optimize at the machine language level. Source code-level optimization is invaluable for developers.

The purpose of code optimization is to improve the time or space complexity of a program without changing its original functionality. The goal is to enhance execution efficiency, thereby making better use of time and hardware resources. For example, Chen et al. \cite{chen2023supersonic} proposed a Supersonic optimization method for C/C++ source code. They used common LLVM as baselines and found that Supersonic improved the program's runtime by 26.0\%, while GPT-3.5-Turbo only achieved 12.0\%, and GPT-4 achieved 4.0\%.

With the development of large models, automatic code optimization techniques have become feasible in recent years. Large models are initially trained on extensive text datasets to obtain universal language representations. This learned representation can be further refined through supervised fine-tuning for various downstream tasks. The input is the original version of the code snippet, such as functions or classes, and the output is LLVM-optimized code. LLM must ensure that the generated code is correct while improving efficiency in terms of time or space. 
Considering these complex requirements, code optimization is an extremely challenging task, as the code generated by current Large Language Models (LLMs) may not even be syntactically correct, let alone show efficiency improvements.

\begin{table*}[]
\caption{The datasets used to evaluate the performance of LLM code optimization.}
\begin{tabular}{|c|c|c|c|}
\hline
Type & Datasets & Languages & Intention \\ \hline
Single programming language & PIE \cite{chen2021evaluating} & C++  & \begin{tabular}[c]{@{}c@{}}To learning performance-improve code\\ edits in C++ \end{tabular} \\ \hline
\multirow{2}{*}{multiprogramming language}  & Supersonic\cite{chen2023supersonic} & C/C++ &  \begin{tabular}[c]{@{}c@{}}To generate source code
level optimizations\\ while retaining significant similarity\\ with the original program\end{tabular} \\ \cline{2-4} 
 & CodeScope \cite{yan2023codescope} & \begin{tabular}[c]{@{}c@{}}43 programming \\ languages\end{tabular} & \begin{tabular}[c]{@{}c@{}}To bridge the gap between existing benchmarks \\ and the expectations of multi-language \\ programming environments\end{tabular} \\ \hline
\end{tabular}
\label{code optimization datasets}
\end{table*}

\subsubsection{Dataset}
There are relevant open-source datasets concerning code optimization, including monolingual datasets (typically for C/C++ due to their high performance, commonly used for low-level algorithm implementations) and multilingual datasets.

In the CodeScope dataset, Yan et al. \cite{yan2023codescope} selected thirty programming tasks for each of four popular programming languages: Python 3, C\#, C, and C++. To ensure diversity in algorithmic and source code syntax solutions for each task, they evaluated the performance of different solutions across various test cases. They chose problem samples with more than 10 correct submissions and over 20 test cases.

Supersonic is a dataset \cite{chen2023supersonic} used data from Codeforces, initially from the CodeNet dataset, and AIZU and AtCoder data originally from CodeNet to construct their target dataset. To ensure relevance in the dataset, they collected submissions from the same author at different times, where each iteration improved the runtime or memory usage compared to the previous one.

Given a problem, programmers often write an initial solution and iteratively improve it. Alexander et al. \cite{madaan2023learning} constructed a Performance Improvement Edits (PIE) dataset based on the CodeNet dataset. They tracked the content submitted by individual programmers over time, filtering for edit sequences corresponding to performance improvements. Finally, they obtained 77,967 pairs for the training set from 1,474 problems, 2,544 pairs for the validation set from 77 problems, and 982 pairs for the test set from 41 problems. For each pair in the test set, they also recorded the fastest human submission execution time for that problem.

\subsubsection{Evaluation Criteria.}

\textbf{Optimization Percentage (\%OPT)}: The percentage of optimized programs in the test set. The programs are optimized if there are improvements in the runtime and memory usage. 

\textbf{Performance Improvement (PI)}: The average improvement in runtime or memory usage of the best-optimized program predicted. 

\textbf{The single-sample optimization percentage, OPT@K}, is based on the assumption that a code sample has the potential for efficiency improvement if the execution efficiency of any optimized code sample exceeds that of the original sample in K optimization attempts. Since the code optimized by LLM may not necessarily compile successfully, the setting of the K value also affects the actual output.

%% file: text/CodeTranslation.tex
\subsection{Code Translation}
The task of code translation involves converting source code from one programming language to another to enhance software portability and maintainability. Manual code translation is highly labor-intensive and expensive. For example, the Commonwealth Bank of Australia spent approximately five years and 750 million dollars to migrate COBOL to Java. With the continuous development of deep learning, efforts have been made towards automatic code translation tasks, requiring models to recognize the features, APIs, and specifications of different programming languages, and assisting programmers in code translation tasks. The input includes source code in a specific language and the target programming language, and the output is code that implements the same functionality in the target programming language. Code translation is a challenging task, even the most sophisticated tools, such as Copilot, have faced criticism due to their robustness, as minor syntax differences can lead to changes in program behavior \cite{10.1145/3491101.3519665}.

\begin{table*}[]
\caption{The datasets used to evaluate the performance of LLM code translation.}
\begin{tabular}{|c|c|c|c|}
\hline
Type & Datasets & Languages & Intention \\ \hline
\tabincell{c}{One-to-one \\code translation} & CodeTrans \cite{lu2021codexglue} & Java/C\#  & \begin{tabular}[c]{@{}c@{}} To foster machine learning research \\for program understanding and generation \end{tabular} \\ \cline{2-4} 
& AVATAR \cite{yang2023assessing} & Java/Python  & \begin{tabular}[c]{@{}c@{}}To assess the effectiveness of CoTR \end{tabular} \\ \hline
\multirow{4}{*}{\tabincell{c}{One-to-many\\ code translation}}  & XCODEEVAL \cite{khan2023xcodeeval} & \begin{tabular}[c]{@{}c@{}}11 programming \\ languages\end{tabular} &  \begin{tabular}[c]{@{}c@{}}To address the challenge of balancing the distributions\\ of text-code samples over multiple attributes\\ in validation/test sets\end{tabular} \\ \cline{2-4} 
& G-TransEval \cite{10298408}  & \begin{tabular}[c]{@{}c@{}}5 programming \\ languages\end{tabular} & \begin{tabular}[c]{@{}c@{}} A benchmark dataset to foster machine learning research\\ for program understanding and generation\end{tabular} \\ \cline{2-4}
& CodeTransOcean \cite{yan2023codetransocean}  & \begin{tabular}[c]{@{}c@{}}24 programming \\ languages\end{tabular} & \begin{tabular}[c]{@{}c@{}}To advance research on code translation and meet\\
diverse requirements of real-world applications\end{tabular} \\ \cline{2-4}
& XLCoST \cite{yan2023codetransocean}  & \begin{tabular}[c]{@{}c@{}}7 programming \\ languages\end{tabular} & \begin{tabular}[c]{@{}c@{}}Facilitate the development
and validation of \\new methods for cross-lingual code intelligenc\end{tabular} \\ \hline

\end{tabular}
\label{code translation datasets}
\end{table*}
\subsubsection{Dataset}
The dataset for code translation comprises implementations of the same functionality in various programming languages. 

Lu et al. \cite{lu2021codexglue} introduced CodeTrans, consisting of code snippets between Java and C\#. Following Nguyen et al. \cite{7372046} and Chen et al. \cite{Chen2018TreetotreeNN}, they gathered data from four open-source projects: Lucene, POI, JGit, and Antlr. These projects were originally developed in Java and later ported to C\#.

Khan et al. introduced XCODEEVAL \cite{khan2023xcodeeval}, containing around 25 million document-level code examples from approximately 7,500 unique problems, covering 11 programming languages with execution-level parallelism. XCODEEVAL includes seven tasks involving code comprehension, generation, translation, and retrieval. It also provides the multi-language code execution engine ExecEval, supporting the execution of unit tests for 11 languages.

Jiao et al. \cite{10298408} proposed a classification, dividing code translation tasks into four main types based on their complexity and knowledge dependency: Token-level (Type 1), Syntax-level (Type 2), Library-level (Type 3), and Algorithm-level (Type 4). Due to challenges faced by existing methods in Type 3 and Type 4, they manually collected translation pairs for these types and constructed a new benchmark, G-TransEval, including unit test cases.

Yan et al. \cite{yan2023codetransocean} introduced the CodeTransOcean dataset, consisting of three novel multilingual datasets: MultilingualTrans for translation between popular programming languages, NicheTrans for translation between less common and popular languages, and LLMTrans for evaluating the executability of code translated by Large Language Models (LLM). CodeTransOcean also includes the cross-framework dataset DLTrans for translating between different deep-learning frameworks.

Zhu et al. \cite{zhu2022xlcost} presented XLCoST, whose data was collected from GeeksForGeeks, a website containing thousands of data structures and algorithm problems. Each problem in XLCoST includes solutions in seven different programming languages: C++, Java, Python, C\#, JavaScript, PHP, and C.
\subsubsection{Evaluation Criteria}

\textbf{Error Rate}: as shown in Table \ref{Taxonomy of code translation errors}, this metric includes translation, language, spurious, code, and documentation omission errors. SLOC represents the total number of lines of code. The minimum value for error rate is 0, but it may exceed 1, as each source code line may have multiple errors \cite{10.1145/3490099.3511157}.

\begin{table*}[]
\caption{Taxonomy of code translation errors \cite{10.1145/3490099.3511157}}
\begin{tabular}{|c|p{10cm}|}
\hline
Error & Description \\ \hline
Translation Error (TE) & Errors in translating code statements, including mistakes in assignment statements, conditional statements, loop conditions, array lookups, spaces, or other logical statements. \\ \hline
Language Error (LE) & 
included Java code snippets within Python or failed to appropriately translate Java language idioms into Pythonic expressions. \\ \hline
Spurious Error (SE) & Participant introduced functionality that was not part of the original Java program, such as by defining new methods. \\ \hline
Code Omission Error (COE) & neglected to translate a method or code statements within a method, providing a trivial implementation (e.g., using pass, return None, print('not implemented'), etc.).\\ \hline

Documentation Omission Error (DOE) & 
failed to provide translation for a function's documentation (e.g., Javadoc comment).\\ \hline

Correctness Error (CE) & translation of a method was inaccurate (e.g., did not pass unit tests).\\
\hline
\end{tabular}
\label{Taxonomy of code translation errors}
\end{table*}

\begin{equation}
\text{Error Rate} = \dfrac{N_{TE} + N_{LE} + N_{SE} + N_{COE} + N_{DOE}}{\text{SLOC}}
\label{error rate}
\end{equation}

\textbf{The proportion of correctly translated methods (PCM)}, is measured based on the number of errors to assess the quality of code translation. Simply put, PCM represents the proportion of methods in the program output that pass unit tests without modification.

\begin{equation}
\text{PCM} = \dfrac{N_{methods} - N_{CE}}{N_{methods}}
\label{CM}
\end{equation}

\subsubsection{Method}
Pan et al. \cite{pan2023stelocoder} introduced SteloCoder, a pure decoder model based on StarCoder \cite{li2023starcoder}. SteloCoder modifies the StarCoder model architecture using the Mixture-of-Experts (MoE) technique to achieve code translation from multiple programming languages to Python. SteloCoder can perform code conversion from languages such as C++, C\#, JavaScript, Java, or PHP to Python without specifying the input programming language.

The study by Yang et al. \cite{yang2023assessing} identified the current limitations of pre-trained models, including large language models, in the task of code translation. CoTR was proposed to enhance the robustness of pre-trained models in code translation tasks.

Pan et al. \cite{pan2023understanding} conducted a groundbreaking study that classified errors introduced by Large Language Models (LLMs) during the code translation process. They identified 14 common error categories present in unsuccessful translations. The researchers proposed heuristic approaches to provide appropriate context to LLMs, aiming to improve the efficiency of their code translation.

However, other related works \cite{lu2021codexglue, khan2023xcodeeval, 10298408, yan2023codetransocean, zhu2022xlcost} opt for assessing the latest popular LLMs (Large Language Models). For instance, models like ChatGPT, Starcoderbase, CodeLlama, codeT5, and others are chosen for evaluation, their main contribution is the introduction of datasets for the code translation task, encompassing both one-to-one and one-to-many datasets. Similar to Code Optimization, code translation remains a challenging task in the current context.

%% file: text/Challenge.tex
\section{Challenge}
\label{challenges}


\textbf{Requirement Generation:} using Large Language Models (LLMs) for requirement generation is an innovative field aimed at integrating advanced natural language processing technology into the domain of requirements engineering. The goal is to enhance automation, alleviate human workload, and provide more accurate and professional requirement texts. We have summarized some challenges faced by this technology, including:
\begin{itemize}
\item [1)] Challenges in Data and Annotation: Existing evaluation datasets are predominantly created in fictional scenarios, limiting LLMs to generating requirements in simulated settings. Consequently, the understanding capabilities of LLMs for requirement documents cannot be accurately validated across different studies and scenarios.
\item [2)] Uncertainty in Handling Technical Information: In practical requirements engineering, LLMs need to process a significant volume of technical documents and requirements in the validation stage of large complex systems. However, uncertainty remains regarding the extent to which LLMs can accurately handle various levels of technical information.
\item [3)] Research on Prompt Guideline Development: Current prompt guidelines include domain-specific and general guidelines. A distinction can be made between those that are more general and those specific to particular domains. However, leveraging more general guidelines in developing guidelines or approaches for specific task domains may reduce susceptibility to fallacies, as the initial prompt guidelines have already laid the foundation. However, integrating general, non-domain-specific prompt guidelines into new prompt guidelines for LLMs tasks in specific domains, such as requirements engineering, remains a precise question.
\end{itemize}
To overcome these challenges and further advance research on LLMs in the requirement generation domain, here are some future research directions:
\begin{itemize}
\item [1)] Explore and Propose New Datasets: Explore more realistic and diverse datasets for requirement documents better to simulate the complexity and diversity in natural engineering environments.
\item [2)] Investigate and Enhance LLMs' Handling of Technical Information: Conduct detailed empirical research to evaluate the practical capabilities of LLMs in handling technical documents and requirements within large complex systems. Additionally, explore fine-tuning strategies to enhance LLMs' ability to process technical information in specific domains.
\item [3)] In-Depth Research on Prompt Guideline Development: Delve into the applicability of general guidelines in different domains to determine how they can be better integrated into domain-specific LLM tasks. Formulate best practices for prompt guideline development to ensure that new prompt guidelines can better withstand fallacies when faced with tasks in different domains.
\end{itemize}


\textbf{Code Generation:} using pre-trained models or LLMs for code generation is a field with potential and challenges. Existing techniques can be categorized into (1) guided approaches: using various techniques and methods to guide the learning and generation process of LLMs, improving the accuracy, performance, and quality of code generation, and (2) error-feedback approaches: providing error feedback to LLMs to help them discover and improve defects in the generated code. This includes self-editing and self-collaboration methods, using execution results and test reports to guide LLMs in code debugging and improvement. We summarize some challenges faced by these techniques, including but not limited to
\begin{enumerate}
    \item [1)] Data and annotation challenges: To train and evaluate the code generation capability of LLMs, existing evaluation datasets require manually constructed test cases, which require a large amount of high-quality data and corresponding annotations. Acquiring and annotating such data can be time-consuming and challenging tasks. Lack of domain knowledge: Although LLMs perform well in language understanding and generation, they may lack in-depth knowledge of specific domains, such as the usage of specific APIs and functions, handling specific exceptions, etc. For domain-specific code generation tasks, LLMs may require additional training and guidance to ensure that the generated code complies with domain best practices and constraints.
    \item [2)] Lack of domain knowledge: Although LLMs perform well in language understanding and generation, they may lack in-depth knowledge of specific domains, such as the usage of specific APIs and functions, handling specific exceptions, etc. For domain-specific code generation tasks, LLMs may require additional training and guidance to ensure that the generated code complies with domain best practices and constraints.
    \item [3)] Assurance of code quality and generation robustness: The generated code may have quality and performance issues, such as vulnerable code. Improving the quality of generated code and ensuring generation robustness are important challenges.
\end{enumerate}

To overcome these challenges and further advance research in LLMs for code generation, here are some feasible future research directions: 
\begin{enumerate}
    \item [1)] Data augmentation and automatic annotation: Develop more effective methods to acquire and annotate large-scale code generation data. This can include rule-based data augmentation techniques, code transformations, and automatic annotation methods to alleviate the burden of data acquisition and annotation.
    \item [2)] Introduction of domain knowledge: Explore how to incorporate domain-specific knowledge into the training and generation process of LLMs to improve code quality and compliance with domain requirements. This can include guidance from domain experts, domain-specific training data, and the construction of knowledge bases.
    \item [3)] Quality and performance assurance: Develop techniques and methods to ensure that the generated code has high quality and good performance. This can involve the application of code analysis and verification techniques, as well as automatic correction and rewriting based on constraints and best practices.
\end{enumerate}

 These research directions contribute to overcoming the current challenges in LLM code generation and further advancing the field.

\subsubsection{Challenge and Future}
\textbf{Challenge}\\
\textbf{Language Specificity:} Each programming language has its own syntax, semantics, and coding conventions. Models trained on datasets for one language may not generalize well to others due to differences in language structure and usage patterns. Thus, separate datasets are required for each language, increasing the demand for diverse datasets across multiple programming languages.\\

\textbf{Project Specificity:} Software projects are project-specific and may involve specialized terminology, libraries, or frameworks. Generating accurate summaries for such code requires models that can understand and incorporate project-specific knowledge. Adapting generic code summarization techniques to unique project while ensuring relevance and accuracy is a non-trivial task. \\

\textbf{Evaluation Metrics:} Assessing the quality of generated code summaries remains an open challenge. Traditional metrics like BLEU, ROUGE and METEOR used in natural language generation tasks may not fully capture the effectiveness of code summaries in aiding developer understanding and productivity. Developing robust evaluation metrics that account for the unique characteristics of code summarization is essential for advancing the field.

\textbf{Future}
\textbf{Customized Summarization: }Since different developers may have varying preferences, expertise levels, and information needs, personalized code summarization techniques can help tailor summaries to individual users. Future research could explore methods for incorporating user feedback, preferences, and historical interactions to personalize the summarization process.  \\

\textbf{Multi-Granularity Summarization: }Codebases consist of code at various levels of granularity, including functions, classes, modules, and entire projects. Generating summaries that are informative and coherent across different levels of granularity poses a challenge. Future research could investigate techniques for multi-granularity summarization, where summaries are generated at different levels of abstraction and detail to cater to different information needs and use cases. This could involve hierarchical summarization approaches, abstraction mechanisms, and content selection strategies tailored to different levels of code granularity.

\textbf{Explainable Summarization: }Investigating methods for generating explainable summaries that not only describe what the code does but also provide insights into why certain decisions were made during its implementation. This involves providing explanations or rationales alongside code summaries to aid developer understanding.

\textbf{Test Generation: }The current challenges of using large language models (LLM) in the field of test generation mainly include the following aspects:
\begin{itemize}
\item [1)] Generation quality: The test cases generated by LLM may not be accurate enough to cover all functions and boundary conditions, thus affecting the effectiveness and comprehensiveness of the test.

\item [2)] Interpretability: The internal working principle of LLM is very complex, and it is difficult to understand the basis and logic for generating test cases, which may make it difficult to judge the effectiveness and accuracy of test cases.

\item [3)] Training data: In order for LLM to perform well in the field of test generation, a large amount of high-quality training data is required. However, collecting and collating this data can be time-consuming and costly.

\item [4)] Computing resources: Training and running large language models requires a large amount of computing resources, which may limit their widespread application in the field of test generation.

\item [5)] Security and reliability: Since LLM may generate test cases that do not comply with specifications, it is necessary to ensure that the test cases it generates will not have a negative impact on the system under test.
\end{itemize}

In order to overcome these challenges and further promote LLM research in the field of test generation, feasible research directions can be explored from the following aspects in the future:
\begin{itemize}
\item [1)] Improve generation quality: Research more advanced generation strategies and algorithms to improve the quality and coverage of generated test cases. For example, you can try to incorporate domain knowledge and expert experience into the model to generate test cases that are more in line with actual needs.

\item [2)] Improve interpretability: Research methods to improve model interpretability so that users can better understand and evaluate generated test cases. For example, you can try to use interpretable machine learning methods or provide additional explanation information for the generated test cases.

\item [3)] Improve training data: Research methods to automatically collect and organize high-quality training data to reduce the difficulty and cost of data preparation. For example, you can try to leverage an existing test case library or automatically extract test cases from open-source projects.

\item [4)] Reduce computing resource requirements: Research methods to reduce model complexity and computing resource requirements to make the application of LLM in the field of test generation more universal. For example, you can try to use knowledge distillation or model compression techniques to reduce the model size.

\item [5)] Improve safety and reliability: Research methods to ensure the safety and reliability of generated test cases and reduce risks to the system under test. For example, you can try to use formal methods to verify whether the generated test cases comply with the specification, or use fault injection techniques to evaluate the effectiveness of the test cases.

\item [6)] Combine with other testing technologies: Combine LLM with other testing technologies (such as model checking, symbolic execution, etc.) to give full play to their respective advantages and improve the effect of test generation. For example, LLM can be used to generate more targeted test cases based on symbolic execution.

\end{itemize}


\textbf{Code optimization}: LLM faces several challenges in code optimization, and addressing these challenges is the primary difficulty in achieving effective code optimization:
\begin{enumerate}
    \item [1)] Increased Complexity: With the continuous development of software systems, the scale and complexity of the code are increasing. Managing large codebases and ensuring optimizations do not introduce errors is a complex task. 
    \item [2)] Uncertainty in Performance Prediction: When optimizing, accurately predicting the impact of different optimization strategies on the final performance is challenging. Sometimes, certain optimizations may be effective in specific scenarios but may not work in other cases.
    \item [3)] Dynamic Nature of Code: Code for some applications may be dynamically generated or modified at runtime, adding to the complexity of optimization. Optimizations may need to consider this dynamic nature.
\end{enumerate}

To address the challenges that LLMs face in code optimization, the following research directions can be considered:
\begin{enumerate}
    \item [1)] Knowledge Fusion: LLMs can enhance their understanding of code by fusing knowledge from multiple domains. This may involve incorporating insights from static analysis, dynamic execution information, and domain-specific expertise.
    \item [2)] Adaptive Optimization Strategies: Given that different code segments may exhibit varying performance on different environments and hardware, LLMs can integrate adaptive optimization strategies. This allows generated code to adjust optimizations based on different conditions automatically.
    \item [3)] Generating Maintainable Code: In addition to optimizing for performance, LLMs can also focus on generating code that is more readable and maintainable. This helps alleviate the burden on developers and reduces the cost of code maintenance.
\end{enumerate}

\textbf{Code translation:} there are the following challenges to consider:
\begin{enumerate}
    \item [1)] The diversity of programming languages presents a complex challenge in accurately translating code, as each language has its own syntax, semantics, and idiomatic expressions. Accurately translating between different languages is an intricate task.
    \item [2)] Handling domain-specific constructs: Code often includes domain-specific constructs and libraries that may not have direct equivalents in the target language. Adjusting such structures during the translation process requires an in-depth understanding of specific domains.
    \item [3)] Dealing with dynamic features: Some programming languages support dynamic features such as reflection, runtime code generation, and metaprogramming. Accurately translating such dynamic features poses a complex challenge.

\end{enumerate}

To address the challenges that LLMs face in code translation, the following research directions can be considered:
\begin{enumerate}
    \item [1)] Addressing Dynamic Features: Considering the dynamic features of certain programming languages, research is needed on how to better support the translation of dynamic code, including runtime code generation and reflection.
    \item [2)] Fine-grained Domain Support: Developing models to support a broader range of programming languages while providing more precise translation within specific domains, such as embedded systems or web development. Integrating domain-specific knowledge to enhance the model's understanding of domain conventions and best practices can improve translation quality.
\end{enumerate}

%% file: text/relatedwork.tex
\section{Related work}
\label{relatedwork}

Mosel et.al. \cite{Mosel} revealed the ability of the Transformer-based pre-trained model to understand words and sentences in the context of software engineering. They compared the BERT Transformer model trained based on software engineering data with the Transformer based on general data: their vocabulary, their ability to understand which words are missing, and their performance in classification tasks. The results show that it is valuable to use software engineering data for tasks that need to understand the background of software engineering. At the same time, when the pre-training model is used for SE tasks, they recommend using a large amount of SE data for pre-training large NLP models.

Niu et.al. \cite{niu2022deep} outlines the development and use of pre-trained code models to enable various SE tasks to achieve the most advanced results, and suggests future research directions. They outlined 20 CodePTMs (code pre-trained models) in the SE community and compared their similarities and differences, as well as their comparative advantages and disadvantages. Relevant research includes the design, development, maintenance, testing, and evolution of software systems. They classify each task according to two dimensions: (1) whether the task involves understanding or generation; and (2) the input type and output type (I-O) of the task. This survey will stimulate the interest of the following groups (1) NLP researchers, especially those who focus on text summary and generation, because many SE tasks (such as code summaries) involve NL (natural language) generation; (2) applied machine learning researchers, because the development of these models may have a great impact on SE. At the same time, the research may also arouse the high interest of SE technology providers and raise their awareness of the added value that artificial intelligence technology may have in enhancing SE tools.

Li et.al. \cite{Liyao} carried out the first comprehensive empirical study on the over-interpretation of PLM applied in SE tasks. Over-interpretation is when a model confidently makes a decision without salient features, or a model finds some irrelevant relationships between the final decision and the dataset. The research consists of two parts, task-oriented and model-oriented. In the task-oriented section, they studied three SE tasks, code search, code summary, and duplicate error report detection. In the model-oriented part, they studied three famous PLMs, generating pre-training (GPT), BERT, and XLNet. The experimental results show that the pre-trained language model is insensitive to the given input, that is, over-interpretation. They also studied two ways to reduce over-interpretation: whole word mask strategy and ensembling. These two methods can enrich model learning to mitigate overinterpretation.

Hou et.al. \cite{hou2023large} carried out a systematic literature survey on LLM4SE, paying special attention to how to use LLM to optimize the process and results. They collected and analyzed 229 research papers from 2017 to 2023, focusing on the following four key research issues: (1) Different LLM categories used in SE tasks, describing their unique characteristics and uses. ( 2) Analyze the methods used for data collection, pre-processing, and application (3) investigate the strategies for optimizing and evaluating LLM performance in SE. (4) Review the specific SE tasks of LLM's success so far, indicating their contribution in this field. From these surveys, they discussed the most advanced technologies and trends, identified the gaps in existing research and the key challenges encountered in the field of using LLM in the field of SE, and proposed several potential research directions for LLM4SE.

Zheng et.al. \cite{zheng2023understanding} proposed a comprehensive review of the research and products combining LLMs with software engineering and explored whether LLM can help better perform the current software engineering tasks. They divided software engineering tasks into seven types. For each category, they provided application examples of LLM, exposing its capabilities and limitations to help researchers better identify and solve potential challenges when applying LLM to software engineering tasks. They also revealed some future directions worth discussing: (1) LLM's current performance on some SE tasks is not good enough or unstable; (2) Most of the current evaluations are based on general large models, such as ChatGPT, but lack a detailed evaluation of large code-centric models such as Codex; (3) Is it necessary to customize large models for specific software engineering tasks?

Zheng et.al. \cite{zheng2023survey} discussed the performance and value of LLM related to software engineering. They collected and screened 134 works related to code LLM, revealing the relationship between code LLM and general LLM. In addition, they made a comprehensive analysis of the performance of general LLM and code LLM in software engineering tasks, and made a detailed analysis of their performance in different sub-tasks. Their research not only helps Code LLM developers choose basic models for the development of more advanced LLM but also provides practitioners with insights to better understand the key improvement direction of Code LLM.

Shin et.al. \cite{shin2023prompt} investigated the effectiveness of the most advanced LLM (i.e. GPT-4). They used different prompt methods (such as basic, context learning, and specific task prompts) to evaluate GPT-4 on three typical SE tasks (i.e. code generation, code summary, and code translation) and compare it with the existing 18 fine-tuned LLMs. They also conducted user studies on 27 participants from academia and 10 participants from the industry to evaluate the quality and usefulness of the GPT-4 response and how participants designed dialogue prompts to guide the model. The results show that GPT-4 can be better than the baseline in comment generation and code translation from C\# to Java, but not as good as the baseline in code generation and code translation from Java to C\# when using basic prompts. They also found that task-specific prompts can significantly improve the performance of GPT-4, and the effect of context learning is mixed. In addition, participants tend to request improvements, add more context, or provide specific instructions as dialogue prompts, which can help GPT-4 produce a better response. In general, GPT-4 has great potential in software engineering tasks, but it also requires careful verification and explanation by manual evaluators.

%% file: text/conclusion.tex
\section{conclusion}
\label{conclusion}
This article comprehensively reviews the development stage of generative tasks in software engineering from the pre-trained model to LLM. We divide the software engineering generation task into 7 sub-tasks: requirements generation, code generation, code summarization, test generation, patch generation, code optimization, and code translation. We have collected and screened works and literature related to these subtasks. It summarizes the development stage and the existing methods, datasets, and evaluation metrics. We also summarize the existing challenges of each subtask and put forward some feasible future research directions for different subtasks.

%% file: main.bbl

\begin{thebibliography}{126}


\ifx \showCODEN    \undefined \def \showCODEN     #1{\unskip}     \fi
\ifx \showDOI      \undefined \def \showDOI       #1{#1}\fi
\ifx \showISBNx    \undefined \def \showISBNx     #1{\unskip}     \fi
\ifx \showISBNxiii \undefined \def \showISBNxiii  #1{\unskip}     \fi
\ifx \showISSN     \undefined \def \showISSN      #1{\unskip}     \fi
\ifx \showLCCN     \undefined \def \showLCCN      #1{\unskip}     \fi
\ifx \shownote     \undefined \def \shownote      #1{#1}          \fi
\ifx \showarticletitle \undefined \def \showarticletitle #1{#1}   \fi
\ifx \showURL      \undefined \def \showURL       {\relax}        \fi
\providecommand\bibfield[2]{#2}
\providecommand\bibinfo[2]{#2}
\providecommand\natexlab[1]{#1}
\providecommand\showeprint[2][]{arXiv:#2}

\bibitem[Arora et~al\mbox{.}(2023)]%
        {arora2023advancing}
\bibfield{author}{\bibinfo{person}{Chetan Arora}, \bibinfo{person}{John Grundy}, {and} \bibinfo{person}{Mohamed Abdelrazek}.} \bibinfo{year}{2023}\natexlab{}.
\newblock \showarticletitle{Advancing Requirements Engineering through Generative AI: Assessing the Role of LLMs}.
\newblock \bibinfo{journal}{\emph{arXiv preprint arXiv:2310.13976}} (\bibinfo{year}{2023}).
\newblock


\bibitem[Arulmohan et~al\mbox{.}(2023)]%
        {arulmohan2023extracting}
\bibfield{author}{\bibinfo{person}{Sathurshan Arulmohan}, \bibinfo{person}{Marie-Jean Meurs}, {and} \bibinfo{person}{S{\'e}bastien Mosser}.} \bibinfo{year}{2023}\natexlab{}.
\newblock \showarticletitle{Extracting domain models from textual requirements in the era of large language models}. In \bibinfo{booktitle}{\emph{2023 ACM/IEEE International Conference on Model Driven Engineering Languages and Systems Companion (MODELS-C)}}. IEEE, \bibinfo{pages}{580--587}.
\newblock


\bibitem[Austin et~al\mbox{.}(2021)]%
        {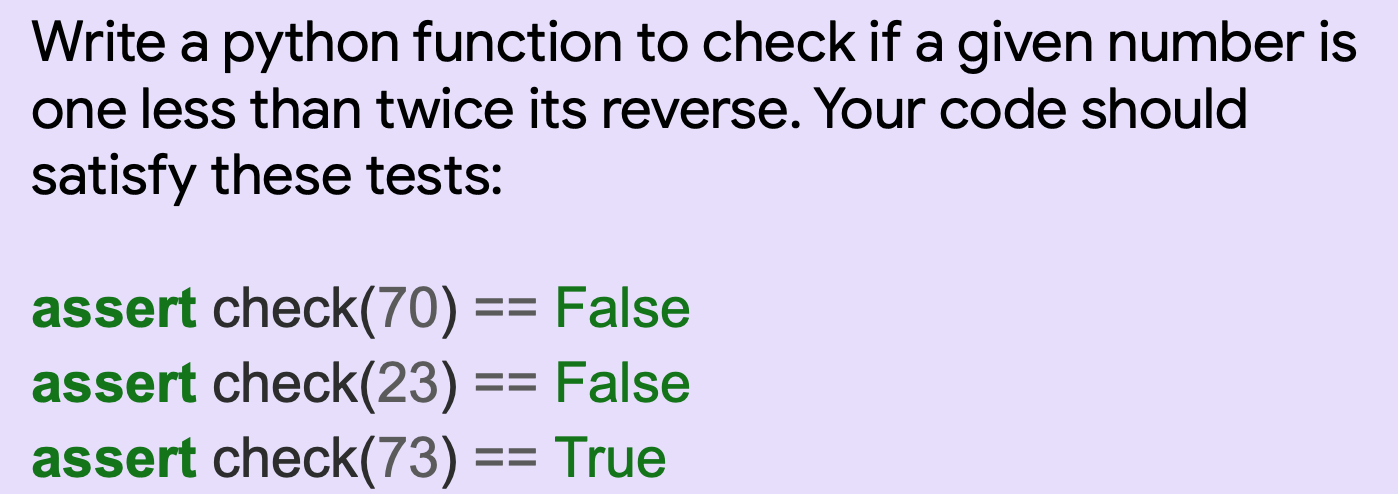}
\bibfield{author}{\bibinfo{person}{Jacob Austin}, \bibinfo{person}{Augustus Odena}, \bibinfo{person}{Maxwell Nye}, \bibinfo{person}{Maarten Bosma}, \bibinfo{person}{Henryk Michalewski}, \bibinfo{person}{David Dohan}, \bibinfo{person}{Ellen Jiang}, \bibinfo{person}{Carrie Cai}, \bibinfo{person}{Michael Terry}, \bibinfo{person}{Quoc Le}, {and} \bibinfo{person}{Charles Sutton}.} \bibinfo{year}{2021}\natexlab{}.
\newblock \bibinfo{title}{Program Synthesis with Large Language Models}.
\newblock
\newblock
\showeprint[arxiv]{2108.07732}~[cs.PL]


\bibitem[Banerjee and Lavie(2005)]%
        {2005METEOR}
\bibfield{author}{\bibinfo{person}{Satanjeev Banerjee} {and} \bibinfo{person}{Alon Lavie}.} \bibinfo{year}{2005}\natexlab{}.
\newblock \showarticletitle{{METEOR:} An Automatic Metric for {MT} Evaluation with Improved Correlation with Human Judgments}. In \bibinfo{booktitle}{\emph{Proceedings of the Workshop on Intrinsic and Extrinsic Evaluation Measures for Machine Translation and/or Summarization@ACL 2005, Ann Arbor, Michigan, USA, June 29, 2005}}, \bibfield{editor}{\bibinfo{person}{Jade Goldstein}, \bibinfo{person}{Alon Lavie}, \bibinfo{person}{Chin{-}Yew Lin}, {and} \bibinfo{person}{Clare~R. Voss}} (Eds.). \bibinfo{publisher}{Association for Computational Linguistics}, \bibinfo{pages}{65--72}.
\newblock


\bibitem[Barei{\ss} et~al\mbox{.}(2022)]%
        {bareiss2022code}
\bibfield{author}{\bibinfo{person}{Patrick Barei{\ss}}, \bibinfo{person}{Beatriz Souza}, \bibinfo{person}{Marcelo d'Amorim}, {and} \bibinfo{person}{Michael Pradel}.} \bibinfo{year}{2022}\natexlab{}.
\newblock \showarticletitle{Code generation tools (almost) for free? a study of few-shot, pre-trained language models on code}.
\newblock \bibinfo{journal}{\emph{arXiv preprint arXiv:2206.01335}} (\bibinfo{year}{2022}).
\newblock


\bibitem[Brown et~al\mbox{.}(2020)]%
        {brown2020language}
\bibfield{author}{\bibinfo{person}{Tom~B. Brown}, \bibinfo{person}{Benjamin Mann}, \bibinfo{person}{Nick Ryder}, \bibinfo{person}{Melanie Subbiah}, \bibinfo{person}{Jared Kaplan}, \bibinfo{person}{Prafulla Dhariwal}, \bibinfo{person}{Arvind Neelakantan}, \bibinfo{person}{Pranav Shyam}, \bibinfo{person}{Girish Sastry}, \bibinfo{person}{Amanda Askell}, \bibinfo{person}{Sandhini Agarwal}, \bibinfo{person}{Ariel Herbert-Voss}, \bibinfo{person}{Gretchen Krueger}, \bibinfo{person}{Tom Henighan}, \bibinfo{person}{Rewon Child}, \bibinfo{person}{Aditya Ramesh}, \bibinfo{person}{Daniel~M. Ziegler}, \bibinfo{person}{Jeffrey Wu}, \bibinfo{person}{Clemens Winter}, \bibinfo{person}{Christopher Hesse}, \bibinfo{person}{Mark Chen}, \bibinfo{person}{Eric Sigler}, \bibinfo{person}{Mateusz Litwin}, \bibinfo{person}{Scott Gray}, \bibinfo{person}{Benjamin Chess}, \bibinfo{person}{Jack Clark}, \bibinfo{person}{Christopher Berner}, \bibinfo{person}{Sam McCandlish}, \bibinfo{person}{Alec Radford}, \bibinfo{person}{Ilya Sutskever},
  {and} \bibinfo{person}{Dario Amodei}.} \bibinfo{year}{2020}\natexlab{}.
\newblock \bibinfo{title}{Language Models are Few-Shot Learners}.
\newblock
\newblock
\showeprint[arxiv]{2005.14165}~[cs.CL]


\bibitem[Bui et~al\mbox{.}(2021)]%
        {bui2021self}
\bibfield{author}{\bibinfo{person}{Nghi~DQ Bui}, \bibinfo{person}{Yijun Yu}, {and} \bibinfo{person}{Lingxiao Jiang}.} \bibinfo{year}{2021}\natexlab{}.
\newblock \showarticletitle{Self-supervised contrastive learning for code retrieval and summarization via semantic-preserving transformations}. In \bibinfo{booktitle}{\emph{Proceedings of the 44th International ACM SIGIR Conference on Research and Development in Information Retrieval}}. \bibinfo{pages}{511--521}.
\newblock


\bibitem[Cassano et~al\mbox{.}(2023)]%
        {mul-e}
\bibfield{author}{\bibinfo{person}{Federico Cassano}, \bibinfo{person}{John Gouwar}, \bibinfo{person}{Daniel Nguyen}, \bibinfo{person}{Sydney Nguyen}, \bibinfo{person}{Luna Phipps-Costin}, \bibinfo{person}{Donald Pinckney}, \bibinfo{person}{Ming-Ho Yee}, \bibinfo{person}{Yangtian Zi}, \bibinfo{person}{Carolyn~Jane Anderson}, \bibinfo{person}{Molly~Q Feldman}, \bibinfo{person}{Arjun Guha}, \bibinfo{person}{Michael Greenberg}, {and} \bibinfo{person}{Abhinav Jangda}.} \bibinfo{year}{2023}\natexlab{}.
\newblock \showarticletitle{MultiPL-E: A Scalable and Polyglot Approach to Benchmarking Neural Code Generation}.
\newblock \bibinfo{journal}{\emph{IEEE Transactions on Software Engineering}} \bibinfo{volume}{49}, \bibinfo{number}{7} (\bibinfo{year}{2023}), \bibinfo{pages}{3675--3691}.
\newblock
\urldef\tempurl%
\url{https://doi.org/10.1109/TSE.2023.3267446}
\showDOI{\tempurl}


\bibitem[Chen et~al\mbox{.}(2023b)]%
        {chen2023codet}
\bibfield{author}{\bibinfo{person}{Bei Chen}, \bibinfo{person}{Fengji Zhang}, \bibinfo{person}{Anh Nguyen}, \bibinfo{person}{Daoguang Zan}, \bibinfo{person}{Zeqi Lin}, \bibinfo{person}{Jian-Guang Lou}, {and} \bibinfo{person}{Weizhu Chen}.} \bibinfo{year}{2023}\natexlab{b}.
\newblock \showarticletitle{CodeT: Code Generation with Generated Tests}. In \bibinfo{booktitle}{\emph{The Eleventh International Conference on Learning Representations}}.
\newblock
\urldef\tempurl%
\url{https://openreview.net/forum?id=ktrw68Cmu9c}
\showURL{%
\tempurl}


\bibitem[Chen et~al\mbox{.}(2021)]%
        {chen2021evaluating}
\bibfield{author}{\bibinfo{person}{Mark Chen}, \bibinfo{person}{Jerry Tworek}, \bibinfo{person}{Heewoo Jun}, \bibinfo{person}{Qiming Yuan}, \bibinfo{person}{Henrique~Ponde de Oliveira~Pinto}, \bibinfo{person}{Jared Kaplan}, \bibinfo{person}{Harri Edwards}, \bibinfo{person}{Yuri Burda}, \bibinfo{person}{Nicholas Joseph}, \bibinfo{person}{Greg Brockman}, \bibinfo{person}{Alex Ray}, \bibinfo{person}{Raul Puri}, \bibinfo{person}{Gretchen Krueger}, \bibinfo{person}{Michael Petrov}, \bibinfo{person}{Heidy Khlaaf}, \bibinfo{person}{Girish Sastry}, \bibinfo{person}{Pamela Mishkin}, \bibinfo{person}{Brooke Chan}, \bibinfo{person}{Scott Gray}, \bibinfo{person}{Nick Ryder}, \bibinfo{person}{Mikhail Pavlov}, \bibinfo{person}{Alethea Power}, \bibinfo{person}{Lukasz Kaiser}, \bibinfo{person}{Mohammad Bavarian}, \bibinfo{person}{Clemens Winter}, \bibinfo{person}{Philippe Tillet}, \bibinfo{person}{Felipe~Petroski Such}, \bibinfo{person}{Dave Cummings}, \bibinfo{person}{Matthias Plappert}, \bibinfo{person}{Fotios Chantzis},
  \bibinfo{person}{Elizabeth Barnes}, \bibinfo{person}{Ariel Herbert-Voss}, \bibinfo{person}{William~Hebgen Guss}, \bibinfo{person}{Alex Nichol}, \bibinfo{person}{Alex Paino}, \bibinfo{person}{Nikolas Tezak}, \bibinfo{person}{Jie Tang}, \bibinfo{person}{Igor Babuschkin}, \bibinfo{person}{Suchir Balaji}, \bibinfo{person}{Shantanu Jain}, \bibinfo{person}{William Saunders}, \bibinfo{person}{Christopher Hesse}, \bibinfo{person}{Andrew~N. Carr}, \bibinfo{person}{Jan Leike}, \bibinfo{person}{Josh Achiam}, \bibinfo{person}{Vedant Misra}, \bibinfo{person}{Evan Morikawa}, \bibinfo{person}{Alec Radford}, \bibinfo{person}{Matthew Knight}, \bibinfo{person}{Miles Brundage}, \bibinfo{person}{Mira Murati}, \bibinfo{person}{Katie Mayer}, \bibinfo{person}{Peter Welinder}, \bibinfo{person}{Bob McGrew}, \bibinfo{person}{Dario Amodei}, \bibinfo{person}{Sam McCandlish}, \bibinfo{person}{Ilya Sutskever}, {and} \bibinfo{person}{Wojciech Zaremba}.} \bibinfo{year}{2021}\natexlab{}.
\newblock \bibinfo{title}{Evaluating Large Language Models Trained on Code}.
\newblock
\newblock
\showeprint[arxiv]{2107.03374}~[cs.LG]


\bibitem[Chen et~al\mbox{.}(2018)]%
        {Chen2018TreetotreeNN}
\bibfield{author}{\bibinfo{person}{Xinyun Chen}, \bibinfo{person}{Chang Liu}, {and} \bibinfo{person}{Dawn~Xiaodong Song}.} \bibinfo{year}{2018}\natexlab{}.
\newblock \showarticletitle{Tree-to-tree Neural Networks for Program Translation}. In \bibinfo{booktitle}{\emph{Neural Information Processing Systems}}.
\newblock
\urldef\tempurl%
\url{https://api.semanticscholar.org/CorpusID:600040}
\showURL{%
\tempurl}


\bibitem[Chen et~al\mbox{.}(2023a)]%
        {chen2023supersonic}
\bibfield{author}{\bibinfo{person}{Zimin Chen}, \bibinfo{person}{Sen Fang}, {and} \bibinfo{person}{Martin Monperrus}.} \bibinfo{year}{2023}\natexlab{a}.
\newblock \showarticletitle{Supersonic: Learning to generate source code optimisations in c/c++}.
\newblock \bibinfo{journal}{\emph{arXiv preprint arXiv:2309.14846}} (\bibinfo{year}{2023}).
\newblock
\urldef\tempurl%
\url{https://doi.org/10.48550/arXiv.2309.14846}
\showURL{%
\tempurl}


\bibitem[Deng et~al\mbox{.}(2023)]%
        {deng2023large}
\bibfield{author}{\bibinfo{person}{Yinlin Deng}, \bibinfo{person}{Chunqiu~Steven Xia}, \bibinfo{person}{Haoran Peng}, \bibinfo{person}{Chenyuan Yang}, {and} \bibinfo{person}{Lingming Zhang}.} \bibinfo{year}{2023}\natexlab{}.
\newblock \showarticletitle{Large language models are zero-shot fuzzers: Fuzzing deep-learning libraries via large language models}. In \bibinfo{booktitle}{\emph{Proceedings of the 32nd ACM SIGSOFT international symposium on software testing and analysis}}. \bibinfo{pages}{423--435}.
\newblock


\bibitem[Devlin et~al\mbox{.}(2019)]%
        {devlin2019bert}
\bibfield{author}{\bibinfo{person}{Jacob Devlin}, \bibinfo{person}{Ming-Wei Chang}, \bibinfo{person}{Kenton Lee}, {and} \bibinfo{person}{Kristina Toutanova}.} \bibinfo{year}{2019}\natexlab{}.
\newblock \bibinfo{title}{BERT: Pre-training of Deep Bidirectional Transformers for Language Understanding}.
\newblock
\newblock
\showeprint[arxiv]{1810.04805}~[cs.CL]


\bibitem[Dinella et~al\mbox{.}(2022)]%
        {dinella2022toga}
\bibfield{author}{\bibinfo{person}{Elizabeth Dinella}, \bibinfo{person}{Gabriel Ryan}, \bibinfo{person}{Todd Mytkowicz}, {and} \bibinfo{person}{Shuvendu~K Lahiri}.} \bibinfo{year}{2022}\natexlab{}.
\newblock \showarticletitle{Toga: A neural method for test oracle generation}. In \bibinfo{booktitle}{\emph{Proceedings of the 44th International Conference on Software Engineering}}. \bibinfo{pages}{2130--2141}.
\newblock


\bibitem[Ding et~al\mbox{.}(2023)]%
        {Ding2023ParameterefficientFO}
\bibfield{author}{\bibinfo{person}{Ning Ding}, \bibinfo{person}{Yujia Qin}, \bibinfo{person}{Guang Yang}, \bibinfo{person}{Fu Wei}, \bibinfo{person}{Zonghan Yang}, \bibinfo{person}{Yusheng Su}, \bibinfo{person}{Shengding Hu}, \bibinfo{person}{Yulin Chen}, \bibinfo{person}{Chi-Min Chan}, \bibinfo{person}{Weize Chen}, \bibinfo{person}{Jing Yi}, \bibinfo{person}{Weilin Zhao}, \bibinfo{person}{Xiaozhi Wang}, \bibinfo{person}{Zhiyuan Liu}, \bibinfo{person}{Haitao Zheng}, \bibinfo{person}{Jianfei Chen}, \bibinfo{person}{Y. Liu}, \bibinfo{person}{Jie Tang}, \bibinfo{person}{Juanzi Li}, {and} \bibinfo{person}{Maosong Sun}.} \bibinfo{year}{2023}\natexlab{}.
\newblock \showarticletitle{Parameter-efficient fine-tuning of large-scale pre-trained language models}.
\newblock \bibinfo{journal}{\emph{Nature Machine Intelligence}}  \bibinfo{volume}{5} (\bibinfo{year}{2023}), \bibinfo{pages}{220--235}.
\newblock
\urldef\tempurl%
\url{https://api.semanticscholar.org/CorpusID:257316425}
\showURL{%
\tempurl}


\bibitem[Dong et~al\mbox{.}(2023)]%
        {selfcollaboration}
\bibfield{author}{\bibinfo{person}{Yihong Dong}, \bibinfo{person}{Xue Jiang}, \bibinfo{person}{Zhi Jin}, {and} \bibinfo{person}{Ge Li}.} \bibinfo{year}{2023}\natexlab{}.
\newblock \bibinfo{title}{Self-collaboration Code Generation via ChatGPT}.
\newblock
\newblock
\showeprint[arxiv]{2304.07590}~[cs.SE]


\bibitem[Du et~al\mbox{.}(2023)]%
        {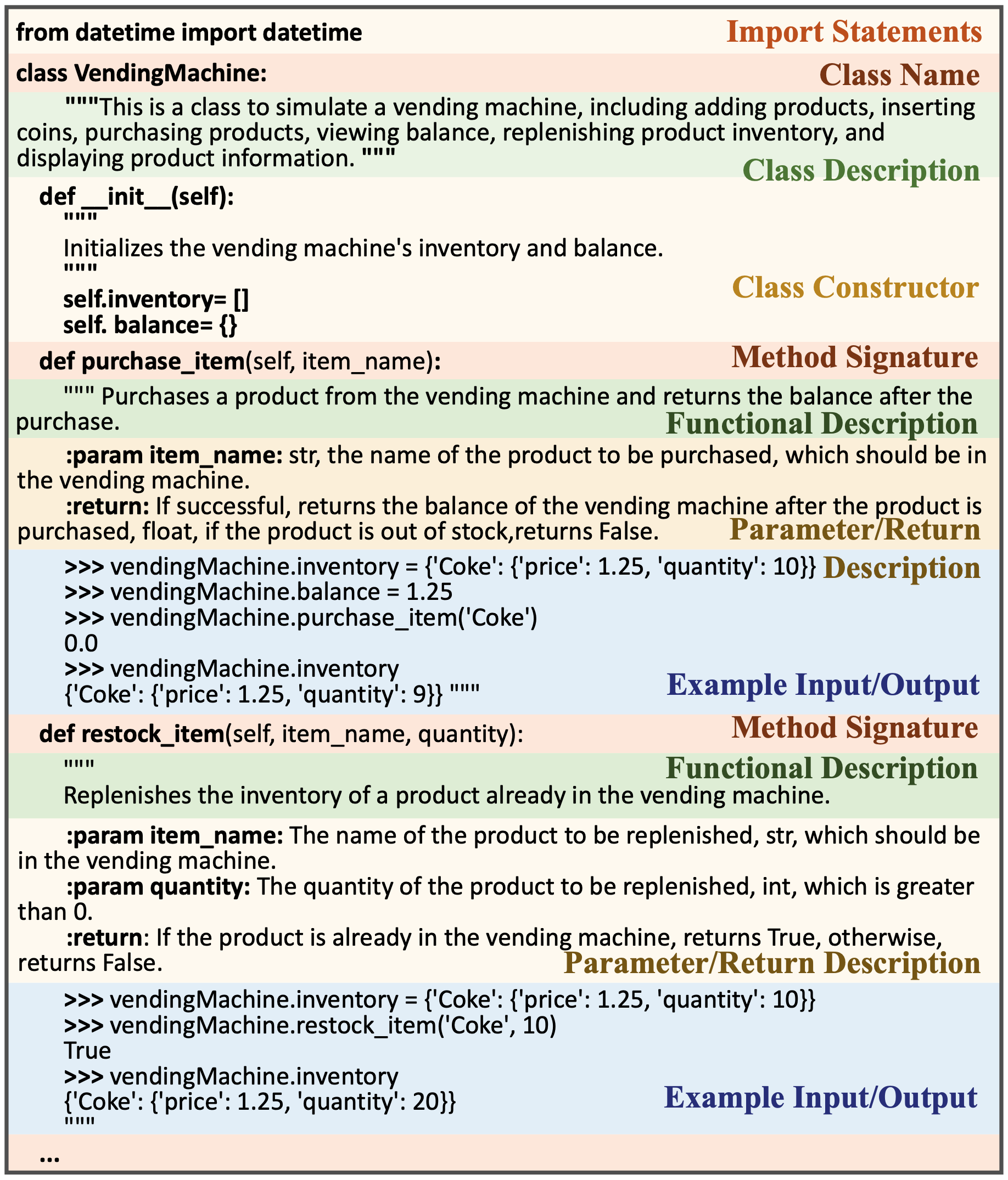}
\bibfield{author}{\bibinfo{person}{Xueying Du}, \bibinfo{person}{Mingwei Liu}, \bibinfo{person}{Kaixin Wang}, \bibinfo{person}{Hanlin Wang}, \bibinfo{person}{Junwei Liu}, \bibinfo{person}{Yixuan Chen}, \bibinfo{person}{Jiayi Feng}, \bibinfo{person}{Chaofeng Sha}, \bibinfo{person}{Xin Peng}, {and} \bibinfo{person}{Yiling Lou}.} \bibinfo{year}{2023}\natexlab{}.
\newblock \bibinfo{title}{ClassEval: A Manually-Crafted Benchmark for Evaluating LLMs on Class-level Code Generation}.
\newblock
\newblock
\showeprint[arxiv]{2308.01861}~[cs.CL]


\bibitem[Eckerdal et~al\mbox{.}(2005)]%
        {prothink}
\bibfield{author}{\bibinfo{person}{Anna Eckerdal}, \bibinfo{person}{Michael Thun\'{e}}, {and} \bibinfo{person}{Anders Berglund}.} \bibinfo{year}{2005}\natexlab{}.
\newblock \showarticletitle{What Does It Take to Learn 'Programming Thinking'?}. In \bibinfo{booktitle}{\emph{Proceedings of the First International Workshop on Computing Education Research}} (Seattle, WA, USA) \emph{(\bibinfo{series}{ICER '05})}. \bibinfo{publisher}{Association for Computing Machinery}, \bibinfo{address}{New York, NY, USA}, \bibinfo{pages}{135–142}.
\newblock
\showISBNx{1595930434}
\urldef\tempurl%
\url{https://doi.org/10.1145/1089786.1089799}
\showDOI{\tempurl}


\bibitem[Fan et~al\mbox{.}(2023)]%
        {fan2023automated}
\bibfield{author}{\bibinfo{person}{Zhiyu Fan}, \bibinfo{person}{Xiang Gao}, \bibinfo{person}{Martin Mirchev}, \bibinfo{person}{Abhik Roychoudhury}, {and} \bibinfo{person}{Shin~Hwei Tan}.} \bibinfo{year}{2023}\natexlab{}.
\newblock \showarticletitle{Automated repair of programs from large language models}. In \bibinfo{booktitle}{\emph{2023 IEEE/ACM 45th International Conference on Software Engineering (ICSE)}}. IEEE, \bibinfo{pages}{1469--1481}.
\newblock


\bibitem[F.Dalpiaz({[n.\,d.]})]%
        {Dalpiaz2018dataset}
\bibfield{author}{\bibinfo{person}{F.Dalpiaz}.} \bibinfo{year}{[n.\,d.]}\natexlab{}.
\newblock
\newblock


\bibitem[Feng et~al\mbox{.}(2020)]%
        {feng2020codebert}
\bibfield{author}{\bibinfo{person}{Zhangyin Feng}, \bibinfo{person}{Daya Guo}, \bibinfo{person}{Duyu Tang}, \bibinfo{person}{Nan Duan}, \bibinfo{person}{Xiaocheng Feng}, \bibinfo{person}{Ming Gong}, \bibinfo{person}{Linjun Shou}, \bibinfo{person}{Bing Qin}, \bibinfo{person}{Ting Liu}, \bibinfo{person}{Daxin Jiang}, {and} \bibinfo{person}{Ming Zhou}.} \bibinfo{year}{2020}\natexlab{}.
\newblock \showarticletitle{CodeBERT: {A} Pre-Trained Model for Programming and Natural Languages}. In \bibinfo{booktitle}{\emph{Findings of the Association for Computational Linguistics: {EMNLP} 2020, Online Event, 16-20 November 2020}} \emph{(\bibinfo{series}{Findings of {ACL}}, Vol.~\bibinfo{volume}{{EMNLP} 2020})}, \bibfield{editor}{\bibinfo{person}{Trevor Cohn}, \bibinfo{person}{Yulan He}, {and} \bibinfo{person}{Yang Liu}} (Eds.). \bibinfo{publisher}{Association for Computational Linguistics}, \bibinfo{pages}{1536--1547}.
\newblock


\bibitem[Fried et~al\mbox{.}(2022)]%
        {fried2022incoder}
\bibfield{author}{\bibinfo{person}{Daniel Fried}, \bibinfo{person}{Armen Aghajanyan}, \bibinfo{person}{Jessy Lin}, \bibinfo{person}{Sida Wang}, \bibinfo{person}{Eric Wallace}, \bibinfo{person}{Freda Shi}, \bibinfo{person}{Ruiqi Zhong}, \bibinfo{person}{Wen-tau Yih}, \bibinfo{person}{Luke Zettlemoyer}, {and} \bibinfo{person}{Mike Lewis}.} \bibinfo{year}{2022}\natexlab{}.
\newblock \showarticletitle{Incoder: A generative model for code infilling and synthesis}.
\newblock \bibinfo{journal}{\emph{arXiv preprint arXiv:2204.05999}} (\bibinfo{year}{2022}).
\newblock


\bibitem[Gao et~al\mbox{.}(2023)]%
        {gao2023makes}
\bibfield{author}{\bibinfo{person}{Shuzheng Gao}, \bibinfo{person}{Xin-Cheng Wen}, \bibinfo{person}{Cuiyun Gao}, \bibinfo{person}{Wenxuan Wang}, \bibinfo{person}{Hongyu Zhang}, {and} \bibinfo{person}{Michael~R Lyu}.} \bibinfo{year}{2023}\natexlab{}.
\newblock \showarticletitle{What Makes Good In-Context Demonstrations for Code Intelligence Tasks with LLMs?}. In \bibinfo{booktitle}{\emph{2023 38th IEEE/ACM International Conference on Automated Software Engineering (ASE)}}. IEEE Computer Society, \bibinfo{pages}{761--773}.
\newblock


\bibitem[Guo et~al\mbox{.}(2022)]%
        {guo2022unixcoder}
\bibfield{author}{\bibinfo{person}{Daya Guo}, \bibinfo{person}{Shuai Lu}, \bibinfo{person}{Nan Duan}, \bibinfo{person}{Yanlin Wang}, \bibinfo{person}{Ming Zhou}, {and} \bibinfo{person}{Jian Yin}.} \bibinfo{year}{2022}\natexlab{}.
\newblock \showarticletitle{Unixcoder: Unified cross-modal pre-training for code representation}.
\newblock \bibinfo{journal}{\emph{arXiv preprint arXiv:2203.03850}} (\bibinfo{year}{2022}).
\newblock


\bibitem[Guo et~al\mbox{.}(2020)]%
        {guo2020graphcodebert}
\bibfield{author}{\bibinfo{person}{Daya Guo}, \bibinfo{person}{Shuo Ren}, \bibinfo{person}{Shuai Lu}, \bibinfo{person}{Zhangyin Feng}, \bibinfo{person}{Duyu Tang}, \bibinfo{person}{Shujie Liu}, \bibinfo{person}{Long Zhou}, \bibinfo{person}{Nan Duan}, \bibinfo{person}{Alexey Svyatkovskiy}, \bibinfo{person}{Shengyu Fu}, {et~al\mbox{.}}} \bibinfo{year}{2020}\natexlab{}.
\newblock \showarticletitle{Graphcodebert: Pre-training code representations with data flow}.
\newblock \bibinfo{journal}{\emph{arXiv preprint arXiv:2009.08366}} (\bibinfo{year}{2020}).
\newblock


\bibitem[Hanam et~al\mbox{.}(2016)]%
        {hanam2016discovering}
\bibfield{author}{\bibinfo{person}{Quinn Hanam}, \bibinfo{person}{Fernando S de~M Brito}, {and} \bibinfo{person}{Ali Mesbah}.} \bibinfo{year}{2016}\natexlab{}.
\newblock \showarticletitle{Discovering bug patterns in JavaScript}. In \bibinfo{booktitle}{\emph{Proceedings of the 2016 24th ACM SIGSOFT international symposium on foundations of software engineering}}. \bibinfo{pages}{144--156}.
\newblock


\bibitem[He(2019)]%
        {he2019understanding}
\bibfield{author}{\bibinfo{person}{Hao He}.} \bibinfo{year}{2019}\natexlab{}.
\newblock \showarticletitle{Understanding source code comments at large-scale}. In \bibinfo{booktitle}{\emph{Proceedings of the {ACM} Joint Meeting on European Software Engineering Conference and Symposium on the Foundations of Software Engineering, {ESEC/SIGSOFT} {FSE} 2019, Tallinn, Estonia, August 26-30, 2019}}, \bibfield{editor}{\bibinfo{person}{Marlon Dumas}, \bibinfo{person}{Dietmar Pfahl}, \bibinfo{person}{Sven Apel}, {and} \bibinfo{person}{Alessandra Russo}} (Eds.). \bibinfo{publisher}{{ACM}}, \bibinfo{pages}{1217--1219}.
\newblock


\bibitem[Hey et~al\mbox{.}(2020)]%
        {hey2020norbert}
\bibfield{author}{\bibinfo{person}{Tobias Hey}, \bibinfo{person}{Jan Keim}, \bibinfo{person}{Anne Koziolek}, {and} \bibinfo{person}{Walter~F Tichy}.} \bibinfo{year}{2020}\natexlab{}.
\newblock \showarticletitle{Norbert: Transfer learning for requirements classification}. In \bibinfo{booktitle}{\emph{2020 IEEE 28th International Requirements Engineering Conference (RE)}}. IEEE, \bibinfo{pages}{169--179}.
\newblock


\bibitem[Hou et~al\mbox{.}(2023)]%
        {hou2023large}
\bibfield{author}{\bibinfo{person}{Xinyi Hou}, \bibinfo{person}{Yanjie Zhao}, \bibinfo{person}{Yue Liu}, \bibinfo{person}{Zhou Yang}, \bibinfo{person}{Kailong Wang}, \bibinfo{person}{Li Li}, \bibinfo{person}{Xiapu Luo}, \bibinfo{person}{David Lo}, \bibinfo{person}{John Grundy}, {and} \bibinfo{person}{Haoyu Wang}.} \bibinfo{year}{2023}\natexlab{}.
\newblock \bibinfo{title}{Large Language Models for Software Engineering: A Systematic Literature Review}.
\newblock
\newblock
\showeprint[arxiv]{2308.10620}~[cs.SE]


\bibitem[Hu et~al\mbox{.}(2018)]%
        {ijcai2018p314}
\bibfield{author}{\bibinfo{person}{Xing Hu}, \bibinfo{person}{Ge Li}, \bibinfo{person}{Xin Xia}, \bibinfo{person}{David Lo}, \bibinfo{person}{Shuai Lu}, {and} \bibinfo{person}{Zhi Jin}.} \bibinfo{year}{2018}\natexlab{}.
\newblock \showarticletitle{Summarizing Source Code with Transferred API Knowledge}. In \bibinfo{booktitle}{\emph{Proceedings of the Twenty-Seventh International Joint Conference on Artificial Intelligence, {IJCAI-18}}}. \bibinfo{publisher}{International Joint Conferences on Artificial Intelligence Organization}, \bibinfo{pages}{2269--2275}.
\newblock
\urldef\tempurl%
\url{https://doi.org/10.24963/ijcai.2018/314}
\showDOI{\tempurl}


\bibitem[Husain et~al\mbox{.}(2019)]%
        {husain2019codesearchnet}
\bibfield{author}{\bibinfo{person}{Hamel Husain}, \bibinfo{person}{Ho-Hsiang Wu}, \bibinfo{person}{Tiferet Gazit}, \bibinfo{person}{Miltiadis Allamanis}, {and} \bibinfo{person}{Marc Brockschmidt}.} \bibinfo{year}{2019}\natexlab{}.
\newblock \showarticletitle{Codesearchnet challenge: Evaluating the state of semantic code search}.
\newblock \bibinfo{journal}{\emph{arXiv preprint arXiv:1909.09436}} (\bibinfo{year}{2019}).
\newblock


\bibitem[ISO(2011)]%
        {iso2011ieee}
\bibfield{author}{\bibinfo{person}{IEC ISO}.} \bibinfo{year}{2011}\natexlab{}.
\newblock \bibinfo{booktitle}{\emph{IEEE. 29148: 2011-Systems and software engineering-Requirements engineering}}.
\newblock \bibinfo{type}{{T}echnical {R}eport}. \bibinfo{institution}{Technical report}.
\newblock


\bibitem[J.~Cleland-Huang and Port(2007)]%
        {nfr2007dataset}
\bibfield{author}{\bibinfo{person}{H.~Liguo J.~Cleland-Huang, S.~Mazrouee} {and} \bibinfo{person}{D. Port}.} \bibinfo{year}{2007}\natexlab{}.
\newblock \bibinfo{title}{nfr}.
\newblock
\newblock
\urldef\tempurl%
\url{https://doi.org/10.5281/zenodo.268542}
\showURL{%
\tempurl}
\newblock
\shownote{[Online]}.


\bibitem[Jha and Mahmoud(2019)]%
        {jha2019mining}
\bibfield{author}{\bibinfo{person}{Nishant Jha} {and} \bibinfo{person}{Anas Mahmoud}.} \bibinfo{year}{2019}\natexlab{}.
\newblock \showarticletitle{Mining non-functional requirements from app store reviews}.
\newblock \bibinfo{journal}{\emph{Empirical Software Engineering}}  \bibinfo{volume}{24} (\bibinfo{year}{2019}), \bibinfo{pages}{3659--3695}.
\newblock


\bibitem[Jiang et~al\mbox{.}(2023)]%
        {jiang2023impact}
\bibfield{author}{\bibinfo{person}{Nan Jiang}, \bibinfo{person}{Kevin Liu}, \bibinfo{person}{Thibaud Lutellier}, {and} \bibinfo{person}{Lin Tan}.} \bibinfo{year}{2023}\natexlab{}.
\newblock \showarticletitle{Impact of code language models on automated program repair}.
\newblock  (\bibinfo{year}{2023}), \bibinfo{pages}{1430--1442}.
\newblock


\bibitem[Jiang et~al\mbox{.}(2021)]%
        {jiang2021cure}
\bibfield{author}{\bibinfo{person}{Nan Jiang}, \bibinfo{person}{Thibaud Lutellier}, {and} \bibinfo{person}{Lin Tan}.} \bibinfo{year}{2021}\natexlab{}.
\newblock \showarticletitle{Cure: Code-aware neural machine translation for automatic program repair}. In \bibinfo{booktitle}{\emph{2021 IEEE/ACM 43rd International Conference on Software Engineering (ICSE)}}. IEEE, \bibinfo{pages}{1161--1173}.
\newblock


\bibitem[Jiao et~al\mbox{.}(2023)]%
        {10298408}
\bibfield{author}{\bibinfo{person}{Mingsheng Jiao}, \bibinfo{person}{Tingrui Yu}, \bibinfo{person}{Xuan Li}, \bibinfo{person}{Guanjie Qiu}, \bibinfo{person}{Xiaodong Gu}, {and} \bibinfo{person}{Beijun Shen}.} \bibinfo{year}{2023}\natexlab{}.
\newblock \showarticletitle{On the Evaluation of Neural Code Translation: Taxonomy and Benchmark}. In \bibinfo{booktitle}{\emph{2023 38th IEEE/ACM International Conference on Automated Software Engineering (ASE)}}. \bibinfo{pages}{1529--1541}.
\newblock
\urldef\tempurl%
\url{https://doi.org/10.1109/ASE56229.2023.00114}
\showDOI{\tempurl}


\bibitem[Jin et~al\mbox{.}(2023)]%
        {jin2023binary}
\bibfield{author}{\bibinfo{person}{Xin Jin}, \bibinfo{person}{Jonathan Larson}, \bibinfo{person}{Weiwei Yang}, {and} \bibinfo{person}{Zhiqiang Lin}.} \bibinfo{year}{2023}\natexlab{}.
\newblock \showarticletitle{Binary code summarization: Benchmarking chatgpt/gpt-4 and other large language models}.
\newblock \bibinfo{journal}{\emph{arXiv preprint arXiv:2312.09601}} (\bibinfo{year}{2023}).
\newblock


\bibitem[Just et~al\mbox{.}(2014)]%
        {just2014defects4j}
\bibfield{author}{\bibinfo{person}{Ren{\'e} Just}, \bibinfo{person}{Darioush Jalali}, {and} \bibinfo{person}{Michael~D Ernst}.} \bibinfo{year}{2014}\natexlab{}.
\newblock \showarticletitle{Defects4J: A database of existing faults to enable controlled testing studies for Java programs}. In \bibinfo{booktitle}{\emph{Proceedings of the 2014 international symposium on software testing and analysis}}. \bibinfo{pages}{437--440}.
\newblock


\bibitem[Kang et~al\mbox{.}(2023)]%
        {kang2023large}
\bibfield{author}{\bibinfo{person}{Sungmin Kang}, \bibinfo{person}{Juyeon Yoon}, {and} \bibinfo{person}{Shin Yoo}.} \bibinfo{year}{2023}\natexlab{}.
\newblock \showarticletitle{Large language models are few-shot testers: Exploring llm-based general bug reproduction}. In \bibinfo{booktitle}{\emph{2023 IEEE/ACM 45th International Conference on Software Engineering (ICSE)}}. IEEE, \bibinfo{pages}{2312--2323}.
\newblock


\bibitem[Khan et~al\mbox{.}(2023)]%
        {khan2023xcodeeval}
\bibfield{author}{\bibinfo{person}{Mohammad Abdullah~Matin Khan}, \bibinfo{person}{M~Saiful Bari}, \bibinfo{person}{Xuan~Long Do}, \bibinfo{person}{Weishi Wang}, \bibinfo{person}{Md~Rizwan Parvez}, {and} \bibinfo{person}{Shafiq Joty}.} \bibinfo{year}{2023}\natexlab{}.
\newblock \showarticletitle{xCodeEval: A Large Scale Multilingual Multitask Benchmark for Code Understanding, Generation, Translation and Retrieval}.
\newblock \bibinfo{journal}{\emph{arXiv preprint arXiv:2303.03004}} (\bibinfo{year}{2023}).
\newblock
\urldef\tempurl%
\url{https://doi.org/10.48550/arXiv.2303.03004}
\showURL{%
\tempurl}


\bibitem[Kitchenham and Charters(2007)]%
        {kitchenham}
\bibfield{author}{\bibinfo{person}{Barbara Kitchenham} {and} \bibinfo{person}{Stuart Charters}.} \bibinfo{year}{2007}\natexlab{}.
\newblock \showarticletitle{Guidelines for performing Systematic Literature Reviews in Software Engineering}.
\newblock   \bibinfo{volume}{2} (\bibinfo{date}{01} \bibinfo{year}{2007}).
\newblock


\bibitem[Kitchenham and Pfleeger(2008)]%
        {kitchenham2008personal}
\bibfield{author}{\bibinfo{person}{Barbara~A Kitchenham} {and} \bibinfo{person}{Shari~L Pfleeger}.} \bibinfo{year}{2008}\natexlab{}.
\newblock \showarticletitle{Personal opinion surveys}.
\newblock In \bibinfo{booktitle}{\emph{Guide to advanced empirical software engineering}}. \bibinfo{publisher}{Springer}, \bibinfo{pages}{63--92}.
\newblock


\bibitem[Kolak et~al\mbox{.}(2022)]%
        {kolak2022patch}
\bibfield{author}{\bibinfo{person}{Sophia~D Kolak}, \bibinfo{person}{Ruben Martins}, \bibinfo{person}{Claire Le~Goues}, {and} \bibinfo{person}{Vincent~Josua Hellendoorn}.} \bibinfo{year}{2022}\natexlab{}.
\newblock \showarticletitle{Patch generation with language models: Feasibility and scaling behavior}. In \bibinfo{booktitle}{\emph{Deep Learning for Code Workshop}}.
\newblock


\bibitem[Kotti et~al\mbox{.}(2023)]%
        {10.1145/3572905}
\bibfield{author}{\bibinfo{person}{Zoe Kotti}, \bibinfo{person}{Rafaila Galanopoulou}, {and} \bibinfo{person}{Diomidis Spinellis}.} \bibinfo{year}{2023}\natexlab{}.
\newblock \showarticletitle{Machine Learning for Software Engineering: A Tertiary Study}.
\newblock \bibinfo{journal}{\emph{ACM Comput. Surv.}} \bibinfo{volume}{55}, \bibinfo{number}{12}, Article \bibinfo{articleno}{256} (\bibinfo{date}{mar} \bibinfo{year}{2023}), \bibinfo{numpages}{39}~pages.
\newblock
\showISSN{0360-0300}
\urldef\tempurl%
\url{https://doi.org/10.1145/3572905}
\showDOI{\tempurl}


\bibitem[Kulal et~al\mbox{.}(2019)]%
        {passk}
\bibfield{author}{\bibinfo{person}{Sumith Kulal}, \bibinfo{person}{Panupong Pasupat}, \bibinfo{person}{Kartik Chandra}, \bibinfo{person}{Mina Lee}, \bibinfo{person}{Oded Padon}, \bibinfo{person}{Alex Aiken}, {and} \bibinfo{person}{Percy Liang}.} \bibinfo{year}{2019}\natexlab{}.
\newblock \bibinfo{booktitle}{\emph{SPoC: Search-Based Pseudocode to Code}}.
\newblock \bibinfo{publisher}{Curran Associates Inc.}, \bibinfo{address}{Red Hook, NY, USA}.
\newblock


\bibitem[Lahiri et~al\mbox{.}(2022)]%
        {lahiri2022interactive}
\bibfield{author}{\bibinfo{person}{Shuvendu~K Lahiri}, \bibinfo{person}{Aaditya Naik}, \bibinfo{person}{Georgios Sakkas}, \bibinfo{person}{Piali Choudhury}, \bibinfo{person}{Curtis von Veh}, \bibinfo{person}{Madanlal Musuvathi}, \bibinfo{person}{Jeevana~Priya Inala}, \bibinfo{person}{Chenglong Wang}, {and} \bibinfo{person}{Jianfeng Gao}.} \bibinfo{year}{2022}\natexlab{}.
\newblock \showarticletitle{Interactive code generation via test-driven user-intent formalization}.
\newblock \bibinfo{journal}{\emph{arXiv preprint arXiv:2208.05950}} (\bibinfo{year}{2022}).
\newblock


\bibitem[Le~Goues et~al\mbox{.}(2015)]%
        {le2015manybugs}
\bibfield{author}{\bibinfo{person}{Claire Le~Goues}, \bibinfo{person}{Neal Holtschulte}, \bibinfo{person}{Edward~K Smith}, \bibinfo{person}{Yuriy Brun}, \bibinfo{person}{Premkumar Devanbu}, \bibinfo{person}{Stephanie Forrest}, {and} \bibinfo{person}{Westley Weimer}.} \bibinfo{year}{2015}\natexlab{}.
\newblock \showarticletitle{The ManyBugs and IntroClass benchmarks for automated repair of C programs}.
\newblock \bibinfo{journal}{\emph{IEEE Transactions on Software Engineering}} \bibinfo{volume}{41}, \bibinfo{number}{12} (\bibinfo{year}{2015}), \bibinfo{pages}{1236--1256}.
\newblock


\bibitem[LeClair et~al\mbox{.}(2019)]%
        {leclair2019neural}
\bibfield{author}{\bibinfo{person}{Alexander LeClair}, \bibinfo{person}{Siyuan Jiang}, {and} \bibinfo{person}{Collin McMillan}.} \bibinfo{year}{2019}\natexlab{}.
\newblock \showarticletitle{A neural model for generating natural language summaries of program subroutines}. In \bibinfo{booktitle}{\emph{2019 IEEE/ACM 41st International Conference on Software Engineering (ICSE)}}. IEEE, \bibinfo{pages}{795--806}.
\newblock


\bibitem[Lemieux et~al\mbox{.}(2023)]%
        {lemieux2023codamosa}
\bibfield{author}{\bibinfo{person}{Caroline Lemieux}, \bibinfo{person}{Jeevana~Priya Inala}, \bibinfo{person}{Shuvendu~K Lahiri}, {and} \bibinfo{person}{Siddhartha Sen}.} \bibinfo{year}{2023}\natexlab{}.
\newblock \showarticletitle{CODAMOSA: Escaping coverage plateaus in test generation with pre-trained large language models}. In \bibinfo{booktitle}{\emph{International conference on software engineering (ICSE)}}.
\newblock


\bibitem[Lethbridge et~al\mbox{.}(2005)]%
        {lethbridge2005studying}
\bibfield{author}{\bibinfo{person}{Timothy~C Lethbridge}, \bibinfo{person}{Susan~Elliott Sim}, {and} \bibinfo{person}{Janice Singer}.} \bibinfo{year}{2005}\natexlab{}.
\newblock \showarticletitle{Studying software engineers: Data collection techniques for software field studies}.
\newblock \bibinfo{journal}{\emph{Empirical software engineering}}  \bibinfo{volume}{10} (\bibinfo{year}{2005}), \bibinfo{pages}{311--341}.
\newblock


\bibitem[Li et~al\mbox{.}(2023b)]%
        {TIP}
\bibfield{author}{\bibinfo{person}{Jia Li}, \bibinfo{person}{Ge Li}, \bibinfo{person}{Yongming Li}, {and} \bibinfo{person}{Zhi Jin}.} \bibinfo{year}{2023}\natexlab{b}.
\newblock \showarticletitle{Enabling Programming Thinking in Large Language Models Toward Code Generation}.
\newblock \bibinfo{journal}{\emph{ArXiv}}  \bibinfo{volume}{abs/2305.06599} (\bibinfo{year}{2023}).
\newblock
\urldef\tempurl%
\url{https://api.semanticscholar.org/CorpusID:263896057}
\showURL{%
\tempurl}


\bibitem[Li et~al\mbox{.}(2023c)]%
        {skcoder}
\bibfield{author}{\bibinfo{person}{Jia Li}, \bibinfo{person}{Yongmin Li}, \bibinfo{person}{Ge Li}, \bibinfo{person}{Zhi Jin}, \bibinfo{person}{Yiyang Hao}, {and} \bibinfo{person}{Xing Hu}.} \bibinfo{year}{2023}\natexlab{c}.
\newblock \showarticletitle{SkCoder: A Sketch-based Approach for Automatic Code Generation}. In \bibinfo{booktitle}{\emph{2023 IEEE/ACM 45th International Conference on Software Engineering (ICSE)}}. \bibinfo{pages}{2124--2135}.
\newblock
\urldef\tempurl%
\url{https://doi.org/10.1109/ICSE48619.2023.00179}
\showDOI{\tempurl}


\bibitem[Li et~al\mbox{.}(2023a)]%
        {li2023starcoder}
\bibfield{author}{\bibinfo{person}{Raymond Li}, \bibinfo{person}{Loubna~Ben Allal}, \bibinfo{person}{Yangtian Zi}, \bibinfo{person}{Niklas Muennighoff}, \bibinfo{person}{Denis Kocetkov}, \bibinfo{person}{Chenghao Mou}, \bibinfo{person}{Marc Marone}, \bibinfo{person}{Christopher Akiki}, \bibinfo{person}{Jia Li}, \bibinfo{person}{Jenny Chim}, {et~al\mbox{.}}} \bibinfo{year}{2023}\natexlab{a}.
\newblock \showarticletitle{StarCoder: may the source be with you!}
\newblock \bibinfo{journal}{\emph{arXiv preprint arXiv:2305.06161}} (\bibinfo{year}{2023}).
\newblock
\urldef\tempurl%
\url{https://doi.org/10.48550/arXiv.2305.06161}
\showURL{%
\tempurl}


\bibitem[Li et~al\mbox{.}(2022)]%
        {li2022competition}
\bibfield{author}{\bibinfo{person}{Yujia Li}, \bibinfo{person}{David Choi}, \bibinfo{person}{Junyoung Chung}, \bibinfo{person}{Nate Kushman}, \bibinfo{person}{Julian Schrittwieser}, \bibinfo{person}{R{\'e}mi Leblond}, \bibinfo{person}{Tom Eccles}, \bibinfo{person}{James Keeling}, \bibinfo{person}{Felix Gimeno}, \bibinfo{person}{Agustin Dal~Lago}, {et~al\mbox{.}}} \bibinfo{year}{2022}\natexlab{}.
\newblock \showarticletitle{Competition-level code generation with alphacode}.
\newblock \bibinfo{journal}{\emph{Science}} \bibinfo{volume}{378}, \bibinfo{number}{6624} (\bibinfo{year}{2022}), \bibinfo{pages}{1092--1097}.
\newblock


\bibitem[Li et~al\mbox{.}(2023d)]%
        {Liyao}
\bibfield{author}{\bibinfo{person}{Yao Li}, \bibinfo{person}{Tao Zhang}, \bibinfo{person}{Xiapu Luo}, \bibinfo{person}{Haipeng Cai}, \bibinfo{person}{Sen Fang}, {and} \bibinfo{person}{Dawei Yuan}.} \bibinfo{year}{2023}\natexlab{d}.
\newblock \showarticletitle{Do Pretrained Language Models Indeed Understand Software Engineering Tasks?}
\newblock \bibinfo{journal}{\emph{IEEE Transactions on Software Engineering}} \bibinfo{volume}{49}, \bibinfo{number}{10} (\bibinfo{year}{2023}), \bibinfo{pages}{4639--4655}.
\newblock
\urldef\tempurl%
\url{https://doi.org/10.1109/TSE.2023.3308952}
\showDOI{\tempurl}


\bibitem[Lin(2004)]%
        {lin2004rouge}
\bibfield{author}{\bibinfo{person}{Chin-Yew Lin}.} \bibinfo{year}{2004}\natexlab{}.
\newblock \showarticletitle{Rouge: A package for automatic evaluation of summaries}. In \bibinfo{booktitle}{\emph{Text summarization branches out}}. \bibinfo{pages}{74--81}.
\newblock


\bibitem[Lin et~al\mbox{.}(2017)]%
        {lin2017quixbugs}
\bibfield{author}{\bibinfo{person}{Derrick Lin}, \bibinfo{person}{James Koppel}, \bibinfo{person}{Angela Chen}, {and} \bibinfo{person}{Armando Solar-Lezama}.} \bibinfo{year}{2017}\natexlab{}.
\newblock \showarticletitle{QuixBugs: A multi-lingual program repair benchmark set based on the Quixey Challenge}. In \bibinfo{booktitle}{\emph{Proceedings Companion of the 2017 ACM SIGPLAN international conference on systems, programming, languages, and applications: software for humanity}}. \bibinfo{pages}{55--56}.
\newblock


\bibitem[Liu et~al\mbox{.}(2023)]%
        {codegen4lib}
\bibfield{author}{\bibinfo{person}{Mingwei Liu}, \bibinfo{person}{Tianyong Yang}, \bibinfo{person}{Yiling Lou}, \bibinfo{person}{Xueying Du}, \bibinfo{person}{Ying Wang}, {and} \bibinfo{person}{Xin Peng}.} \bibinfo{year}{2023}\natexlab{}.
\newblock \showarticletitle{CodeGen4Libs: A Two-Stage Approach for Library-Oriented Code Generation}. In \bibinfo{booktitle}{\emph{2023 38th IEEE/ACM International Conference on Automated Software Engineering (ASE)}}. \bibinfo{pages}{434--445}.
\newblock
\urldef\tempurl%
\url{https://doi.org/10.1109/ASE56229.2023.00159}
\showDOI{\tempurl}


\bibitem[Lu et~al\mbox{.}(2021)]%
        {lu2021codexglue}
\bibfield{author}{\bibinfo{person}{Shuai Lu}, \bibinfo{person}{Daya Guo}, \bibinfo{person}{Shuo Ren}, \bibinfo{person}{Junjie Huang}, \bibinfo{person}{Alexey Svyatkovskiy}, \bibinfo{person}{Ambrosio Blanco}, \bibinfo{person}{Colin Clement}, \bibinfo{person}{Dawn Drain}, \bibinfo{person}{Daxin Jiang}, \bibinfo{person}{Duyu Tang}, {et~al\mbox{.}}} \bibinfo{year}{2021}\natexlab{}.
\newblock \showarticletitle{Codexglue: A machine learning benchmark dataset for code understanding and generation}.
\newblock \bibinfo{journal}{\emph{arXiv preprint arXiv:2102.04664}} (\bibinfo{year}{2021}).
\newblock
\urldef\tempurl%
\url{https://doi.org/10.48550/arXiv.2102.04664}
\showURL{%
\tempurl}


\bibitem[Madaan et~al\mbox{.}(2023)]%
        {madaan2023learning}
\bibfield{author}{\bibinfo{person}{Aman Madaan}, \bibinfo{person}{Alexander Shypula}, \bibinfo{person}{Uri Alon}, \bibinfo{person}{Milad Hashemi}, \bibinfo{person}{Parthasarathy Ranganathan}, \bibinfo{person}{Yiming Yang}, \bibinfo{person}{Graham Neubig}, {and} \bibinfo{person}{Amir Yazdanbakhsh}.} \bibinfo{year}{2023}\natexlab{}.
\newblock \showarticletitle{Learning performance-improving code edits}.
\newblock \bibinfo{journal}{\emph{arXiv preprint arXiv:2302.07867}} (\bibinfo{year}{2023}).
\newblock
\urldef\tempurl%
\url{https://doi.org/10.48550/arXiv.2302.07867}
\showURL{%
\tempurl}


\bibitem[Mashhadi and Hemmati(2021)]%
        {mashhadi2021applying}
\bibfield{author}{\bibinfo{person}{Ehsan Mashhadi} {and} \bibinfo{person}{Hadi Hemmati}.} \bibinfo{year}{2021}\natexlab{}.
\newblock \showarticletitle{Applying codebert for automated program repair of java simple bugs}. In \bibinfo{booktitle}{\emph{2021 IEEE/ACM 18th International Conference on Mining Software Repositories (MSR)}}. IEEE, \bibinfo{pages}{505--509}.
\newblock


\bibitem[Mastropaolo et~al\mbox{.}(2021)]%
        {mastropaolo2021studying}
\bibfield{author}{\bibinfo{person}{Antonio Mastropaolo}, \bibinfo{person}{Simone Scalabrino}, \bibinfo{person}{Nathan Cooper}, \bibinfo{person}{David~Nader Palacio}, \bibinfo{person}{Denys Poshyvanyk}, \bibinfo{person}{Rocco Oliveto}, {and} \bibinfo{person}{Gabriele Bavota}.} \bibinfo{year}{2021}\natexlab{}.
\newblock \showarticletitle{Studying the usage of text-to-text transfer transformer to support code-related tasks}. In \bibinfo{booktitle}{\emph{2021 IEEE/ACM 43rd International Conference on Software Engineering (ICSE)}}. IEEE, \bibinfo{pages}{336--347}.
\newblock


\bibitem[Mu et~al\mbox{.}(2023)]%
        {clarifygpt}
\bibfield{author}{\bibinfo{person}{Fangwen Mu}, \bibinfo{person}{Lin Shi}, \bibinfo{person}{Song Wang}, \bibinfo{person}{Zhuohao Yu}, \bibinfo{person}{Binquan Zhang}, \bibinfo{person}{Chenxue Wang}, \bibinfo{person}{Shichao Liu}, {and} \bibinfo{person}{Qing Wang}.} \bibinfo{year}{2023}\natexlab{}.
\newblock \bibinfo{title}{ClarifyGPT: Empowering LLM-based Code Generation with Intention Clarification}.
\newblock
\newblock
\showeprint[arxiv]{2310.10996}~[cs.SE]


\bibitem[Murr et~al\mbox{.}(2023)]%
        {murr}
\bibfield{author}{\bibinfo{person}{Lincoln Murr}, \bibinfo{person}{Morgan Grainger}, {and} \bibinfo{person}{David Gao}.} \bibinfo{year}{2023}\natexlab{}.
\newblock \showarticletitle{Testing LLMs on Code Generation with Varying Levels of Prompt Specificity}.
\newblock \bibinfo{journal}{\emph{ArXiv}}  \bibinfo{volume}{abs/2311.07599} (\bibinfo{year}{2023}).
\newblock
\urldef\tempurl%
\url{https://api.semanticscholar.org/CorpusID:265157726}
\showURL{%
\tempurl}


\bibitem[Nashid et~al\mbox{.}(2023)]%
        {nashid2023retrieval}
\bibfield{author}{\bibinfo{person}{Noor Nashid}, \bibinfo{person}{Mifta Sintaha}, {and} \bibinfo{person}{Ali Mesbah}.} \bibinfo{year}{2023}\natexlab{}.
\newblock \showarticletitle{Retrieval-based prompt selection for code-related few-shot learning}. In \bibinfo{booktitle}{\emph{Proceedings of the 45th International Conference on Software Engineering (ICSE’23)}}.
\newblock


\bibitem[Nguyen et~al\mbox{.}(2015)]%
        {7372046}
\bibfield{author}{\bibinfo{person}{Anh~Tuan Nguyen}, \bibinfo{person}{Tung~Thanh Nguyen}, {and} \bibinfo{person}{Tien~N. Nguyen}.} \bibinfo{year}{2015}\natexlab{}.
\newblock \showarticletitle{Divide-and-Conquer Approach for Multi-phase Statistical Migration for Source Code (T)}. In \bibinfo{booktitle}{\emph{2015 30th IEEE/ACM International Conference on Automated Software Engineering (ASE)}}. \bibinfo{pages}{585--596}.
\newblock
\urldef\tempurl%
\url{https://doi.org/10.1109/ASE.2015.74}
\showDOI{\tempurl}


\bibitem[Niu et~al\mbox{.}(2022a)]%
        {niu2022deep}
\bibfield{author}{\bibinfo{person}{Changan Niu}, \bibinfo{person}{Chuanyi Li}, \bibinfo{person}{Bin Luo}, {and} \bibinfo{person}{Vincent Ng}.} \bibinfo{year}{2022}\natexlab{a}.
\newblock \bibinfo{title}{Deep Learning Meets Software Engineering: A Survey on Pre-Trained Models of Source Code}.
\newblock
\newblock
\showeprint[arxiv]{2205.11739}~[cs.SE]


\bibitem[Niu et~al\mbox{.}(2022b)]%
        {niu2022spt}
\bibfield{author}{\bibinfo{person}{Changan Niu}, \bibinfo{person}{Chuanyi Li}, \bibinfo{person}{Vincent Ng}, \bibinfo{person}{Jidong Ge}, \bibinfo{person}{Liguo Huang}, {and} \bibinfo{person}{Bin Luo}.} \bibinfo{year}{2022}\natexlab{b}.
\newblock \showarticletitle{Spt-code: Sequence-to-sequence pre-training for learning source code representations}. In \bibinfo{booktitle}{\emph{Proceedings of the 44th International Conference on Software Engineering}}. \bibinfo{pages}{2006--2018}.
\newblock


\bibitem[Pan et~al\mbox{.}(2023b)]%
        {pan2023stelocoder}
\bibfield{author}{\bibinfo{person}{Jialing Pan}, \bibinfo{person}{Adrien Sad{\'e}}, \bibinfo{person}{Jin Kim}, \bibinfo{person}{Eric Soriano}, \bibinfo{person}{Guillem Sole}, {and} \bibinfo{person}{Sylvain Flamant}.} \bibinfo{year}{2023}\natexlab{b}.
\newblock \showarticletitle{SteloCoder: a Decoder-Only LLM for Multi-Language to Python Code Translation}.
\newblock \bibinfo{journal}{\emph{arXiv preprint arXiv:2310.15539}} (\bibinfo{year}{2023}).
\newblock


\bibitem[Pan et~al\mbox{.}(2023a)]%
        {pan2023understanding}
\bibfield{author}{\bibinfo{person}{Rangeet Pan}, \bibinfo{person}{Ali~Reza Ibrahimzada}, \bibinfo{person}{Rahul Krishna}, \bibinfo{person}{Divya Sankar}, \bibinfo{person}{Lambert~Pouguem Wassi}, \bibinfo{person}{Michele Merler}, \bibinfo{person}{Boris Sobolev}, \bibinfo{person}{Raju Pavuluri}, \bibinfo{person}{Saurabh Sinha}, {and} \bibinfo{person}{Reyhaneh Jabbarvand}.} \bibinfo{year}{2023}\natexlab{a}.
\newblock \showarticletitle{Understanding the Effectiveness of Large Language Models in Code Translation}.
\newblock \bibinfo{journal}{\emph{arXiv preprint arXiv:2308.03109}} (\bibinfo{year}{2023}).
\newblock
\urldef\tempurl%
\url{https://doi.org/10.48550/arXiv.2308.03109}
\showURL{%
\tempurl}


\bibitem[Papineni et~al\mbox{.}(2002)]%
        {papineni2002bleu}
\bibfield{author}{\bibinfo{person}{Kishore Papineni}, \bibinfo{person}{Salim Roukos}, \bibinfo{person}{Todd Ward}, {and} \bibinfo{person}{Wei-Jing Zhu}.} \bibinfo{year}{2002}\natexlab{}.
\newblock \showarticletitle{Bleu: a method for automatic evaluation of machine translation}. In \bibinfo{booktitle}{\emph{Proceedings of the 40th annual meeting of the Association for Computational Linguistics}}. \bibinfo{pages}{311--318}.
\newblock


\bibitem[Prenner et~al\mbox{.}(2022)]%
        {prenner2022can}
\bibfield{author}{\bibinfo{person}{Julian~Aron Prenner}, \bibinfo{person}{Hlib Babii}, {and} \bibinfo{person}{Romain Robbes}.} \bibinfo{year}{2022}\natexlab{}.
\newblock \showarticletitle{Can OpenAI's codex fix bugs? an evaluation on QuixBugs}. In \bibinfo{booktitle}{\emph{Proceedings of the Third International Workshop on Automated Program Repair}}. \bibinfo{pages}{69--75}.
\newblock


\bibitem[Ren et~al\mbox{.}(2023)]%
        {kpc}
\bibfield{author}{\bibinfo{person}{Xiaoxue Ren}, \bibinfo{person}{Xinyuan Ye}, \bibinfo{person}{Dehai Zhao}, \bibinfo{person}{Zhenchang Xing}, {and} \bibinfo{person}{Xiaohu Yang}.} \bibinfo{year}{2023}\natexlab{}.
\newblock \showarticletitle{From Misuse to Mastery: Enhancing Code Generation with Knowledge-Driven AI Chaining}. In \bibinfo{booktitle}{\emph{2023 38th IEEE/ACM International Conference on Automated Software Engineering (ASE)}}. \bibinfo{pages}{976--987}.
\newblock
\urldef\tempurl%
\url{https://doi.org/10.1109/ASE56229.2023.00143}
\showDOI{\tempurl}


\bibitem[Robeer et~al\mbox{.}(2016)]%
        {robeer2016automated}
\bibfield{author}{\bibinfo{person}{Marcel Robeer}, \bibinfo{person}{Garm Lucassen}, \bibinfo{person}{Jan Martijn~EM Van Der~Werf}, \bibinfo{person}{Fabiano Dalpiaz}, {and} \bibinfo{person}{Sjaak Brinkkemper}.} \bibinfo{year}{2016}\natexlab{}.
\newblock \showarticletitle{Automated extraction of conceptual models from user stories via NLP}. In \bibinfo{booktitle}{\emph{2016 IEEE 24th international requirements engineering conference (RE)}}. IEEE, \bibinfo{pages}{196--205}.
\newblock


\bibitem[Rokon et~al\mbox{.}(2020)]%
        {bleu}
\bibfield{author}{\bibinfo{person}{Md~Omar~Faruk Rokon}, \bibinfo{person}{Risul Islam}, \bibinfo{person}{Ahmad Darki}, \bibinfo{person}{Evangelos~E. Papalexakis}, {and} \bibinfo{person}{Michalis Faloutsos}.} \bibinfo{year}{2020}\natexlab{}.
\newblock \showarticletitle{{SourceFinder}: Finding Malware {Source-Code} from Publicly Available Repositories in {GitHub}}. In \bibinfo{booktitle}{\emph{23rd International Symposium on Research in Attacks, Intrusions and Defenses (RAID 2020)}}. \bibinfo{publisher}{USENIX Association}, \bibinfo{address}{San Sebastian}, \bibinfo{pages}{149--163}.
\newblock
\showISBNx{978-1-939133-18-2}
\urldef\tempurl%
\url{https://www.usenix.org/conference/raid2020/presentation/omar}
\showURL{%
\tempurl}


\bibitem[S.~Arulmohan and Meurs(2023)]%
        {Arulmohan2023dataset}
\bibfield{author}{\bibinfo{person}{S.~Mosser S.~Arulmohan} {and} \bibinfo{person}{M.-J. Meurs}.} \bibinfo{year}{2023}\natexlab{}.
\newblock \bibinfo{title}{ace-design/qualified user-stories: Version 1.0}.
\newblock
\newblock
\urldef\tempurl%
\url{https://doi.org/10.5281/zenodo.8136975}
\showURL{%
\tempurl}
\newblock
\shownote{[Online]}.


\bibitem[Sch{\"a}fer et~al\mbox{.}(2023)]%
        {schafer2023empirical}
\bibfield{author}{\bibinfo{person}{Max Sch{\"a}fer}, \bibinfo{person}{Sarah Nadi}, \bibinfo{person}{Aryaz Eghbali}, {and} \bibinfo{person}{Frank Tip}.} \bibinfo{year}{2023}\natexlab{}.
\newblock \showarticletitle{An Empirical Evaluation of Using Large Language Models for Automated Unit Test Generation}.
\newblock \bibinfo{journal}{\emph{arXiv preprint arXiv:2302.06527}} (\bibinfo{year}{2023}).
\newblock


\bibitem[Schneider and Berenbach(2013)]%
        {schneider2013literature}
\bibfield{author}{\bibinfo{person}{Florian Schneider} {and} \bibinfo{person}{Brian Berenbach}.} \bibinfo{year}{2013}\natexlab{}.
\newblock \showarticletitle{A literature survey on international standards for systems requirements engineering}.
\newblock \bibinfo{journal}{\emph{Procedia Computer Science}}  \bibinfo{volume}{16} (\bibinfo{year}{2013}), \bibinfo{pages}{796--805}.
\newblock


\bibitem[Shi et~al\mbox{.}(2023)]%
        {shi2023sotana}
\bibfield{author}{\bibinfo{person}{Ensheng Shi}, \bibinfo{person}{Fengji Zhang}, \bibinfo{person}{Yanlin Wang}, \bibinfo{person}{Bei Chen}, \bibinfo{person}{Lun Du}, \bibinfo{person}{Hongyu Zhang}, \bibinfo{person}{Shi Han}, \bibinfo{person}{Dongmei Zhang}, {and} \bibinfo{person}{Hongbin Sun}.} \bibinfo{year}{2023}\natexlab{}.
\newblock \showarticletitle{SoTaNa: The Open-Source Software Development Assistant}.
\newblock \bibinfo{journal}{\emph{arXiv preprint arXiv:2308.13416}} (\bibinfo{year}{2023}).
\newblock


\bibitem[Shin et~al\mbox{.}(2023)]%
        {shin2023prompt}
\bibfield{author}{\bibinfo{person}{Jiho Shin}, \bibinfo{person}{Clark Tang}, \bibinfo{person}{Tahmineh Mohati}, \bibinfo{person}{Maleknaz Nayebi}, \bibinfo{person}{Song Wang}, {and} \bibinfo{person}{Hadi Hemmati}.} \bibinfo{year}{2023}\natexlab{}.
\newblock \bibinfo{title}{Prompt Engineering or Fine Tuning: An Empirical Assessment of Large Language Models in Automated Software Engineering Tasks}.
\newblock
\newblock
\showeprint[arxiv]{2310.10508}~[cs.SE]


\bibitem[Siddiq et~al\mbox{.}(2023)]%
        {siddiq2023exploring}
\bibfield{author}{\bibinfo{person}{Mohammed~Latif Siddiq}, \bibinfo{person}{Joanna Santos}, \bibinfo{person}{Ridwanul~Hasan Tanvir}, \bibinfo{person}{Noshin Ulfat}, \bibinfo{person}{Fahmid~Al Rifat}, {and} \bibinfo{person}{Vinicius~Carvalho Lopes}.} \bibinfo{year}{2023}\natexlab{}.
\newblock \showarticletitle{Exploring the Effectiveness of Large Language Models in Generating Unit Tests}.
\newblock \bibinfo{journal}{\emph{arXiv preprint arXiv:2305.00418}} (\bibinfo{year}{2023}).
\newblock


\bibitem[Sobania et~al\mbox{.}(2023)]%
        {sobania2023analysis}
\bibfield{author}{\bibinfo{person}{Dominik Sobania}, \bibinfo{person}{Martin Briesch}, \bibinfo{person}{Carol Hanna}, {and} \bibinfo{person}{Justyna Petke}.} \bibinfo{year}{2023}\natexlab{}.
\newblock \showarticletitle{An Analysis of the Automatic Bug Fixing Performance of ChatGPT}. APR workshop.
\newblock


\bibitem[Su and McMillan(2023)]%
        {su2023distilled}
\bibfield{author}{\bibinfo{person}{Chia-Yi Su} {and} \bibinfo{person}{Collin McMillan}.} \bibinfo{year}{2023}\natexlab{}.
\newblock \showarticletitle{Distilled GPT for Source Code Summarization}.
\newblock \bibinfo{journal}{\emph{arXiv preprint arXiv:2308.14731}} (\bibinfo{year}{2023}).
\newblock


\bibitem[Sun et~al\mbox{.}(2023a)]%
        {sun2023prompt}
\bibfield{author}{\bibinfo{person}{Weisong Sun}, \bibinfo{person}{Chunrong Fang}, \bibinfo{person}{Yudu You}, \bibinfo{person}{Yuchen Chen}, \bibinfo{person}{Yi Liu}, \bibinfo{person}{Chong Wang}, \bibinfo{person}{Jian Zhang}, \bibinfo{person}{Quanjun Zhang}, \bibinfo{person}{Hanwei Qian}, \bibinfo{person}{Wei Zhao}, {et~al\mbox{.}}} \bibinfo{year}{2023}\natexlab{a}.
\newblock \showarticletitle{A Prompt Learning Framework for Source Code Summarization}.
\newblock \bibinfo{journal}{\emph{arXiv preprint arXiv:2312.16066}} (\bibinfo{year}{2023}).
\newblock


\bibitem[Sun et~al\mbox{.}(2023b)]%
        {sun2023automatic}
\bibfield{author}{\bibinfo{person}{Weisong Sun}, \bibinfo{person}{Chunrong Fang}, \bibinfo{person}{Yudu You}, \bibinfo{person}{Yun Miao}, \bibinfo{person}{Yi Liu}, \bibinfo{person}{Yuekang Li}, \bibinfo{person}{Gelei Deng}, \bibinfo{person}{Shenghan Huang}, \bibinfo{person}{Yuchen Chen}, \bibinfo{person}{Quanjun Zhang}, {et~al\mbox{.}}} \bibinfo{year}{2023}\natexlab{b}.
\newblock \showarticletitle{Automatic Code Summarization via ChatGPT: How Far Are We?}
\newblock \bibinfo{journal}{\emph{arXiv preprint arXiv:2305.12865}} (\bibinfo{year}{2023}).
\newblock


\bibitem[Tan et~al\mbox{.}(2018)]%
        {10.1007/978-3-030-01424-7_27}
\bibfield{author}{\bibinfo{person}{Chuanqi Tan}, \bibinfo{person}{Fuchun Sun}, \bibinfo{person}{Tao Kong}, \bibinfo{person}{Wenchang Zhang}, \bibinfo{person}{Chao Yang}, {and} \bibinfo{person}{Chunfang Liu}.} \bibinfo{year}{2018}\natexlab{}.
\newblock \showarticletitle{A Survey on Deep Transfer Learning}. In \bibinfo{booktitle}{\emph{Artificial Neural Networks and Machine Learning -- ICANN 2018}}, \bibfield{editor}{\bibinfo{person}{V{\v{e}}ra K{\r{u}}rkov{\'a}}, \bibinfo{person}{Yannis Manolopoulos}, \bibinfo{person}{Barbara Hammer}, \bibinfo{person}{Lazaros Iliadis}, {and} \bibinfo{person}{Ilias Maglogiannis}} (Eds.). \bibinfo{publisher}{Springer International Publishing}, \bibinfo{address}{Cham}, \bibinfo{pages}{270--279}.
\newblock
\showISBNx{978-3-030-01424-7}


\bibitem[Tang et~al\mbox{.}(2023)]%
        {Tang2023TheSO}
\bibfield{author}{\bibinfo{person}{Ruixiang Tang}, \bibinfo{person}{Yu-Neng Chuang}, {and} \bibinfo{person}{Xia Hu}.} \bibinfo{year}{2023}\natexlab{}.
\newblock \showarticletitle{The Science of Detecting LLM-Generated Texts}.
\newblock \bibinfo{journal}{\emph{ArXiv}}  \bibinfo{volume}{abs/2303.07205} (\bibinfo{year}{2023}).
\newblock
\urldef\tempurl%
\url{https://api.semanticscholar.org/CorpusID:257496757}
\showURL{%
\tempurl}


\bibitem[Tufano et~al\mbox{.}(2020)]%
        {tufano2020unit}
\bibfield{author}{\bibinfo{person}{Michele Tufano}, \bibinfo{person}{Dawn Drain}, \bibinfo{person}{Alexey Svyatkovskiy}, \bibinfo{person}{Shao~Kun Deng}, {and} \bibinfo{person}{Neel Sundaresan}.} \bibinfo{year}{2020}\natexlab{}.
\newblock \showarticletitle{Unit test case generation with transformers and focal context}.
\newblock \bibinfo{journal}{\emph{arXiv preprint arXiv:2009.05617}} (\bibinfo{year}{2020}).
\newblock


\bibitem[Tufano et~al\mbox{.}(2022)]%
        {tufano2022generating}
\bibfield{author}{\bibinfo{person}{Michele Tufano}, \bibinfo{person}{Dawn Drain}, \bibinfo{person}{Alexey Svyatkovskiy}, {and} \bibinfo{person}{Neel Sundaresan}.} \bibinfo{year}{2022}\natexlab{}.
\newblock \showarticletitle{Generating accurate assert statements for unit test cases using pretrained transformers}. In \bibinfo{booktitle}{\emph{Proceedings of the 3rd ACM/IEEE International Conference on Automation of Software Test}}. \bibinfo{pages}{54--64}.
\newblock


\bibitem[Vaithilingam et~al\mbox{.}(2022)]%
        {10.1145/3491101.3519665}
\bibfield{author}{\bibinfo{person}{Priyan Vaithilingam}, \bibinfo{person}{Tianyi Zhang}, {and} \bibinfo{person}{Elena~L. Glassman}.} \bibinfo{year}{2022}\natexlab{}.
\newblock \showarticletitle{Expectation vs. Experience: Evaluating the Usability of Code Generation Tools Powered by Large Language Models}. In \bibinfo{booktitle}{\emph{Extended Abstracts of the 2022 CHI Conference on Human Factors in Computing Systems}} (New Orleans, LA, USA) \emph{(\bibinfo{series}{CHI EA '22})}. \bibinfo{publisher}{Association for Computing Machinery}, \bibinfo{address}{New York, NY, USA}, Article \bibinfo{articleno}{332}, \bibinfo{numpages}{7}~pages.
\newblock
\showISBNx{9781450391566}
\urldef\tempurl%
\url{https://doi.org/10.1145/3491101.3519665}
\showDOI{\tempurl}


\bibitem[Van~Lamsweerde(2000)]%
        {van2000requirements}
\bibfield{author}{\bibinfo{person}{Axel Van~Lamsweerde}.} \bibinfo{year}{2000}\natexlab{}.
\newblock \showarticletitle{Requirements engineering in the year 00: A research perspective}. In \bibinfo{booktitle}{\emph{Proceedings of the 22nd international conference on Software engineering}}. \bibinfo{pages}{5--19}.
\newblock


\bibitem[Vaswani et~al\mbox{.}(2023)]%
        {vaswani2023attention}
\bibfield{author}{\bibinfo{person}{Ashish Vaswani}, \bibinfo{person}{Noam Shazeer}, \bibinfo{person}{Niki Parmar}, \bibinfo{person}{Jakob Uszkoreit}, \bibinfo{person}{Llion Jones}, \bibinfo{person}{Aidan~N. Gomez}, \bibinfo{person}{Lukasz Kaiser}, {and} \bibinfo{person}{Illia Polosukhin}.} \bibinfo{year}{2023}\natexlab{}.
\newblock \bibinfo{title}{Attention Is All You Need}.
\newblock
\newblock
\showeprint[arxiv]{1706.03762}~[cs.CL]


\bibitem[von~der Mosel et~al\mbox{.}(2023)]%
        {Mosel}
\bibfield{author}{\bibinfo{person}{Julian von~der Mosel}, \bibinfo{person}{Alexander Trautsch}, {and} \bibinfo{person}{Steffen Herbold}.} \bibinfo{year}{2023}\natexlab{}.
\newblock \showarticletitle{On the Validity of Pre-Trained Transformers for Natural Language Processing in the Software Engineering Domain}.
\newblock \bibinfo{journal}{\emph{IEEE Transactions on Software Engineering}} \bibinfo{volume}{49}, \bibinfo{number}{4} (\bibinfo{year}{2023}), \bibinfo{pages}{1487--1507}.
\newblock
\urldef\tempurl%
\url{https://doi.org/10.1109/TSE.2022.3178469}
\showDOI{\tempurl}


\bibitem[Wang et~al\mbox{.}(2023d)]%
        {recode}
\bibfield{author}{\bibinfo{person}{Shiqi Wang}, \bibinfo{person}{Zheng Li}, \bibinfo{person}{Haifeng Qian}, \bibinfo{person}{Chenghao Yang}, \bibinfo{person}{Zijian Wang}, \bibinfo{person}{Mingyue Shang}, \bibinfo{person}{Varun Kumar}, \bibinfo{person}{Samson Tan}, \bibinfo{person}{Baishakhi Ray}, \bibinfo{person}{Parminder Bhatia}, \bibinfo{person}{Ramesh Nallapati}, \bibinfo{person}{Murali~Krishna Ramanathan}, \bibinfo{person}{Dan Roth}, {and} \bibinfo{person}{Bing Xiang}.} \bibinfo{year}{2023}\natexlab{d}.
\newblock \showarticletitle{{R}e{C}ode: Robustness Evaluation of Code Generation Models}. In \bibinfo{booktitle}{\emph{Proceedings of the 61st Annual Meeting of the Association for Computational Linguistics (Volume 1: Long Papers)}}, \bibfield{editor}{\bibinfo{person}{Anna Rogers}, \bibinfo{person}{Jordan Boyd-Graber}, {and} \bibinfo{person}{Naoaki Okazaki}} (Eds.). \bibinfo{publisher}{Association for Computational Linguistics}, \bibinfo{address}{Toronto, Canada}, \bibinfo{pages}{13818--13843}.
\newblock
\urldef\tempurl%
\url{https://doi.org/10.18653/v1/2023.acl-long.773}
\showDOI{\tempurl}


\bibitem[Wang et~al\mbox{.}(2023b)]%
        {wang2023codet5+}
\bibfield{author}{\bibinfo{person}{Yue Wang}, \bibinfo{person}{Hung Le}, \bibinfo{person}{Akhilesh~Deepak Gotmare}, \bibinfo{person}{Nghi~DQ Bui}, \bibinfo{person}{Junnan Li}, {and} \bibinfo{person}{Steven~CH Hoi}.} \bibinfo{year}{2023}\natexlab{b}.
\newblock \showarticletitle{Codet5+: Open code large language models for code understanding and generation}.
\newblock \bibinfo{journal}{\emph{arXiv preprint arXiv:2305.07922}} (\bibinfo{year}{2023}).
\newblock


\bibitem[Wang et~al\mbox{.}(2021)]%
        {wang2021codet5}
\bibfield{author}{\bibinfo{person}{Yue Wang}, \bibinfo{person}{Weishi Wang}, \bibinfo{person}{Shafiq Joty}, {and} \bibinfo{person}{Steven~CH Hoi}.} \bibinfo{year}{2021}\natexlab{}.
\newblock \showarticletitle{CodeT5: Identifier-aware Unified Pre-trained Encoder-Decoder Models for Code Understanding and Generation}. In \bibinfo{booktitle}{\emph{Proceedings of the 2021 Conference on Empirical Methods in Natural Language Processing}}. \bibinfo{pages}{8696--8708}.
\newblock


\bibitem[Wang et~al\mbox{.}(2023a)]%
        {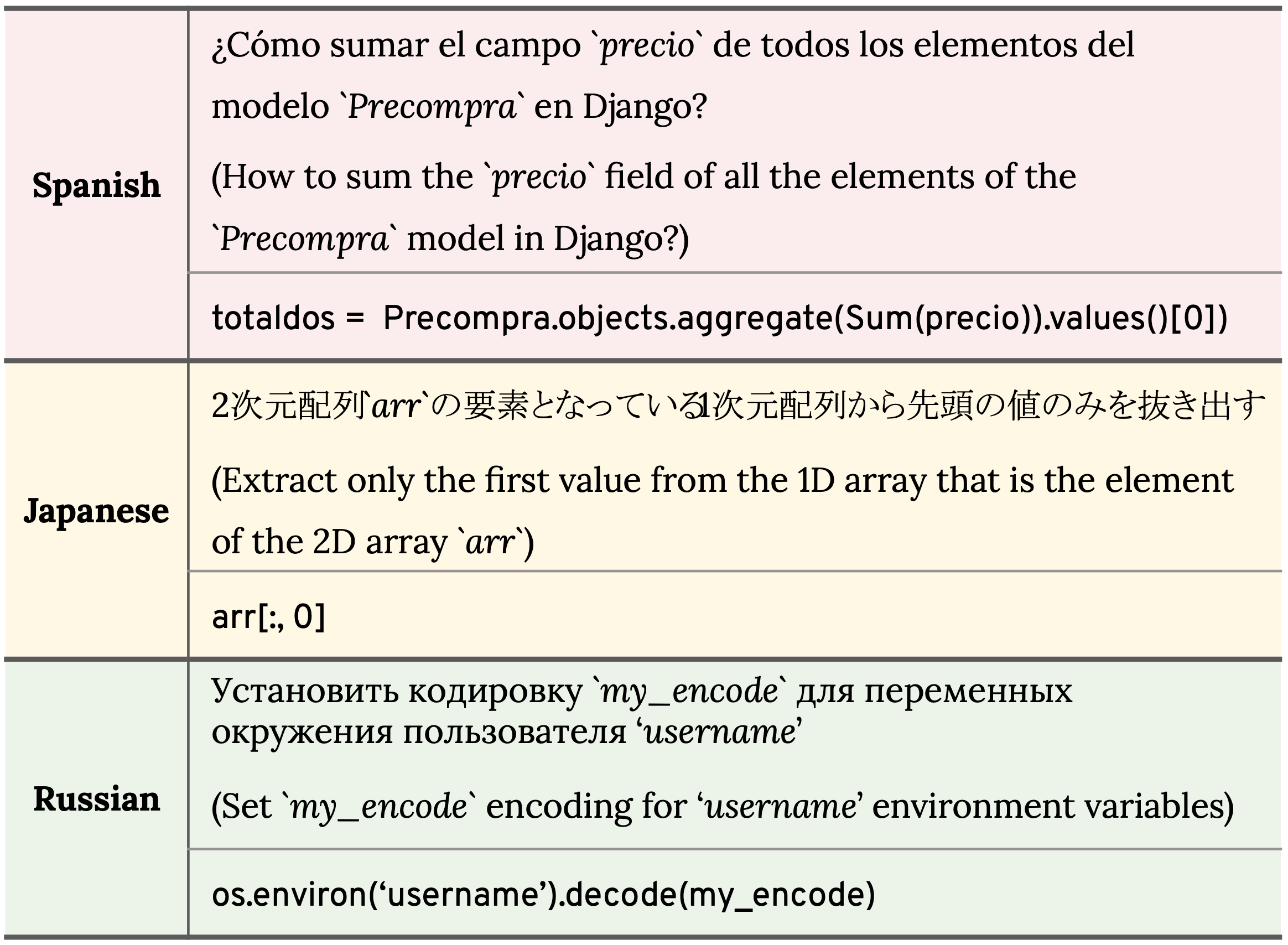}
\bibfield{author}{\bibinfo{person}{Zhiruo Wang}, \bibinfo{person}{Grace Cuenca}, \bibinfo{person}{Shuyan Zhou}, \bibinfo{person}{Frank~F. Xu}, {and} \bibinfo{person}{Graham Neubig}.} \bibinfo{year}{2023}\natexlab{a}.
\newblock \bibinfo{title}{MCoNaLa: A Benchmark for Code Generation from Multiple Natural Languages}.
\newblock
\newblock
\showeprint[arxiv]{2203.08388}~[cs.CL]


\bibitem[Wang et~al\mbox{.}(2023c)]%
        {wang2023chatcoder}
\bibfield{author}{\bibinfo{person}{Zejun Wang}, \bibinfo{person}{Jia Li}, \bibinfo{person}{Ge Li}, {and} \bibinfo{person}{Zhi Jin}.} \bibinfo{year}{2023}\natexlab{c}.
\newblock \showarticletitle{ChatCoder: Chat-based Refine Requirement Improves LLMs' Code Generation}.
\newblock \bibinfo{journal}{\emph{arXiv preprint arXiv:2311.00272}} (\bibinfo{year}{2023}).
\newblock


\bibitem[Waseem et~al\mbox{.}(2023)]%
        {waseem2023using}
\bibfield{author}{\bibinfo{person}{Muhammad Waseem}, \bibinfo{person}{Teerath Das}, \bibinfo{person}{Aakash Ahmad}, \bibinfo{person}{Mahdi Fehmideh}, \bibinfo{person}{Peng Liang}, {and} \bibinfo{person}{Tommi Mikkonen}.} \bibinfo{year}{2023}\natexlab{}.
\newblock \showarticletitle{Using ChatGPT throughout the Software Development Life Cycle by Novice Developers}.
\newblock \bibinfo{journal}{\emph{arXiv preprint arXiv:2310.13648}} (\bibinfo{year}{2023}).
\newblock


\bibitem[Watson et~al\mbox{.}(2020)]%
        {watson2020learning}
\bibfield{author}{\bibinfo{person}{Cody Watson}, \bibinfo{person}{Michele Tufano}, \bibinfo{person}{Kevin Moran}, \bibinfo{person}{Gabriele Bavota}, {and} \bibinfo{person}{Denys Poshyvanyk}.} \bibinfo{year}{2020}\natexlab{}.
\newblock \showarticletitle{On learning meaningful assert statements for unit test cases}. In \bibinfo{booktitle}{\emph{Proceedings of the ACM/IEEE 42nd International Conference on Software Engineering}}. \bibinfo{pages}{1398--1409}.
\newblock


\bibitem[Weisz et~al\mbox{.}(2022)]%
        {10.1145/3490099.3511157}
\bibfield{author}{\bibinfo{person}{Justin~D. Weisz}, \bibinfo{person}{Michael Muller}, \bibinfo{person}{Steven~I. Ross}, \bibinfo{person}{Fernando Martinez}, \bibinfo{person}{Stephanie Houde}, \bibinfo{person}{Mayank Agarwal}, \bibinfo{person}{Kartik Talamadupula}, {and} \bibinfo{person}{John~T. Richards}.} \bibinfo{year}{2022}\natexlab{}.
\newblock \showarticletitle{Better Together? An Evaluation of AI-Supported Code Translation}. In \bibinfo{booktitle}{\emph{27th International Conference on Intelligent User Interfaces}} (Helsinki, Finland) \emph{(\bibinfo{series}{IUI '22})}. \bibinfo{publisher}{Association for Computing Machinery}, \bibinfo{address}{New York, NY, USA}, \bibinfo{pages}{369–391}.
\newblock
\showISBNx{9781450391443}
\urldef\tempurl%
\url{https://doi.org/10.1145/3490099.3511157}
\showDOI{\tempurl}


\bibitem[White et~al\mbox{.}(2023)]%
        {white2023chatgpt}
\bibfield{author}{\bibinfo{person}{Jules White}, \bibinfo{person}{Sam Hays}, \bibinfo{person}{Quchen Fu}, \bibinfo{person}{Jesse Spencer-Smith}, {and} \bibinfo{person}{Douglas~C Schmidt}.} \bibinfo{year}{2023}\natexlab{}.
\newblock \showarticletitle{Chatgpt prompt patterns for improving code quality, refactoring, requirements elicitation, and software design}.
\newblock \bibinfo{journal}{\emph{arXiv preprint arXiv:2303.07839}} (\bibinfo{year}{2023}).
\newblock


\bibitem[White and Krinke(2020)]%
        {white2020reassert}
\bibfield{author}{\bibinfo{person}{Robert White} {and} \bibinfo{person}{Jens Krinke}.} \bibinfo{year}{2020}\natexlab{}.
\newblock \showarticletitle{Reassert: Deep learning for assert generation}.
\newblock \bibinfo{journal}{\emph{arXiv preprint arXiv:2011.09784}} (\bibinfo{year}{2020}).
\newblock


\bibitem[Xia et~al\mbox{.}(2023)]%
        {xia2023automated}
\bibfield{author}{\bibinfo{person}{Chunqiu~Steven Xia}, \bibinfo{person}{Yuxiang Wei}, {and} \bibinfo{person}{Lingming Zhang}.} \bibinfo{year}{2023}\natexlab{}.
\newblock \showarticletitle{Automated program repair in the era of large pre-trained language models}. In \bibinfo{booktitle}{\emph{Proceedings of the 45th International Conference on Software Engineering (ICSE 2023). Association for Computing Machinery}}.
\newblock


\bibitem[Xia and Zhang(2022)]%
        {xia2022less}
\bibfield{author}{\bibinfo{person}{Chunqiu~Steven Xia} {and} \bibinfo{person}{Lingming Zhang}.} \bibinfo{year}{2022}\natexlab{}.
\newblock \showarticletitle{Less training, more repairing please: revisiting automated program repair via zero-shot learning}. In \bibinfo{booktitle}{\emph{Proceedings of the 30th ACM Joint European Software Engineering Conference and Symposium on the Foundations of Software Engineering}}. \bibinfo{pages}{959--971}.
\newblock


\bibitem[Xia and Zhang(2023)]%
        {xia2023keep}
\bibfield{author}{\bibinfo{person}{Chunqiu~Steven Xia} {and} \bibinfo{person}{Lingming Zhang}.} \bibinfo{year}{2023}\natexlab{}.
\newblock \showarticletitle{Keep the Conversation Going: Fixing 162 out of 337 bugs for \$0.42 each using ChatGPT}.
\newblock \bibinfo{journal}{\emph{arXiv preprint arXiv:2304.00385}} (\bibinfo{year}{2023}).
\newblock


\bibitem[Xia et~al\mbox{.}(2018)]%
        {xia2017measuring}
\bibfield{author}{\bibinfo{person}{Xin Xia}, \bibinfo{person}{Lingfeng Bao}, \bibinfo{person}{David Lo}, \bibinfo{person}{Zhenchang Xing}, \bibinfo{person}{Ahmed~E. Hassan}, {and} \bibinfo{person}{Shanping Li}.} \bibinfo{year}{2018}\natexlab{}.
\newblock \showarticletitle{Measuring Program Comprehension: {A} Large-Scale Field Study with Professionals}.
\newblock \bibinfo{journal}{\emph{{IEEE} Trans. Software Eng.}} \bibinfo{volume}{44}, \bibinfo{number}{10} (\bibinfo{year}{2018}), \bibinfo{pages}{951--976}.
\newblock


\bibitem[Yan et~al\mbox{.}(2023a)]%
        {yanrepair}
\bibfield{author}{\bibinfo{person}{Dapeng Yan}, \bibinfo{person}{Zhipeng Gao}, {and} \bibinfo{person}{Zhiming Liu}.} \bibinfo{year}{2023}\natexlab{a}.
\newblock \showarticletitle{A Closer Look at Different Difficulty Levels Code Generation Abilities of ChatGPT}. In \bibinfo{booktitle}{\emph{2023 38th IEEE/ACM International Conference on Automated Software Engineering (ASE)}}. \bibinfo{pages}{1887--1898}.
\newblock
\urldef\tempurl%
\url{https://doi.org/10.1109/ASE56229.2023.00096}
\showDOI{\tempurl}


\bibitem[Yan et~al\mbox{.}(2023c)]%
        {yan2023codescope}
\bibfield{author}{\bibinfo{person}{Weixiang Yan}, \bibinfo{person}{Haitian Liu}, \bibinfo{person}{Yunkun Wang}, \bibinfo{person}{Yunzhe Li}, \bibinfo{person}{Qian Chen}, \bibinfo{person}{Wen Wang}, \bibinfo{person}{Tingyu Lin}, \bibinfo{person}{Weishan Zhao}, \bibinfo{person}{Li Zhu}, \bibinfo{person}{Shuiguang Deng}, {et~al\mbox{.}}} \bibinfo{year}{2023}\natexlab{c}.
\newblock \showarticletitle{CodeScope: An Execution-based Multilingual Multitask Multidimensional Benchmark for Evaluating LLMs on Code Understanding and Generation}.
\newblock \bibinfo{journal}{\emph{arXiv preprint arXiv:2311.08588}} (\bibinfo{year}{2023}).
\newblock


\bibitem[Yan et~al\mbox{.}(2023b)]%
        {codescope}
\bibfield{author}{\bibinfo{person}{Weixiang Yan}, \bibinfo{person}{Haitian Liu}, \bibinfo{person}{Yunkun Wang}, \bibinfo{person}{Yunzhe Li}, \bibinfo{person}{Qian Chen}, \bibinfo{person}{Wen Wang}, \bibinfo{person}{Tingyu Lin}, \bibinfo{person}{Weishan Zhao}, \bibinfo{person}{Li Zhu}, \bibinfo{person}{Shuiguang Deng}, {and} \bibinfo{person}{Hari Sundaram}.} \bibinfo{year}{2023}\natexlab{b}.
\newblock \bibinfo{title}{CodeScope: An Execution-based Multilingual Multitask Multidimensional Benchmark for Evaluating LLMs on Code Understanding and Generation}.
\newblock
\newblock
\showeprint[arxiv]{2311.08588}~[cs.CL]


\bibitem[Yan et~al\mbox{.}(2023d)]%
        {yan2023codetransocean}
\bibfield{author}{\bibinfo{person}{Weixiang Yan}, \bibinfo{person}{Yuchen Tian}, \bibinfo{person}{Yunzhe Li}, \bibinfo{person}{Qian Chen}, {and} \bibinfo{person}{Wen Wang}.} \bibinfo{year}{2023}\natexlab{d}.
\newblock \showarticletitle{CodeTransOcean: A Comprehensive Multilingual Benchmark for Code Translation}.
\newblock \bibinfo{journal}{\emph{arXiv preprint arXiv:2310.04951}} (\bibinfo{year}{2023}).
\newblock
\urldef\tempurl%
\url{https://doi.org/10.48550/arXiv.2310.04951}
\showURL{%
\tempurl}


\bibitem[Yang et~al\mbox{.}(2023)]%
        {yang2023assessing}
\bibfield{author}{\bibinfo{person}{Guang Yang}, \bibinfo{person}{Yu Zhou}, \bibinfo{person}{Xiangyu Zhang}, \bibinfo{person}{Xiang Chen}, \bibinfo{person}{Tingting Han}, {and} \bibinfo{person}{Taolue Chen}.} \bibinfo{year}{2023}\natexlab{}.
\newblock \showarticletitle{Assessing and Improving Syntactic Adversarial Robustness of Pre-trained Models for Code Translation}.
\newblock \bibinfo{journal}{\emph{arXiv preprint arXiv:2310.18587}} (\bibinfo{year}{2023}).
\newblock
\urldef\tempurl%
\url{https://doi.org/10.48550/arXiv.2310.18587}
\showURL{%
\tempurl}


\bibitem[Ye et~al\mbox{.}(2022)]%
        {ye2022neural}
\bibfield{author}{\bibinfo{person}{He Ye}, \bibinfo{person}{Matias Martinez}, {and} \bibinfo{person}{Martin Monperrus}.} \bibinfo{year}{2022}\natexlab{}.
\newblock \showarticletitle{Neural program repair with execution-based backpropagation}. In \bibinfo{booktitle}{\emph{Proceedings of the 44th International Conference on Software Engineering}}. \bibinfo{pages}{1506--1518}.
\newblock


\bibitem[Yu et~al\mbox{.}(2023)]%
        {yu2023llm}
\bibfield{author}{\bibinfo{person}{Shengcheng Yu}, \bibinfo{person}{Chunrong Fang}, \bibinfo{person}{Yuchen Ling}, \bibinfo{person}{Chentian Wu}, {and} \bibinfo{person}{Zhenyu Chen}.} \bibinfo{year}{2023}\natexlab{}.
\newblock \showarticletitle{Llm for test script generation and migration: Challenges, capabilities, and opportunities}.
\newblock \bibinfo{journal}{\emph{arXiv preprint arXiv:2309.13574}} (\bibinfo{year}{2023}).
\newblock


\bibitem[Yuan et~al\mbox{.}(2022)]%
        {yuan2022circle}
\bibfield{author}{\bibinfo{person}{Wei Yuan}, \bibinfo{person}{Quanjun Zhang}, \bibinfo{person}{Tieke He}, \bibinfo{person}{Chunrong Fang}, \bibinfo{person}{Nguyen Quoc~Viet Hung}, \bibinfo{person}{Xiaodong Hao}, {and} \bibinfo{person}{Hongzhi Yin}.} \bibinfo{year}{2022}\natexlab{}.
\newblock \showarticletitle{CIRCLE: Continual repair across programming languages}. In \bibinfo{booktitle}{\emph{Proceedings of the 31st ACM SIGSOFT International Symposium on Software Testing and Analysis}}. \bibinfo{pages}{678--690}.
\newblock


\bibitem[Zan et~al\mbox{.}(2023)]%
        {zan-etal-2023-large}
\bibfield{author}{\bibinfo{person}{Daoguang Zan}, \bibinfo{person}{Bei Chen}, \bibinfo{person}{Fengji Zhang}, \bibinfo{person}{Dianjie Lu}, \bibinfo{person}{Bingchao Wu}, \bibinfo{person}{Bei Guan}, \bibinfo{person}{Wang Yongji}, {and} \bibinfo{person}{Jian-Guang Lou}.} \bibinfo{year}{2023}\natexlab{}.
\newblock \showarticletitle{Large Language Models Meet {NL}2{C}ode: A Survey}. In \bibinfo{booktitle}{\emph{Proceedings of the 61st Annual Meeting of the Association for Computational Linguistics (Volume 1: Long Papers)}}, \bibfield{editor}{\bibinfo{person}{Anna Rogers}, \bibinfo{person}{Jordan Boyd-Graber}, {and} \bibinfo{person}{Naoaki Okazaki}} (Eds.). \bibinfo{publisher}{Association for Computational Linguistics}, \bibinfo{address}{Toronto, Canada}, \bibinfo{pages}{7443--7464}.
\newblock
\urldef\tempurl%
\url{https://doi.org/10.18653/v1/2023.acl-long.411}
\showDOI{\tempurl}


\bibitem[Zhang et~al\mbox{.}({[n.\,d.]})]%
        {zhang4450322evaluation}
\bibfield{author}{\bibinfo{person}{Jianzhang Zhang}, \bibinfo{person}{Yiyang Chen}, \bibinfo{person}{Nan Niu}, {and} \bibinfo{person}{Chuang Liu}.} \bibinfo{year}{[n.\,d.]}\natexlab{}.
\newblock \showarticletitle{Evaluation of Chatgpt on Requirements Information Retrieval Under Zero-Shot Setting}.
\newblock \bibinfo{journal}{\emph{Available at SSRN 4450322}} (\bibinfo{year}{[n.\,d.]}).
\newblock


\bibitem[Zhang et~al\mbox{.}(2023b)]%
        {selfedit}
\bibfield{author}{\bibinfo{person}{Kechi Zhang}, \bibinfo{person}{Zhuo Li}, \bibinfo{person}{Jia Li}, \bibinfo{person}{Ge Li}, {and} \bibinfo{person}{Zhi Jin}.} \bibinfo{year}{2023}\natexlab{b}.
\newblock \showarticletitle{Self-Edit: Fault-Aware Code Editor for Code Generation}. In \bibinfo{booktitle}{\emph{Proceedings of the 61st Annual Meeting of the Association for Computational Linguistics (Volume 1: Long Papers)}}, \bibfield{editor}{\bibinfo{person}{Anna Rogers}, \bibinfo{person}{Jordan Boyd-Graber}, {and} \bibinfo{person}{Naoaki Okazaki}} (Eds.). \bibinfo{publisher}{Association for Computational Linguistics}, \bibinfo{address}{Toronto, Canada}, \bibinfo{pages}{769--787}.
\newblock
\urldef\tempurl%
\url{https://doi.org/10.18653/v1/2023.acl-long.45}
\showDOI{\tempurl}


\bibitem[Zhang et~al\mbox{.}(2023c)]%
        {zhang2023algo}
\bibfield{author}{\bibinfo{person}{Kexun Zhang}, \bibinfo{person}{Danqing Wang}, \bibinfo{person}{Jingtao Xia}, \bibinfo{person}{William~Yang Wang}, {and} \bibinfo{person}{Lei Li}.} \bibinfo{year}{2023}\natexlab{c}.
\newblock \showarticletitle{ALGO: Synthesizing Algorithmic Programs with Generated Oracle Verifiers}.
\newblock \bibinfo{journal}{\emph{arXiv preprint arXiv:2305.14591}} (\bibinfo{year}{2023}).
\newblock


\bibitem[Zhang et~al\mbox{.}(2023a)]%
        {zhang2023gamma}
\bibfield{author}{\bibinfo{person}{Quanjun Zhang}, \bibinfo{person}{Chunrong Fang}, \bibinfo{person}{Tongke Zhang}, \bibinfo{person}{Bowen Yu}, \bibinfo{person}{Weisong Sun}, {and} \bibinfo{person}{Zhenyu Chen}.} \bibinfo{year}{2023}\natexlab{a}.
\newblock \showarticletitle{Gamma: Revisiting Template-Based Automated Program Repair Via Mask Prediction}. In \bibinfo{booktitle}{\emph{2023 38th IEEE/ACM International Conference on Automated Software Engineering (ASE)}}. IEEE Computer Society, \bibinfo{pages}{535--547}.
\newblock


\bibitem[Zheng et~al\mbox{.}(2023a)]%
        {zheng2023understanding}
\bibfield{author}{\bibinfo{person}{Zibin Zheng}, \bibinfo{person}{Kaiwen Ning}, \bibinfo{person}{Jiachi Chen}, \bibinfo{person}{Yanlin Wang}, \bibinfo{person}{Wenqing Chen}, \bibinfo{person}{Lianghong Guo}, {and} \bibinfo{person}{Weicheng Wang}.} \bibinfo{year}{2023}\natexlab{a}.
\newblock \bibinfo{title}{Towards an Understanding of Large Language Models in Software Engineering Tasks}.
\newblock
\newblock
\showeprint[arxiv]{2308.11396}~[cs.SE]


\bibitem[Zheng et~al\mbox{.}(2023b)]%
        {zheng2023survey}
\bibfield{author}{\bibinfo{person}{Zibin Zheng}, \bibinfo{person}{Kaiwen Ning}, \bibinfo{person}{Yanlin Wang}, \bibinfo{person}{Jingwen Zhang}, \bibinfo{person}{Dewu Zheng}, \bibinfo{person}{Mingxi Ye}, {and} \bibinfo{person}{Jiachi Chen}.} \bibinfo{year}{2023}\natexlab{b}.
\newblock \bibinfo{title}{A Survey of Large Language Models for Code: Evolution, Benchmarking, and Future Trends}.
\newblock
\newblock
\showeprint[arxiv]{2311.10372}~[cs.SE]


\bibitem[Zhu et~al\mbox{.}(2022)]%
        {zhu2022xlcost}
\bibfield{author}{\bibinfo{person}{Ming Zhu}, \bibinfo{person}{Aneesh Jain}, \bibinfo{person}{Karthik Suresh}, \bibinfo{person}{Roshan Ravindran}, \bibinfo{person}{Sindhu Tipirneni}, {and} \bibinfo{person}{Chandan~K Reddy}.} \bibinfo{year}{2022}\natexlab{}.
\newblock \showarticletitle{Xlcost: A benchmark dataset for cross-lingual code intelligence}.
\newblock \bibinfo{journal}{\emph{arXiv preprint arXiv:2206.08474}} (\bibinfo{year}{2022}).
\newblock
\urldef\tempurl%
\url{https://doi.org/10.48550/arXiv.2206.08474}
\showURL{%
\tempurl}


\bibitem[Zhu et~al\mbox{.}(2021)]%
        {zhu2021syntax}
\bibfield{author}{\bibinfo{person}{Qihao Zhu}, \bibinfo{person}{Zeyu Sun}, \bibinfo{person}{Yuan-an Xiao}, \bibinfo{person}{Wenjie Zhang}, \bibinfo{person}{Kang Yuan}, \bibinfo{person}{Yingfei Xiong}, {and} \bibinfo{person}{Lu Zhang}.} \bibinfo{year}{2021}\natexlab{}.
\newblock \showarticletitle{A syntax-guided edit decoder for neural program repair}. In \bibinfo{booktitle}{\emph{Proceedings of the 29th ACM Joint Meeting on European Software Engineering Conference and Symposium on the Foundations of Software Engineering}}. \bibinfo{pages}{341--353}.
\newblock


\end{thebibliography}
